%% file: main.tex
\RequirePackage{amsmath}
\documentclass{jfphack}
\usepackage{tikz-cd}
\pgfrealjobname{main}
\usepackage{quoting}
\usepackage{bigfoot} 
\usepackage{hyperref}
\usepackage[utf8]{inputenc}

\makeatletter
\@ifundefined{lhs2tex.lhs2tex.sty.read}%
  {\@namedef{lhs2tex.lhs2tex.sty.read}{}%
   \newcommand\SkipToFmtEnd{}%
   \newcommand\EndFmtInput{}%
   \long\def\SkipToFmtEnd#1\EndFmtInput{}%
  }\SkipToFmtEnd

\newcommand\ReadOnlyOnce[1]{\@ifundefined{#1}{\@namedef{#1}{}}\SkipToFmtEnd}
\usepackage{amstext}
\usepackage{amssymb}
\usepackage{stmaryrd}
\DeclareFontFamily{OT1}{cmtex}{}
\DeclareFontShape{OT1}{cmtex}{m}{n}
  {<5><6><7><8>cmtex8
   <9>cmtex9
   <10><10.95><12><14.4><17.28><20.74><24.88>cmtex10}{}
\DeclareFontShape{OT1}{cmtex}{m}{it}
  {<-> ssub * cmtt/m/it}{}

\DeclareFontShape{OT1}{cmtt}{bx}{n}
  {<5><6><7><8>cmtt8
   <9>cmbtt9
   <10><10.95><12><14.4><17.28><20.74><24.88>cmbtt10}{}
\DeclareFontShape{OT1}{cmtex}{bx}{n}
  {<-> ssub * cmtt/bx/n}{}

\newcommand{\Conid}[1]{\mathit{#1}}
\newcommand{\Varid}[1]{\mathit{#1}}
\newcommand{\anonymous}{\kern0.06em \vbox{\hrule\@width.5em}}

\newcommand{\bind}{\mathbin{>\!\!\!>\mkern-6.7mu=}}

\usepackage{polytable}

\@ifundefined{mathindent}%
  {\newdimen\mathindent\mathindent\leftmargini}%
  {}%

\def\resethooks{%
  \global\let\SaveRestoreHook\empty
  \global\let\ColumnHook\empty}
\newcommand*{\savecolumns}[1][default]%
  {\g@addto@macro\SaveRestoreHook{\savecolumns[#1]}}
\newcommand*{\restorecolumns}[1][default]%
  {\g@addto@macro\SaveRestoreHook{\restorecolumns[#1]}}
\newcommand*{\aligncolumn}[2]%
  {\g@addto@macro\ColumnHook{\column{#1}{#2}}}

\resethooks

\newcommand{\onelinecommentchars}{\quad-{}- }
\newcommand{\commentbeginchars}{\enskip\{-}
\newcommand{\commentendchars}{-\}\enskip}

\newcommand{\visiblecomments}{%
  \let\onelinecomment=\onelinecommentchars
  \let\commentbegin=\commentbeginchars
  \let\commentend=\commentendchars}

\newcommand{\invisiblecomments}{%
  \let\onelinecomment=\empty
  \let\commentbegin=\empty
  \let\commentend=\empty}

\visiblecomments

\newlength{\blanklineskip}
\newcommand{\hsindent}[1]{\quad}
\let\hspre\empty
\let\hspost\empty

\EndFmtInput
\makeatother
\ReadOnlyOnce{polycode.fmt}%
\makeatletter

\newcommand{\hsnewpar}[1]%
  {{\parskip=0pt\parindent=0pt\par\vskip #1\noindent}}

\newcommand{\hscodestyle}{}

\newcommand{\sethscode}[1]%
  {\expandafter\let\expandafter\hscode\csname #1\endcsname
   \expandafter\let\expandafter\endhscode\csname end#1\endcsname}

  {\par\noindent
   \advance\leftskip\mathindent
   \hscodestyle
   \let\\=\@normalcr
   \let\hspre\(\let\hspost\)%
   \pboxed}%
  {\endpboxed\)%
   \par\noindent
   \ignorespacesafterend}

  {\hsnewpar\abovedisplayskip
   \advance\leftskip\mathindent
   \hscodestyle
   \let\hspre\(\let\hspost\)%
   \pboxed}%
  {\endpboxed%
   \hsnewpar\belowdisplayskip
   \ignorespacesafterend}

  {\hsnewpar\abovedisplayskip
   \advance\leftskip\mathindent
   \hscodestyle
   \let\\=\@normalcr
   \(\pboxed}%
  {\endpboxed\)%
   \hsnewpar\belowdisplayskip
   \ignorespacesafterend}

\newcommand{\plainhs}{\sethscode{plainhscode}}

\plainhs

  {\hsnewpar\abovedisplayskip
   \advance\leftskip\mathindent
   \hscodestyle
   \let\\=\@normalcr
   \(\parray}%
  {\endparray\)%
   \hsnewpar\belowdisplayskip
   \ignorespacesafterend}

  {\parray}{\endparray}

  {\(\parray}{\endparray\)}

\def\codeframewidth{\arrayrulewidth}
\RequirePackage{calc}

  {\parskip=\abovedisplayskip\par\noindent
   \hscodestyle
   \arrayrulewidth=\codeframewidth
   \tabular{@{}|p{\linewidth-2\arraycolsep-2\arrayrulewidth-2pt}|@{}}%
   \hline\framedhslinecorrect\\{-1.5ex}%
   \let\endoflinesave=\\
   \let\\=\@normalcr
   \(\pboxed}%
  {\endpboxed\)%
   \framedhslinecorrect\endoflinesave{.5ex}\hline
   \endtabular
   \parskip=\belowdisplayskip\par\noindent
   \ignorespacesafterend}

\newcommand{\framedhslinecorrect}[2]%
  {#1[#2]}

  {\(\def\column##1##2{}%
   \let\>\undefined\let\<\undefined\let\\\undefined
   \newcommand\>[1][]{}\newcommand\<[1][]{}\newcommand\\[1][]{}%
   \def\fromto##1##2##3{##3}%
   }{\) }%

\newenvironment{joincode}%
  {\let\orighscode=\hscode
   \let\origendhscode=\endhscode
   \def\endhscode{\def\hscode{\endgroup\def\@currenvir{hscode}\\}\begingroup}
   \orighscode\def\hscode{\endgroup\def\@currenvir{hscode}}}%
  {\origendhscode
   \global\let\hscode=\orighscode
   \global\let\endhscode=\origendhscode}%

\makeatother
\EndFmtInput
\def\doubleequals{\mathrel{\unitlength 0.01em
  \begin{picture}(78,40)
    \put(7,34){\line(1,0){25}} \put(45,34){\line(1,0){25}}
    \put(7,14){\line(1,0){25}} \put(45,14){\line(1,0){25}}
  \end{picture}}}

\input{macros.TeX}

\begin{document}

\journaltitle{JFP}
\cpr{The Author(s),}
\doival{10.1017/xxxxx}

\newcommand{\shorttitle}{Extensional equality preservation and verified generic programming}
\lefttitle{Botta, Brede, Jansson and Richter}
\righttitle{\shorttitle}
\totalpg{\pageref{lastpage01}}
\jnlDoiYr{2021}

\title[\shorttitle]{Extensional equality preservation and\\verified generic programming}

\begin{authgrp}
\author{NICOLA BOTTA}
\affiliation{Potsdam Institute for Climate Impact Research, Potsdam, Germany, \\
  Chalmers University of Technology, Göteborg, Sweden.
  (\email{botta@pik-potsdam.de})}
\author{NURIA BREDE}
\affiliation{Potsdam Institute for Climate Impact Research, Potsdam,
  Germany.
  (\email{nubrede@pik-potsdam.de})}
\author{PATRIK JANSSON}
\affiliation{Chalmers University of Technology, Göteborg, Sweden.
  (\email{patrikj@chalmers.se})}
\author{TIM RICHTER}
\affiliation{Potsdam University, Potsdam, Germany.
  (\email{tim.richter@uni-potsdam.de})}
\end{authgrp}

\begin{abstract}
%
In verified generic programming, one cannot exploit the structure of
concrete data types but has to rely on \emph{well chosen} sets of
specifications or abstract data types (ADTs).
Functors and monads are at the core of many applications of functional
programming. This raises the question of what useful ADTs for
verified functors and monads could look like.
The functorial map of many important monads preserves extensional
equality. For instance, if \ensuremath{\Varid{f},\Varid{g}\ \mathop{:}\ \Conid{A}\,\to\,\Conid{B}} are extensionally equal, that
is, \ensuremath{\forall \Varid{x}\in \Conid{A},\phantom{x} \Varid{f}\;\Varid{x}\mathrel{=}\Varid{g}\;\Varid{x}}, then \ensuremath{\Varid{map}\;\Varid{f}\ \mathop{:}\ \Conid{List}\;\Conid{A}\,\to\,\Conid{List}\;\Conid{B}} and \ensuremath{\Varid{map}\;\Varid{g}}
are also extensionally equal.
This suggests that \emph{preservation of extensional equality} could be
a useful principle in verified generic programming.
We explore this possibility with a minimalist approach: we deal with
(the lack of) extensional equality in Martin-Löf's intensional type
theories without extending the theories or using full-fledged setoids.
Perhaps surprisingly, this minimal approach turns out to be extremely
useful. It allows one to derive simple generic proofs of monadic laws
but also verified, generic results in dynamical systems and control
theory.
In turn, these results avoid tedious code duplication and ad-hoc
proofs. Thus, our work is a contribution towards pragmatic, verified
generic programming.
\end{abstract}

\maketitle


\setlength{\mathindent}{1em}
\newcommand{\fixlengths}{\setlength{\abovedisplayskip}{6pt plus 1pt minus 1pt}\setlength{\belowdisplayskip}{6pt plus 1pt minus 1pt}}
\renewcommand{\hscodestyle}{\small\fixlengths}
\fixlengths


\section{Introduction}
\label{section:about}

This paper is about \emph{extensional equality preservation} in
dependently typed languages like Idris~\citep{idrisbook}, 
Agda~\citep{norell2007thesis} and Coq~\citep{CoqProofAssistant} that
implement 
Martin-Löf's intensional type
theory \citep{martinlof1984}. 
%
We discuss Idris code but the results could be translated to other
languages.
Extensional equality is a property of functions, stating that they are
``pointwise equal'':
%
\begin{hscode}\SaveRestoreHook
\column{B}{@{}>{\hspre}l<{\hspost}@{}}%
\column{3}{@{}>{\hspre}l<{\hspost}@{}}%
\column{E}{@{}>{\hspre}l<{\hspost}@{}}%
\>[3]{}(\doteq)\ \mathop{:}\ \{\mskip1.5mu \Conid{A},\Conid{B}\ \mathop{:}\ \Conid{Type}\mskip1.5mu\}\,\to\,(\Conid{A}\,\to\,\Conid{B})\,\to\,(\Conid{A}\,\to\,\Conid{B})\,\to\,\Conid{Type}{}\<[E]%
\\
\>[3]{}(\doteq)\;\{\mskip1.5mu \Conid{A}\mskip1.5mu\}\;\Varid{f}\;\Varid{g}\mathrel{=}(\Varid{x}\ \mathop{:}\ \Conid{A})\,\to\,\Varid{f}\;\Varid{x}\mathrel{=}\Varid{g}\;\Varid{x}{}\<[E]%
\ColumnHook
\end{hscode}\resethooks
Note that the definition of extensional equality \ensuremath{(\doteq)} depends on another equality \ensuremath{(\mathrel{=})}.


\paragraph*{Different flavours of equality.}
\begin{quote}
  ``All animals are equal, but some animals are more equal than others'' [Animal Farm, Orwell (1946)]
\end{quote}

There are several kinds of ``equality'' relevant for programming.
Programming languages usually offer a Boolean equality check operator and in Idris it is written \ensuremath{(\doubleequals)}, has type \ensuremath{\{\mskip1.5mu \Conid{A}\ \mathop{:}\ \Conid{Type}\mskip1.5mu\}\rightarrow\Conid{Eq}\;\Conid{A}\Rightarrow \Conid{A}\rightarrow\Conid{A}\rightarrow\Conid{Bool}} and is packaged in the interface \ensuremath{\Conid{Eq}}.
This is an ``ad-hoc'' equality, computing whatever the programmer supplies as an implementation.
This paper is not about value level Boolean equality.

On the type level, the dependently typed languages we consider in this paper provide
a notion of \emph{intensional equality}, also referred to as an ``equality type'',
which is an inductively defined family of types, usually written infix: \ensuremath{(\Varid{a}\mathrel{=}\Varid{b})\ \mathop{:}\ \Conid{Type}} for \ensuremath{\Varid{a}\ \mathop{:}\ \Conid{A}} and \ensuremath{\Varid{b}\ \mathop{:}\ \Conid{B}}. It has
just one constructor \ensuremath{\Conid{Refl}\ \mathop{:}\ \Varid{a}\mathrel{=}\Varid{a}}.
The resulting notion is not as boring as it may look at first. We have \ensuremath{\Conid{Refl}\ \mathop{:}\ \Varid{a}\mathrel{=}\Varid{b}} not only if \ensuremath{\Varid{a}} and \ensuremath{\Varid{b}} are identical, but
also if they \emph{reduce} to identical expressions. Builtin reduction rules normally include alpha-conversion (capture-free
renaming of bound variables), beta-reduction (using substitution) and eta-reduction: \ensuremath{\Varid{f}\mathrel{=}\lambda \Varid{x}\Rightarrow \Varid{f}\;\Varid{x}}.
So, for example, we have \ensuremath{\Conid{Refl}\ \mathop{:}\ \Varid{id}\;\Varid{x}\mathrel{=}\Varid{x}}.
Furthermore, user-defined equations are also used for reduction.
A typical example is addition of natural numbers: with \ensuremath{\mathbin{+}} defined by pattern matching on the first argument, we have e.g.
\ensuremath{\Conid{Refl}\ \mathop{:}\ \mathrm{1}\mathbin{+}\mathrm{1}\mathrel{=}\mathrm{2}}. However, while for a variable \ensuremath{\Varid{n}\ \mathop{:}\ \mathbb{N}} we have \ensuremath{\Conid{Refl}\ \mathop{:}\ \mathrm{0}\mathbin{+}\Varid{n}\mathrel{=}\Varid{n}}, we do not have \ensuremath{\Conid{Refl}\ \mathop{:}\ \Varid{n}\mathbin{+}\mathrm{0}\mathrel{=}\Varid{n}}.

One very useful property of intensional equality is that it is a congruence with respect to any function.
In other words, all functions preserve intensional equality.
The proof uses pattern matching, which is straightforward here because
\ensuremath{\Conid{Refl}} is the only constructor:
\begin{hscode}\SaveRestoreHook
\column{B}{@{}>{\hspre}l<{\hspost}@{}}%
\column{3}{@{}>{\hspre}l<{\hspost}@{}}%
\column{11}{@{}>{\hspre}l<{\hspost}@{}}%
\column{E}{@{}>{\hspre}l<{\hspost}@{}}%
\>[3]{}\Varid{cong}\ \mathop{:}\ {}\<[11]%
\>[11]{}\{\mskip1.5mu \Conid{A},\Conid{B}\ \mathop{:}\ \Conid{Type}\mskip1.5mu\}\,\to\,\{\mskip1.5mu \Varid{f}\ \mathop{:}\ \Conid{A}\,\to\,\Conid{B}\mskip1.5mu\}\,\to\,\{\mskip1.5mu \Varid{a},\Varid{a'}\ \mathop{:}\ \Conid{A}\mskip1.5mu\}\,\to\,\Varid{a}\mathrel{=}\Varid{a'}\,\to\,\Varid{f}\;\Varid{a}\mathrel{=}\Varid{f}\;\Varid{a'}{}\<[E]%
\\
\>[3]{}\Varid{cong}\;\Conid{Refl}\mathrel{=}\Conid{Refl}{}\<[E]%
\ColumnHook
\end{hscode}\resethooks
In a similar way, one can prove that \ensuremath{(\mathrel{=})} is an equivalence relation:
reflexivity is directly implemented by \ensuremath{\Conid{Refl}}, while symmetry and
transitivity can be proven by pattern matching.

\paragraph*{Extensional equality.}

As one would expect, extensional equality is an equivalence relation
\begin{hscode}\SaveRestoreHook
\column{B}{@{}>{\hspre}l<{\hspost}@{}}%
\column{3}{@{}>{\hspre}l<{\hspost}@{}}%
\column{12}{@{}>{\hspre}l<{\hspost}@{}}%
\column{16}{@{}>{\hspre}l<{\hspost}@{}}%
\column{E}{@{}>{\hspre}l<{\hspost}@{}}%
\>[3]{}\Varid{reflEE}{}\<[12]%
\>[12]{}\ \mathop{:}\ \{\mskip1.5mu \Conid{A},\Conid{B}\ \mathop{:}\ \Conid{Type}\mskip1.5mu\}\,\to\,\{\mskip1.5mu \Varid{f}\ \mathop{:}\ \Conid{A}\,\to\,\Conid{B}\mskip1.5mu\}\,\to\,\Varid{f}\doteq\Varid{f}{}\<[E]%
\\
\>[3]{}\Varid{symEE}{}\<[12]%
\>[12]{}\ \mathop{:}\ \{\mskip1.5mu \Conid{A},\Conid{B}\ \mathop{:}\ \Conid{Type}\mskip1.5mu\}\,\to\,\{\mskip1.5mu \Varid{f},\Varid{g}\ \mathop{:}\ \Conid{A}\,\to\,\Conid{B}\mskip1.5mu\}\,\to\,\Varid{f}\doteq\Varid{g}\,\to\,\Varid{g}\doteq\Varid{f}{}\<[E]%
\\
\>[3]{}\Varid{transEE}{}\<[12]%
\>[12]{}\ \mathop{:}\ \{\mskip1.5mu \Conid{A},\Conid{B}\ \mathop{:}\ \Conid{Type}\mskip1.5mu\}\,\to\,\{\mskip1.5mu \Varid{f},\Varid{g},\Varid{h}\ \mathop{:}\ \Conid{A}\,\to\,\Conid{B}\mskip1.5mu\}\,\to\,\Varid{f}\doteq\Varid{g}\,\to\,\Varid{g}\doteq\Varid{h}\,\to\,\Varid{f}\doteq\Varid{h}{}\<[E]%
\\[\blanklineskip]%
\>[3]{}\Varid{reflEE}{}\<[16]%
\>[16]{}\mathrel{=}\lambda \Varid{x}\Rightarrow \Conid{Refl}{}\<[E]%
\\
\>[3]{}\Varid{symEE}\;\Varid{p}{}\<[16]%
\>[16]{}\mathrel{=}\lambda \Varid{x}\Rightarrow \Varid{sym}\;(\Varid{p}\;\Varid{x}){}\<[E]%
\\
\>[3]{}\Varid{transEE}\;\Varid{p}\;\Varid{q}{}\<[16]%
\>[16]{}\mathrel{=}\lambda \Varid{x}\Rightarrow \Varid{trans}\;(\Varid{p}\;\Varid{x})\;(\Varid{q}\;\Varid{x}){}\<[E]%
\ColumnHook
\end{hscode}\resethooks
In general, we can lift any (type-valued) binary relation on a type
\ensuremath{\Conid{B}} to a binary relation on function types with co-domain \ensuremath{\Conid{B}}.
%
\begin{hscode}\SaveRestoreHook
\column{B}{@{}>{\hspre}l<{\hspost}@{}}%
\column{3}{@{}>{\hspre}l<{\hspost}@{}}%
\column{E}{@{}>{\hspre}l<{\hspost}@{}}%
\>[3]{}\Varid{extify}\ \mathop{:}\ \{\mskip1.5mu \Conid{A},\Conid{B}\ \mathop{:}\ \Conid{Type}\mskip1.5mu\}\,\to\,(\Conid{B}\,\to\,\Conid{B}\,\to\,\Conid{Type})\,\to\,((\Conid{A}\,\to\,\Conid{B})\,\to\,(\Conid{A}\,\to\,\Conid{B})\,\to\,\Conid{Type}){}\<[E]%
\\
\>[3]{}\Varid{extify}\;\{\mskip1.5mu \Conid{A}\mskip1.5mu\}\;\Varid{relB}\;\Varid{f}\;\Varid{g}\mathrel{=}(\Varid{a}\ \mathop{:}\ \Conid{A})\,\to\,\Varid{relB}\;(\Varid{f}\;\Varid{a})\;(\Varid{g}\;\Varid{a}){}\<[E]%
\ColumnHook
\end{hscode}\resethooks
The \ensuremath{\Varid{extify}} combinator maps equivalence relations to equivalence relations.
Using it we can redefine \ensuremath{(\doteq)\mathrel{=}\Varid{extify}\;(\mathrel{=})} and we can easily continue to quantify
over more arguments: \ensuremath{(\stackrel{..}{=})\mathrel{=}\Varid{extify}\;(\doteq)}, etc.
In this paper our main focus is equality on functions, and we will explore in some detail the relationship between \ensuremath{\Varid{f}\mathrel{=}\Varid{g}} and \ensuremath{\Varid{f}\doteq\Varid{g}}.

In Martin-Löf's intensional type theory, and thus in Idris,
extensional equality is strictly weaker than intensional equality.
More concretely, we can implement
\begin{hscode}\SaveRestoreHook
\column{B}{@{}>{\hspre}l<{\hspost}@{}}%
\column{3}{@{}>{\hspre}l<{\hspost}@{}}%
\column{E}{@{}>{\hspre}l<{\hspost}@{}}%
\>[3]{}\Conid{IEqImplEE}\ \mathop{:}\ \{\mskip1.5mu \Conid{A},\Conid{B}\ \mathop{:}\ \Conid{Type}\mskip1.5mu\}\,\to\,(\Varid{f},\Varid{g}\ \mathop{:}\ \Conid{A}\,\to\,\Conid{B})\,\to\,\Varid{f}\mathrel{=}\Varid{g}\,\to\,\Varid{f}\doteq\Varid{g}{}\<[E]%
\\
\>[3]{}\Conid{IEqImplEE}\;\Varid{f}\;\Varid{f}\;\Conid{Refl}\mathrel{=}\lambda \Varid{x}\Rightarrow \Conid{Refl}{}\<[E]%
\ColumnHook
\end{hscode}\resethooks
but not the converse, normally referred to as \emph{function extensionality}:
\begin{hscode}\SaveRestoreHook
\column{B}{@{}>{\hspre}l<{\hspost}@{}}%
\column{3}{@{}>{\hspre}l<{\hspost}@{}}%
\column{69}{@{}>{\hspre}l<{\hspost}@{}}%
\column{E}{@{}>{\hspre}l<{\hspost}@{}}%
\>[3]{}\Conid{EEqImplIE}\ \mathop{:}\ \{\mskip1.5mu \Conid{A},\Conid{B}\ \mathop{:}\ \Conid{Type}\mskip1.5mu\}\,\to\,(\Varid{f},\Varid{g}\ \mathop{:}\ \Conid{A}\,\to\,\Conid{B})\,\to\,\Varid{f}\doteq\Varid{g}\,\to\,\Varid{f}\mathrel{=}\Varid{g}{}\<[69]%
\>[69]{}\mbox{\onelinecomment  not implementable}{}\<[E]%
\ColumnHook
\end{hscode}\resethooks
When working with functions, extensional equality is often the notion of
interest and libraries of formalized mathematics typically provide
definitions like \ensuremath{\doteq} and basic results like \ensuremath{\Conid{IEqImplEE}}.
See for example \ensuremath{\Varid{homot}} and \ensuremath{\Varid{eqtohomot}} in Part A of the UniMath
library \citep{UniMath} or \ensuremath{\Varid{eqfun}} in Coq \citep{CoqProofAssistant}.

In reasoning about generic programs in the style of the Algebra of
Programming \citep{DBLP:books/daglib/0096998,mu2009algebra} and, more
generally, in pen and paper proofs, the principle of function
extensionality is often taken for granted.

\paragraph*{EE preservation.}
Preservation of extensional equality is a property of higher order
functions: we say that, for fixed, non-function types \ensuremath{\Conid{A}}, \ensuremath{\Conid{B}}, \ensuremath{\Conid{C}}
and \ensuremath{\Conid{D}}, a function \ensuremath{\Varid{h}\ \mathop{:}\ (\Conid{A}\,\to\,\Conid{B})\,\to\,(\Conid{C}\,\to\,\Conid{D})} preserves extensional
equality (in one argument) if \ensuremath{\Varid{f}\doteq\Varid{g}} implies \ensuremath{\Varid{h}\;\Varid{f}\doteq\Varid{h}\;\Varid{g}}.

Higher order functions are a distinguished trait of functional
programming languages \citep{bird2014thinking} and many well known
function combinators can be shown to preserve extensional equality.
%
In particular, the arrow-function \ensuremath{\Varid{map}} for \ensuremath{\Conid{Identity}}, \ensuremath{\Conid{List}}, \ensuremath{\Conid{Maybe}}
and for many other polymorphic data types preserves extensional
equality. The standard libraries of Agda and Coq provide several
instances\footnote{e.g. Agda for \ensuremath{\Conid{Maybe}}: \url{https://agda.github.io/agda-stdlib/Data.Maybe.Properties.html}}
\footnote{e.g. Coq for \ensuremath{\Conid{List}}: \url{https://coq.inria.fr/distrib/current/stdlib/Coq.Lists.List.html}}.

Similarly, if \ensuremath{\Varid{h}} takes two function arguments it preserves
extensional equality (in two arguments) if \ensuremath{\Varid{f}_{1}\doteq\Varid{g}_{1}} and \ensuremath{\Varid{f}_{2}\doteq\Varid{g}_{2}}
implies \ensuremath{\Varid{h}\;\Varid{f}_{1}\;\Varid{f}_{2}\doteq\Varid{h}\;\Varid{g}_{1}\;\Varid{g}_{2}}, etc.
%
To illustrate the Idris notation for equational reasoning we show the
lemma \ensuremath{\Varid{compPresEE}} proving that function composition satisfies the
two-argument version of extensional equality preservation:
\begin{hscode}\SaveRestoreHook
\column{B}{@{}>{\hspre}l<{\hspost}@{}}%
\column{3}{@{}>{\hspre}l<{\hspost}@{}}%
\column{6}{@{}>{\hspre}c<{\hspost}@{}}%
\column{6E}{@{}l@{}}%
\column{9}{@{}>{\hspre}l<{\hspost}@{}}%
\column{15}{@{}>{\hspre}c<{\hspost}@{}}%
\column{15E}{@{}l@{}}%
\column{18}{@{}>{\hspre}l<{\hspost}@{}}%
\column{22}{@{}>{\hspre}l<{\hspost}@{}}%
\column{25}{@{}>{\hspre}l<{\hspost}@{}}%
\column{67}{@{}>{\hspre}c<{\hspost}@{}}%
\column{67E}{@{}l@{}}%
\column{E}{@{}>{\hspre}l<{\hspost}@{}}%
\>[3]{}\Varid{compPresEE}{}\<[15]%
\>[15]{}\ \mathop{:}\ {}\<[15E]%
\>[18]{}\{\mskip1.5mu \Conid{A},\Conid{B},\Conid{C}\ \mathop{:}\ \Conid{Type}\mskip1.5mu\}\,\to\,\{\mskip1.5mu \Varid{g},\Varid{g'}\ \mathop{:}\ \Conid{B}\,\to\,\Conid{C}\mskip1.5mu\}\,\to\,\{\mskip1.5mu \Varid{f},\Varid{f'}\ \mathop{:}\ \Conid{A}\,\to\,\Conid{B}\mskip1.5mu\}\,\to\,{}\<[E]%
\\
\>[18]{}\Varid{g}\doteq\Varid{g'}\,\to\,\Varid{f}\doteq\Varid{f'}\,\to\,\Varid{g}\mathbin{\circ}\Varid{f}\doteq\Varid{g'}\mathbin{\circ}\Varid{f'}{}\<[E]%
\\
\>[3]{}\Varid{compPresEE}\;\{\mskip1.5mu \Varid{g}\mskip1.5mu\}\;\{\mskip1.5mu \Varid{g'}\mskip1.5mu\}\;\{\mskip1.5mu \Varid{f}\mskip1.5mu\}\;\{\mskip1.5mu \Varid{f'}\mskip1.5mu\}\;\Varid{gExtEq}\;\ \Varid{fExtEq}\;\ \Varid{x}{}\<[67]%
\>[67]{}\mathrel{=}{}\<[67E]%
\\
\>[3]{}\hsindent{3}{}\<[6]%
\>[6]{}({}\<[6E]%
\>[9]{}(\Varid{g}\mathbin{\circ}\Varid{f})\;\Varid{x}{}\<[22]%
\>[22]{}){}\<[25]%
\>[25]{}=\hspace{-3pt}\{\; \Conid{Refl}\;\}\hspace{-3pt}={}\<[E]%
\\
\>[3]{}\hsindent{3}{}\<[6]%
\>[6]{}({}\<[6E]%
\>[9]{}\Varid{g}\;(\Varid{f}\;\Varid{x}){}\<[22]%
\>[22]{}){}\<[25]%
\>[25]{}=\hspace{-3pt}\{\; \Varid{cong}\;(\Varid{fExtEq}\;\Varid{x})\;\}\hspace{-3pt}={}\<[E]%
\\
\>[3]{}\hsindent{3}{}\<[6]%
\>[6]{}({}\<[6E]%
\>[9]{}\Varid{g}\;(\Varid{f'}\;\Varid{x}){}\<[22]%
\>[22]{}){}\<[25]%
\>[25]{}=\hspace{-3pt}\{\; \Varid{gExtEq}\;(\Varid{f'}\;\Varid{x})\;\}\hspace{-3pt}={}\<[E]%
\\
\>[3]{}\hsindent{3}{}\<[6]%
\>[6]{}({}\<[6E]%
\>[9]{}\Varid{g'}\;(\Varid{f'}\;\Varid{x}){}\<[22]%
\>[22]{}){}\<[25]%
\>[25]{}=\hspace{-3pt}\{\; \Conid{Refl}\;\}\hspace{-3pt}={}\<[E]%
\\
\>[3]{}\hsindent{3}{}\<[6]%
\>[6]{}({}\<[6E]%
\>[9]{}(\Varid{g'}\mathbin{\circ}\Varid{f'})\;\Varid{x}{}\<[22]%
\>[22]{})\;{}\<[25]%
\>[25]{}\Conid{QED}{}\<[E]%
\ColumnHook
\end{hscode}\resethooks
The right hand side is a chain of equal expressions connected by the
\ensuremath{=\hspace{-3pt}\{\; } proofs \ensuremath{\;\}\hspace{-3pt}=} of the individual steps within special braces and
ending in \ensuremath{\Conid{QED}}, see ``Preorder reasoning'' in the documentation by
\citet{idrisdocs}.
The steps with \ensuremath{\Conid{Refl}} are just for human readability, they could be
omitted as far as Idris is concerned.

Note that the proof steps are at the level of intensional equality
which all functions preserve as witnessed by \ensuremath{\Varid{cong}}. So one can often
use \ensuremath{\Varid{cong}} in steps where an outer context is unchanged (like \ensuremath{\Varid{g}}
in this example).
A special case of a two-argument version of \ensuremath{\Varid{cong}} shows that composition (like all functions) preserves intensional equality:
\begin{hscode}\SaveRestoreHook
\column{B}{@{}>{\hspre}l<{\hspost}@{}}%
\column{3}{@{}>{\hspre}l<{\hspost}@{}}%
\column{15}{@{}>{\hspre}c<{\hspost}@{}}%
\column{15E}{@{}l@{}}%
\column{18}{@{}>{\hspre}l<{\hspost}@{}}%
\column{25}{@{}>{\hspre}c<{\hspost}@{}}%
\column{25E}{@{}l@{}}%
\column{28}{@{}>{\hspre}l<{\hspost}@{}}%
\column{E}{@{}>{\hspre}l<{\hspost}@{}}%
\>[3]{}\Varid{compPresIE}{}\<[15]%
\>[15]{}\ \mathop{:}\ {}\<[15E]%
\>[18]{}\{\mskip1.5mu \Conid{A},\Conid{B},\Conid{C}\ \mathop{:}\ \Conid{Type}\mskip1.5mu\}\,\to\,\{\mskip1.5mu \Varid{g},\Varid{g'}\ \mathop{:}\ \Conid{B}\,\to\,\Conid{C}\mskip1.5mu\}\,\to\,\{\mskip1.5mu \Varid{f},\Varid{f'}\ \mathop{:}\ \Conid{A}\,\to\,\Conid{B}\mskip1.5mu\}\,\to\,{}\<[E]%
\\
\>[18]{}\Varid{g}\mathrel{=}\Varid{g'}\,\to\,\Varid{f}\mathrel{=}\Varid{f'}\,\to\,\Varid{g}\mathbin{\circ}\Varid{f}\mathrel{=}\Varid{g'}\mathbin{\circ}\Varid{f'}{}\<[E]%
\\
\>[3]{}\Varid{compPresIE}\;\Conid{Refl}\;\Conid{Refl}{}\<[25]%
\>[25]{}\mathrel{=}{}\<[25E]%
\>[28]{}\Conid{Refl}{}\<[E]%
\ColumnHook
\end{hscode}\resethooks
Note that the ``strengths'' of the two equality preservation lemmas
are not comparable: \ensuremath{\Varid{compPresIE}} proves a stronger conclusion, but
from stronger assumptions.

\paragraph*{ADTs and equality preservation.}
Abstract data types are often specified (e.g., via Idris
\emph{interfaces} or Agda \emph{records}) in terms of higher order
functions.
Typical examples are, beside the already mentioned \ensuremath{\Varid{map}} for functors,
bind and Kleisli composition (see section \ref{section:example}) for monads.
This paper is also about ADTs and generic programming.
More specifically, we show how to exploit the notion of extensional
equality preservation to inform the design of ADTs for generic
programming and embedded domain-specific languages (DSLs).
This is exemplified in sections \ref{section:functors} and
\ref{section:monads} with ADTs for functors and monads but we conjecture
that other abstract data types, e.g. for applicatives and arrows, could
also profit from a design informed by the notion of preservation of
extensional equality.

Thus, our work can also be seen as a contribution to the discussion on
verified ADTs initiated by Nicholas Drozd on
\href{https://groups.google.com/forum/#!topic/idris-lang/VZVpi-QUyUc}{idris-lang}.
A caveat is perhaps in place: the discussion on ADTs for functors and
monads in sections \ref{section:functors} and
\ref{section:monads} is not meant to answer the question of ``what
verified interfaces should look like''. Our aim is to demonstrate that,
like preservation of identity functions or preservation of composition,
preservation of extensional equality is a useful principle
for ADT design.

\paragraph*{What this paper is not about.}
Before turning to a first example, let us spend a few words on what this
paper is \emph{not} about.
It is not intended as a contribution to the theoretical study of the
equality type in intensional type theory or the algorithmic content of
the function extensionality principle. 

%
The equality type in intensional type theory and the question of how
to deal with extensional concepts in this context has been the subject
of important research for the last thirty years.
Since Hofmann's seminal work \citep{hofmann1995extensional}, setoids
have been the established, but also often dreaded (who coined the
expression \emph{``setoid hell''}?)  means to deal with extensional
concepts in intensional type theory (see also section
\ref{section:relatedwork}).
Eventually, the study of Martin-Löf's equality type has lead to the
development of Homotopy Type Theory and Voevodsky's Univalent
Foundations program
\citep{DBLP:books/daglib/0067012,DBLP:conf/lics/HofmannS94,hottbook}.
Univalence and recent developments in \emph{Cubical Type Theory}
\citep{cohenetal18:cubical} promise to finally provide developers with a
computational version of function extensionality.



This paper is a contribution towards \emph{pragmatic} verified generic
programming. From this perspective, it might become obsolete when fully
computational notions of function extensionality will become available
in mainstream programming.
From a more mathematical perspective, there are good reasons not to rely
on axioms that are stronger than necessary: there are interesting models
of type theory that refute function extensionality
\citep{streicher1993investigations, vonGlehn_2015,
  10.1145/3018610.3018620}, and our results can be interpreted in these
models.

The paper has been generated from literate Idris files.
These can be type-checked with Idris 1.3.2 and are available at
\url{https://gitlab.pik-potsdam.de/botta/papers}.

%

\section{Equality examples from dynamical systems theory}
\label{section:example}

In dynamical systems theory \citep{Kuznetsov:1998:EAB:289919,
  thomas2012catastrophe}, a prominent notion is that of the flow (or
iteration) of a system.
We start by discussing deterministic systems and then generalize to the
monadic case. A \emph{deterministic} dynamical system on a set \ensuremath{\Conid{X}} is an
endofunction on \ensuremath{\Conid{X}}.  The set \ensuremath{\Conid{X}} is often called the \emph{state space}
of the system.

Given a deterministic system \ensuremath{\Varid{f}\ \mathop{:}\ \Conid{X}\,\to\,\Conid{X}}, its $n$-th iterate or flow,
is typically denoted by \ensuremath{\ensuremath{\Varid{f}^{\Varid{n}}}\ \mathop{:}\ \Conid{X}\,\to\,\Conid{X}} and is defined by induction
on $n$: the base case \ensuremath{\ensuremath{\Varid{f}^{\mathrm{0}}}\mathrel{=}\Varid{id}} is the identity function and \ensuremath{\ensuremath{\Varid{f}^{\Varid{n}\mathbin{+}\mathrm{1}}}} is defined to be either \ensuremath{\Varid{f}\mathbin{\circ}\ensuremath{\Varid{f}^{\Varid{n}}}} or \ensuremath{\ensuremath{\Varid{f}^{\Varid{n}}}\mathbin{\circ}\Varid{f}}.
The two definitions are mathematically equivalent because of
associativity of function composition but what can one prove about \ensuremath{\Varid{f}\mathbin{\circ}\ensuremath{\Varid{f}^{\Varid{n}}}} and \ensuremath{\ensuremath{\Varid{f}^{\Varid{n}}}\mathbin{\circ}\Varid{f}} in intensional type theory?
We define the two variants as \ensuremath{\Varid{flowL}} and \ensuremath{\Varid{flowR}}:
\begin{hscode}\SaveRestoreHook
\column{B}{@{}>{\hspre}l<{\hspost}@{}}%
\column{3}{@{}>{\hspre}l<{\hspost}@{}}%
\column{12}{@{}>{\hspre}c<{\hspost}@{}}%
\column{12E}{@{}l@{}}%
\column{15}{@{}>{\hspre}l<{\hspost}@{}}%
\column{21}{@{}>{\hspre}c<{\hspost}@{}}%
\column{21E}{@{}l@{}}%
\column{24}{@{}>{\hspre}l<{\hspost}@{}}%
\column{E}{@{}>{\hspre}l<{\hspost}@{}}%
\>[3]{}\Varid{flowL}\ \mathop{:}\ \{\mskip1.5mu \Conid{X}\ \mathop{:}\ \Conid{Type}\mskip1.5mu\}\,\to\,(\Conid{X}\,\to\,\Conid{X})\,\to\,\mathbb{N}\,\to\,(\Conid{X}\,\to\,\Conid{X}){}\<[E]%
\\
\>[3]{}\Varid{flowL}\;\Varid{f}\;{}\<[15]%
\>[15]{}\Conid{Z}{}\<[21]%
\>[21]{}\mathrel{=}{}\<[21E]%
\>[24]{}\Varid{id}{}\<[E]%
\\
\>[3]{}\Varid{flowL}\;\Varid{f}\;{}\<[12]%
\>[12]{}({}\<[12E]%
\>[15]{}\Conid{S}\;\Varid{n}){}\<[21]%
\>[21]{}\mathrel{=}{}\<[21E]%
\>[24]{}\Varid{flowL}\;\Varid{f}\;\Varid{n}\mathbin{\circ}\Varid{f}{}\<[E]%
\\[\blanklineskip]%
\>[3]{}\Varid{flowR}\ \mathop{:}\ \{\mskip1.5mu \Conid{X}\ \mathop{:}\ \Conid{Type}\mskip1.5mu\}\,\to\,(\Conid{X}\,\to\,\Conid{X})\,\to\,\mathbb{N}\,\to\,(\Conid{X}\,\to\,\Conid{X}){}\<[E]%
\\
\>[3]{}\Varid{flowR}\;\Varid{f}\;{}\<[15]%
\>[15]{}\Conid{Z}{}\<[21]%
\>[21]{}\mathrel{=}{}\<[21E]%
\>[24]{}\Varid{id}{}\<[E]%
\\
\>[3]{}\Varid{flowR}\;\Varid{f}\;{}\<[12]%
\>[12]{}({}\<[12E]%
\>[15]{}\Conid{S}\;\Varid{n}){}\<[21]%
\>[21]{}\mathrel{=}{}\<[21E]%
\>[24]{}\Varid{f}\mathbin{\circ}\Varid{flowR}\;\Varid{f}\;\Varid{n}{}\<[E]%
\ColumnHook
\end{hscode}\resethooks
The flows \ensuremath{\Varid{flowL}\;\Varid{f}\;\Varid{n}} and \ensuremath{\Varid{flowR}\;\Varid{f}\;\Varid{n}} are intensionally equal:
\begin{hscode}\SaveRestoreHook
\column{B}{@{}>{\hspre}l<{\hspost}@{}}%
\column{3}{@{}>{\hspre}l<{\hspost}@{}}%
\column{E}{@{}>{\hspre}l<{\hspost}@{}}%
\>[3]{}\Varid{flowLemma}\ \mathop{:}\ \{\mskip1.5mu \Conid{X}\ \mathop{:}\ \Conid{Type}\mskip1.5mu\}\,\to\,(\Varid{f}\ \mathop{:}\ \Conid{X}\,\to\,\Conid{X})\,\to\,(\Varid{n}\ \mathop{:}\ \mathbb{N})\,\to\,\Varid{flowL}\;\Varid{f}\;\Varid{n}\mathrel{=}\Varid{flowR}\;\Varid{f}\;\Varid{n}{}\<[E]%
\ColumnHook
\end{hscode}\resethooks
With \ensuremath{\Varid{compPresIE}} from section~\ref{section:about}, one can implement
\ensuremath{\Varid{flowLemma}} by induction on the number of iterations~\ensuremath{\Varid{n}}.
The base case is trivial
\begin{hscode}\SaveRestoreHook
\column{B}{@{}>{\hspre}l<{\hspost}@{}}%
\column{3}{@{}>{\hspre}l<{\hspost}@{}}%
\column{19}{@{}>{\hspre}l<{\hspost}@{}}%
\column{25}{@{}>{\hspre}c<{\hspost}@{}}%
\column{25E}{@{}l@{}}%
\column{28}{@{}>{\hspre}l<{\hspost}@{}}%
\column{E}{@{}>{\hspre}l<{\hspost}@{}}%
\>[3]{}\Varid{flowLemma}\;\Varid{f}\;{}\<[19]%
\>[19]{}\Conid{Z}{}\<[25]%
\>[25]{}\mathrel{=}{}\<[25E]%
\>[28]{}\Conid{Refl}{}\<[E]%
\ColumnHook
\end{hscode}\resethooks
For readability, we spell out the proof sketch for the induction step in
full:
\begin{hscode}\SaveRestoreHook
\column{B}{@{}>{\hspre}l<{\hspost}@{}}%
\column{3}{@{}>{\hspre}l<{\hspost}@{}}%
\column{16}{@{}>{\hspre}c<{\hspost}@{}}%
\column{16E}{@{}l@{}}%
\column{19}{@{}>{\hspre}l<{\hspost}@{}}%
\column{25}{@{}>{\hspre}c<{\hspost}@{}}%
\column{25E}{@{}l@{}}%
\column{28}{@{}>{\hspre}l<{\hspost}@{}}%
\column{50}{@{}>{\hspre}l<{\hspost}@{}}%
\column{E}{@{}>{\hspre}l<{\hspost}@{}}%
\>[3]{}\Varid{flowLemma}\;\Varid{f}\;{}\<[16]%
\>[16]{}({}\<[16E]%
\>[19]{}\Conid{S}\;\Varid{n}){}\<[25]%
\>[25]{}\mathrel{=}{}\<[25E]%
\>[28]{}(\Varid{flowL}\;\Varid{f}\;(\Conid{S}\;\Varid{n})){}\<[50]%
\>[50]{}=\hspace{-3pt}\{\; \Conid{Refl}\;\}\hspace{-3pt}={}\<[E]%
\\
\>[28]{}(\Varid{flowL}\;\Varid{f}\;\Varid{n}\mathbin{\circ}\Varid{f}){}\<[50]%
\>[50]{}=\hspace{-3pt}\{\; \Varid{compPresIE}\;(\Varid{flowLemma}\;\Varid{f}\;\Varid{n})\;\Conid{Refl}\;\}\hspace{-3pt}={}\<[E]%
\\
\>[28]{}(\Varid{flowR}\;\Varid{f}\;\Varid{n}\mathbin{\circ}\Varid{f}){}\<[50]%
\>[50]{}=\hspace{-3pt}\{\; \Varid{flowRLemma}\;\Varid{f}\;\Varid{n}\;\}\hspace{-3pt}={}\<[E]%
\\
\>[28]{}(\Varid{f}\mathbin{\circ}\Varid{flowR}\;\Varid{f}\;\Varid{n}){}\<[50]%
\>[50]{}=\hspace{-3pt}\{\; \Conid{Refl}\;\}\hspace{-3pt}={}\<[E]%
\\
\>[28]{}(\Varid{flowR}\;\Varid{f}\;(\Conid{S}\;\Varid{n}))\;{}\<[50]%
\>[50]{}\Conid{QED}{}\<[E]%
\ColumnHook
\end{hscode}\resethooks
First, we apply the definition of \ensuremath{\Varid{flowL}} to deduce \ensuremath{\Varid{flowL}\;\Varid{f}\;(\Conid{S}\;\Varid{n})\mathrel{=}\Varid{flowL}\;\Varid{f}\;\Varid{n}\mathbin{\circ}\Varid{f}}.
Next, we apply \ensuremath{\Varid{compPresIE}} with the induction hypothesis \ensuremath{\Varid{flowLemma}\;\Varid{f}\;\Varid{n}} and deduce \ensuremath{\Varid{flowL}\;\Varid{f}\;\Varid{n}\mathbin{\circ}\Varid{f}\mathrel{=}\Varid{flowR}\;\Varid{f}\;\Varid{n}\mathbin{\circ}\Varid{f}}.
The (almost) final step is to show that \ensuremath{\Varid{flowR}\;\Varid{f}\;\Varid{n}\mathbin{\circ}\Varid{f}\mathrel{=}\Varid{f}\mathbin{\circ}\Varid{flowR}\;\Varid{f}\;\Varid{n}}.
This is obtained via the auxiliary \ensuremath{\Varid{flowRLemma}} where we use
associativity and preservation of intensional equality again.

\begin{joincode}
\begin{hscode}\SaveRestoreHook
\column{B}{@{}>{\hspre}l<{\hspost}@{}}%
\column{3}{@{}>{\hspre}l<{\hspost}@{}}%
\column{E}{@{}>{\hspre}l<{\hspost}@{}}%
\>[3]{}\Varid{flowRLemma}\ \mathop{:}\ \{\mskip1.5mu \Conid{X}\ \mathop{:}\ \Conid{Type}\mskip1.5mu\}\,\to\,(\Varid{f}\ \mathop{:}\ \Conid{X}\,\to\,\Conid{X})\,\to\,(\Varid{n}\ \mathop{:}\ \mathbb{N})\,\to\,\Varid{flowR}\;\Varid{f}\;\Varid{n}\mathbin{\circ}\Varid{f}\mathrel{=}\Varid{f}\mathbin{\circ}\Varid{flowR}\;\Varid{f}\;\Varid{n}{}\<[E]%
\ColumnHook
\end{hscode}\resethooks
\begin{hscode}\SaveRestoreHook
\column{B}{@{}>{\hspre}l<{\hspost}@{}}%
\column{3}{@{}>{\hspre}l<{\hspost}@{}}%
\column{17}{@{}>{\hspre}c<{\hspost}@{}}%
\column{17E}{@{}l@{}}%
\column{20}{@{}>{\hspre}l<{\hspost}@{}}%
\column{26}{@{}>{\hspre}c<{\hspost}@{}}%
\column{26E}{@{}l@{}}%
\column{29}{@{}>{\hspre}l<{\hspost}@{}}%
\column{57}{@{}>{\hspre}l<{\hspost}@{}}%
\column{61}{@{}>{\hspre}l<{\hspost}@{}}%
\column{E}{@{}>{\hspre}l<{\hspost}@{}}%
\>[3]{}\Varid{flowRLemma}\;\Varid{f}\;{}\<[20]%
\>[20]{}\Conid{Z}{}\<[26]%
\>[26]{}\mathrel{=}{}\<[26E]%
\>[29]{}\Conid{Refl}{}\<[E]%
\\
\>[3]{}\Varid{flowRLemma}\;\Varid{f}\;{}\<[17]%
\>[17]{}({}\<[17E]%
\>[20]{}\Conid{S}\;\Varid{n}){}\<[26]%
\>[26]{}\mathrel{=}{}\<[26E]%
\>[29]{}(\Varid{flowR}\;\Varid{f}\;(\Conid{S}\;\Varid{n})\mathbin{\circ}\Varid{f}){}\<[57]%
\>[57]{}=\hspace{-3pt}\{\; {}\<[61]%
\>[61]{}\Conid{Refl}\;\}\hspace{-3pt}={}\<[E]%
\\
\>[29]{}((\Varid{f}\mathbin{\circ}\Varid{flowR}\;\Varid{f}\;\Varid{n})\mathbin{\circ}\Varid{f}){}\<[57]%
\>[57]{}=\hspace{-3pt}\{\; {}\<[61]%
\>[61]{}\Varid{compAssociative}\;\Varid{f}\;(\Varid{flowR}\;\Varid{f}\;\Varid{n})\;\Varid{f}\;\}\hspace{-3pt}={}\<[E]%
\\
\>[29]{}(\Varid{f}\mathbin{\circ}(\Varid{flowR}\;\Varid{f}\;\Varid{n}\mathbin{\circ}\Varid{f})){}\<[57]%
\>[57]{}=\hspace{-3pt}\{\; {}\<[61]%
\>[61]{}\Varid{compPresIE}\;\Conid{Refl}\;(\Varid{flowRLemma}\;\Varid{f}\;\Varid{n})\;\}\hspace{-3pt}={}\<[E]%
\\
\>[29]{}(\Varid{f}\mathbin{\circ}(\Varid{f}\mathbin{\circ}\Varid{flowR}\;\Varid{f}\;\Varid{n})){}\<[57]%
\>[57]{}=\hspace{-3pt}\{\; {}\<[61]%
\>[61]{}\Conid{Refl}\;\}\hspace{-3pt}={}\<[E]%
\\
\>[29]{}(\Varid{f}\mathbin{\circ}\Varid{flowR}\;\Varid{f}\;(\Conid{S}\;\Varid{n}))\;{}\<[57]%
\>[57]{}\Conid{QED}{}\<[E]%
\ColumnHook
\end{hscode}\resethooks
\end{joincode}
%

\noindent
Let's summarize: we have considered the special case of deterministic
dynamical systems, defined the flow (a higher order function) in two
different ways and shown that the two definitions are equivalent in
the sense that \ensuremath{\Varid{e}_{1}\mathrel{=}\Varid{flowR}\;\Varid{f}\;\Varid{n}} and \ensuremath{\Varid{e}_{2}\mathrel{=}\Varid{flowL}\;\Varid{f}\;\Varid{n}} are intensionally
equal for all \ensuremath{\Varid{f}} and \ensuremath{\Varid{n}}.
Before we move on to a more general setting, where intensional
equality does not hold, let's expand a bit on the different levels of
equality relevant for these two functions.
The two expressions \ensuremath{\Varid{e}_{1}} and \ensuremath{\Varid{e}_{2}} denote functions of type \ensuremath{\Conid{X}\,\to\,\Conid{X}} and
thus for any \ensuremath{\Varid{x}\ \mathop{:}\ \Conid{X}} we also have \ensuremath{\Varid{e}_{1}\;\Varid{x}\mathrel{=}\Varid{e}_{2}\;\Varid{x}}.
On the other hand, the quantification over all \ensuremath{\Varid{n}} can be absorbed
into the definition of extensional equality so that we have \ensuremath{\Varid{flowR}\;\Varid{f}\doteq\Varid{flowL}\;\Varid{f}} for all \ensuremath{\Varid{f}}.
And with the two-argument version of extensional equality we get \ensuremath{\Varid{flowR}\stackrel{..}{=}\Varid{flowL}}.

\paragraph*{Monadic systems.}
What about non-deterministic systems, stochastic systems or perhaps
fuzzy systems?
Can we extend our results to the general case of \emph{monadic}
dynamical systems?
Monadic dynamical systems \citep{ionescu2009}
on a set \ensuremath{\Conid{X}} are functions of type \ensuremath{\Conid{X}\,\to\,\Conid{M}\;\Conid{X}} where \ensuremath{\Conid{M}} is a monad.
When \ensuremath{\Conid{M}} is the identity monad, one recovers the deterministic case.
When \ensuremath{\Conid{M}} is \ensuremath{\Conid{List}} one has non-deterministic systems, and finite
probability monads
capture the notion of stochastic systems.
\nocite{10.1017/S0956796805005721,ionescu2009, 2017_Botta_Jansson_Ionescu}
%
\nocite{ionescu2009,giry1981}

One can extend the flow (and, as we will see in
section~\ref{section:applications}, other elementary operations) of
deterministic systems to the general, monadic case by replacing \ensuremath{\Varid{id}}
with \ensuremath{\Varid{pure}} and function composition with Kleisli composition (\ensuremath{\mathbin{\mathbin{>}\hspace{-5.2pt}\mathrel{=}\hspace{-5.5pt}\mathbin{>}}}):
\begin{hscode}\SaveRestoreHook
\column{B}{@{}>{\hspre}l<{\hspost}@{}}%
\column{3}{@{}>{\hspre}l<{\hspost}@{}}%
\column{13}{@{}>{\hspre}c<{\hspost}@{}}%
\column{13E}{@{}l@{}}%
\column{16}{@{}>{\hspre}l<{\hspost}@{}}%
\column{22}{@{}>{\hspre}l<{\hspost}@{}}%
\column{25}{@{}>{\hspre}c<{\hspost}@{}}%
\column{25E}{@{}l@{}}%
\column{28}{@{}>{\hspre}l<{\hspost}@{}}%
\column{34}{@{}>{\hspre}c<{\hspost}@{}}%
\column{34E}{@{}l@{}}%
\column{37}{@{}>{\hspre}l<{\hspost}@{}}%
\column{E}{@{}>{\hspre}l<{\hspost}@{}}%
\>[3]{}\Varid{flowMonL}{}\<[13]%
\>[13]{}\ \mathop{:}\ {}\<[13E]%
\>[16]{}\{\mskip1.5mu \Conid{X}\ \mathop{:}\ \Conid{Type}\mskip1.5mu\}\,\to\,\{\mskip1.5mu \Conid{M}\ \mathop{:}\ \Conid{Type}\,\to\,\Conid{Type}\mskip1.5mu\}\,\to\,\Conid{Monad}\;\Conid{M}\Rightarrow {}\<[E]%
\\
\>[16]{}(\Conid{X}\,\to\,\Conid{M}\;\Conid{X})\,\to\,\mathbb{N}\,\to\,(\Conid{X}\,\to\,\Conid{M}\;\Conid{X}){}\<[E]%
\\
\>[3]{}\Varid{flowMonL}\;\,{}\<[22]%
\>[22]{}\Varid{f}\;{}\<[28]%
\>[28]{}\Conid{Z}{}\<[34]%
\>[34]{}\mathrel{=}{}\<[34E]%
\>[37]{}\Varid{pure}{}\<[E]%
\\
\>[3]{}\Varid{flowMonL}\;{}\<[22]%
\>[22]{}\Varid{f}\;{}\<[25]%
\>[25]{}({}\<[25E]%
\>[28]{}\Conid{S}\;\Varid{n}){}\<[34]%
\>[34]{}\mathrel{=}{}\<[34E]%
\>[37]{}\Varid{flowMonL}\;\Varid{f}\;\Varid{n}\mathbin{\mathbin{>}\hspace{-5.2pt}\mathrel{=}\hspace{-5.5pt}\mathbin{>}}\Varid{f}{}\<[E]%
\\[\blanklineskip]%
\>[3]{}\Varid{flowMonR}{}\<[13]%
\>[13]{}\ \mathop{:}\ {}\<[13E]%
\>[16]{}\{\mskip1.5mu \Conid{X}\ \mathop{:}\ \Conid{Type}\mskip1.5mu\}\,\to\,\{\mskip1.5mu \Conid{M}\ \mathop{:}\ \Conid{Type}\,\to\,\Conid{Type}\mskip1.5mu\}\,\to\,\Conid{Monad}\;\Conid{M}\Rightarrow {}\<[E]%
\\
\>[16]{}(\Conid{X}\,\to\,\Conid{M}\;\Conid{X})\,\to\,\mathbb{N}\,\to\,(\Conid{X}\,\to\,\Conid{M}\;\Conid{X}){}\<[E]%
\\
\>[3]{}\Varid{flowMonR}\;{}\<[22]%
\>[22]{}\Varid{f}\;{}\<[28]%
\>[28]{}\Conid{Z}{}\<[34]%
\>[34]{}\mathrel{=}{}\<[34E]%
\>[37]{}\Varid{pure}{}\<[E]%
\\
\>[3]{}\Varid{flowMonR}\;{}\<[22]%
\>[22]{}\Varid{f}\;{}\<[25]%
\>[25]{}({}\<[25E]%
\>[28]{}\Conid{S}\;\Varid{n}){}\<[34]%
\>[34]{}\mathrel{=}{}\<[34E]%
\>[37]{}\Varid{f}\mathbin{\mathbin{>}\hspace{-5.2pt}\mathrel{=}\hspace{-5.5pt}\mathbin{>}}\Varid{flowMonR}\;\Varid{f}\;\Varid{n}{}\<[E]%
\ColumnHook
\end{hscode}\resethooks
Notice, however, that now the implementations of \ensuremath{\Varid{flowMonL}} and
\ensuremath{\Varid{flowMonR}} depend on \ensuremath{(\mathbin{\mathbin{>}\hspace{-5.2pt}\mathrel{=}\hspace{-5.5pt}\mathbin{>}})}, which is a monad-specific operation.
This means that, in proving properties of the flow of monadic systems,
we can no longer rely on a specific \emph{definition} of \ensuremath{(\mathbin{\mathbin{>}\hspace{-5.2pt}\mathrel{=}\hspace{-5.5pt}\mathbin{>}})}: we
have to derive our proofs on the basis of properties that we know (or
require) \ensuremath{(\mathbin{\mathbin{>}\hspace{-5.2pt}\mathrel{=}\hspace{-5.5pt}\mathbin{>}})} to fulfil -- that is, on its \emph{specification}.

What do we know about Kleisli composition \emph{in general}?
We discuss this question in the next two sections but let us
anticipate that, if we require functors to preserve the extensional
equality of arrows (in addition to identity and composition) and
Kleisli composition to fulfil the specification
\vspace*{-2ex}
\begin{center}
\beginpgfgraphicnamed{ExtEqPres-f1}
\begin{tikzcd}[column sep=large]
  \ensuremath{\Conid{M}\;\Conid{C}}
& \ensuremath{\Conid{M}\;(\Conid{M}\;\Conid{C})}  \arrow[l, "\ensuremath{\Varid{join}}"]
& \ensuremath{\Conid{M}\;\Conid{B}}      \arrow[l, "\ensuremath{\Varid{map}\;\Varid{g}}"]
& \ensuremath{\Conid{A}}        \arrow[l, "\ensuremath{\Varid{f}}"]\arrow[lll, bend right=30, "f
    \mathbin{> \!\! = \!\! >}
    g"]
\end{tikzcd}
\endpgfgraphicnamed
\end{center}
\begin{hscode}\SaveRestoreHook
\column{B}{@{}>{\hspre}l<{\hspost}@{}}%
\column{3}{@{}>{\hspre}l<{\hspost}@{}}%
\column{16}{@{}>{\hspre}c<{\hspost}@{}}%
\column{16E}{@{}l@{}}%
\column{19}{@{}>{\hspre}l<{\hspost}@{}}%
\column{E}{@{}>{\hspre}l<{\hspost}@{}}%
\>[3]{}\Varid{kleisliSpec}{}\<[16]%
\>[16]{}\ \mathop{:}\ {}\<[16E]%
\>[19]{}\{\mskip1.5mu \Conid{A},\Conid{B},\Conid{C}\ \mathop{:}\ \Conid{Type}\mskip1.5mu\}\,\to\,\{\mskip1.5mu \Conid{M}\ \mathop{:}\ \Conid{Type}\,\to\,\Conid{Type}\mskip1.5mu\}\,\to\,\Conid{Monad}\;\Conid{M}\Rightarrow {}\<[E]%
\\
\>[19]{}(\Varid{f}\ \mathop{:}\ \Conid{A}\,\to\,\Conid{M}\;\Conid{B})\,\to\,(\Varid{g}\ \mathop{:}\ \Conid{B}\,\to\,\Conid{M}\;\Conid{C})\,\to\,(\Varid{f}\mathbin{\mathbin{>}\hspace{-5.2pt}\mathrel{=}\hspace{-5.5pt}\mathbin{>}}\Varid{g})\doteq\Varid{join}\mathbin{\circ}\Varid{map}\;\Varid{g}\mathbin{\circ}\Varid{f}{}\<[E]%
\ColumnHook
\end{hscode}\resethooks
then we can derive preservation of extensional equality
\begin{hscode}\SaveRestoreHook
\column{B}{@{}>{\hspre}l<{\hspost}@{}}%
\column{3}{@{}>{\hspre}l<{\hspost}@{}}%
\column{18}{@{}>{\hspre}c<{\hspost}@{}}%
\column{18E}{@{}l@{}}%
\column{21}{@{}>{\hspre}l<{\hspost}@{}}%
\column{E}{@{}>{\hspre}l<{\hspost}@{}}%
\>[3]{}\Varid{kleisliPresEE}{}\<[18]%
\>[18]{}\ \mathop{:}\ {}\<[18E]%
\>[21]{}\{\mskip1.5mu \Conid{A},\Conid{B},\Conid{C}\ \mathop{:}\ \Conid{Type}\mskip1.5mu\}\,\to\,\{\mskip1.5mu \Conid{M}\ \mathop{:}\ \Conid{Type}\,\to\,\Conid{Type}\mskip1.5mu\}\,\to\,\Conid{Monad}\;\Conid{M}\Rightarrow {}\<[E]%
\\
\>[21]{}(\Varid{f},\Varid{f'}\ \mathop{:}\ \Conid{A}\,\to\,\Conid{M}\;\Conid{B})\,\to\,(\Varid{g},\Varid{g'}\ \mathop{:}\ \Conid{B}\,\to\,\Conid{M}\;\Conid{C})\,\to\,{}\<[E]%
\\
\>[21]{}\Varid{f}\doteq\Varid{f'}\,\to\,\Varid{g}\doteq\Varid{g'}\,\to\,(\Varid{f}\mathbin{\mathbin{>}\hspace{-5.2pt}\mathrel{=}\hspace{-5.5pt}\mathbin{>}}\Varid{g})\doteq(\Varid{f'}\mathbin{\mathbin{>}\hspace{-5.2pt}\mathrel{=}\hspace{-5.5pt}\mathbin{>}}\Varid{g'}){}\<[E]%
\ColumnHook
\end{hscode}\resethooks
and associativity of Kleisli composition generically.
\begin{hscode}\SaveRestoreHook
\column{B}{@{}>{\hspre}l<{\hspost}@{}}%
\column{3}{@{}>{\hspre}l<{\hspost}@{}}%
\column{17}{@{}>{\hspre}c<{\hspost}@{}}%
\column{17E}{@{}l@{}}%
\column{20}{@{}>{\hspre}l<{\hspost}@{}}%
\column{E}{@{}>{\hspre}l<{\hspost}@{}}%
\>[3]{}\Varid{kleisliAssoc}{}\<[17]%
\>[17]{}\ \mathop{:}\ {}\<[17E]%
\>[20]{}\{\mskip1.5mu \Conid{A},\Conid{B},\Conid{C},\Conid{D}\ \mathop{:}\ \Conid{Type}\mskip1.5mu\}\,\to\,\{\mskip1.5mu \Conid{M}\ \mathop{:}\ \Conid{Type}\,\to\,\Conid{Type}\mskip1.5mu\}\,\to\,\Conid{Monad}\;\Conid{M}\Rightarrow {}\<[E]%
\\
\>[20]{}(\Varid{f}\ \mathop{:}\ \Conid{A}\,\to\,\Conid{M}\;\Conid{B})\,\to\,(\Varid{g}\ \mathop{:}\ \Conid{B}\,\to\,\Conid{M}\;\Conid{C})\,\to\,(\Varid{h}\ \mathop{:}\ \Conid{C}\,\to\,\Conid{M}\;\Conid{D})\,\to\,{}\<[E]%
\\
\>[20]{}((\Varid{f}\mathbin{\mathbin{>}\hspace{-5.2pt}\mathrel{=}\hspace{-5.5pt}\mathbin{>}}\Varid{g})\mathbin{\mathbin{>}\hspace{-5.2pt}\mathrel{=}\hspace{-5.5pt}\mathbin{>}}\Varid{h})\doteq(\Varid{f}\mathbin{\mathbin{>}\hspace{-5.2pt}\mathrel{=}\hspace{-5.5pt}\mathbin{>}}(\Varid{g}\mathbin{\mathbin{>}\hspace{-5.2pt}\mathrel{=}\hspace{-5.5pt}\mathbin{>}}\Varid{h})){}\<[E]%
\ColumnHook
\end{hscode}\resethooks
From these premises, we can prove the \emph{extensional} equality of
\ensuremath{\Varid{flowMonL}} and \ensuremath{\Varid{flowMonR}} using a similar lemma as in the deterministic case:
\begin{hscode}\SaveRestoreHook
\column{B}{@{}>{\hspre}l<{\hspost}@{}}%
\column{3}{@{}>{\hspre}l<{\hspost}@{}}%
\column{16}{@{}>{\hspre}c<{\hspost}@{}}%
\column{16E}{@{}l@{}}%
\column{19}{@{}>{\hspre}l<{\hspost}@{}}%
\column{E}{@{}>{\hspre}l<{\hspost}@{}}%
\>[3]{}\Varid{flowMonRLem}{}\<[16]%
\>[16]{}\ \mathop{:}\ {}\<[16E]%
\>[19]{}\{\mskip1.5mu \Conid{X}\ \mathop{:}\ \Conid{Type}\mskip1.5mu\}\,\to\,\{\mskip1.5mu \Conid{M}\ \mathop{:}\ \Conid{Type}\,\to\,\Conid{Type}\mskip1.5mu\}\,\to\,\Conid{Monad}\;\Conid{M}\Rightarrow {}\<[E]%
\\
\>[19]{}(\Varid{f}\ \mathop{:}\ \Conid{X}\,\to\,\Conid{M}\;\Conid{X})\,\to\,(\Varid{n}\ \mathop{:}\ \mathbb{N})\,\to\,(\Varid{flowMonR}\;\Varid{f}\;\Varid{n}\mathbin{\mathbin{>}\hspace{-5.2pt}\mathrel{=}\hspace{-5.5pt}\mathbin{>}}\Varid{f})\doteq(\Varid{f}\mathbin{\mathbin{>}\hspace{-5.2pt}\mathrel{=}\hspace{-5.5pt}\mathbin{>}}\Varid{flowMonR}\;\Varid{f}\;\Varid{n}){}\<[E]%
\ColumnHook
\end{hscode}\resethooks
First, notice that the base case of the lemma requires computing
evidence that \ensuremath{\Varid{pure}\mathbin{\mathbin{>}\hspace{-5.2pt}\mathrel{=}\hspace{-5.5pt}\mathbin{>}}\Varid{f}} is extensionally equal to \ensuremath{\Varid{f}\mathbin{\mathbin{>}\hspace{-5.2pt}\mathrel{=}\hspace{-5.5pt}\mathbin{>}}\Varid{pure}}.
This is a consequence of \ensuremath{\Varid{pure}} being a left and a right identity for
Kleisli composition: for \ensuremath{\Varid{f}\ \mathop{:}\ \Conid{A}\,\to\,\Conid{M}\;\Conid{B}} we have
\begin{hscode}\SaveRestoreHook
\column{B}{@{}>{\hspre}l<{\hspost}@{}}%
\column{3}{@{}>{\hspre}l<{\hspost}@{}}%
\column{23}{@{}>{\hspre}l<{\hspost}@{}}%
\column{26}{@{}>{\hspre}l<{\hspost}@{}}%
\column{42}{@{}>{\hspre}l<{\hspost}@{}}%
\column{E}{@{}>{\hspre}l<{\hspost}@{}}%
\>[3]{}\Varid{pureLeftIdKleisli}\;{}\<[23]%
\>[23]{}\Varid{f}{}\<[26]%
\>[26]{}\ \mathop{:}\ (\Varid{pure}\mathbin{\mathbin{>}\hspace{-5.2pt}\mathrel{=}\hspace{-5.5pt}\mathbin{>}}\Varid{f}){}\<[42]%
\>[42]{}\doteq\Varid{f}{}\<[E]%
\\
\>[3]{}\Varid{pureRightIdKleisli}\;{}\<[23]%
\>[23]{}\Varid{f}{}\<[26]%
\>[26]{}\ \mathop{:}\ (\Varid{f}\mathbin{\mathbin{>}\hspace{-5.2pt}\mathrel{=}\hspace{-5.5pt}\mathbin{>}}\Varid{pure}){}\<[42]%
\>[42]{}\doteq\Varid{f}{}\<[E]%
\ColumnHook
\end{hscode}\resethooks
\begin{hscode}\SaveRestoreHook
\column{B}{@{}>{\hspre}l<{\hspost}@{}}%
\column{3}{@{}>{\hspre}l<{\hspost}@{}}%
\column{5}{@{}>{\hspre}l<{\hspost}@{}}%
\column{35}{@{}>{\hspre}l<{\hspost}@{}}%
\column{E}{@{}>{\hspre}l<{\hspost}@{}}%
\>[3]{}\Varid{flowMonRLem}\;\Varid{f}\;\Conid{Z}\;\Varid{x}\mathrel{=}{}\<[E]%
\\
\>[3]{}\hsindent{2}{}\<[5]%
\>[5]{}((\Varid{flowMonR}\;\Varid{f}\;\Conid{Z}\mathbin{\mathbin{>}\hspace{-5.2pt}\mathrel{=}\hspace{-5.5pt}\mathbin{>}}\Varid{f})\;\Varid{x}){}\<[35]%
\>[35]{}=\hspace{-3pt}\{\; \Conid{Refl}\;\}\hspace{-3pt}={}\<[E]%
\\
\>[3]{}\hsindent{2}{}\<[5]%
\>[5]{}((\Varid{pure}\mathbin{\mathbin{>}\hspace{-5.2pt}\mathrel{=}\hspace{-5.5pt}\mathbin{>}}\Varid{f})\;\Varid{x}){}\<[35]%
\>[35]{}=\hspace{-3pt}\{\; \Varid{pureLeftIdKleisli}\;\Varid{f}\;\Varid{x}\;\}\hspace{-3pt}={}\<[E]%
\\
\>[3]{}\hsindent{2}{}\<[5]%
\>[5]{}(\Varid{f}\;\Varid{x}){}\<[35]%
\>[35]{}=\hspace{-3pt}\{\; \Varid{sym}\;(\Varid{pureRightIdKleisli}\;\Varid{f}\;\Varid{x})\;\}\hspace{-3pt}={}\<[E]%
\\
\>[3]{}\hsindent{2}{}\<[5]%
\>[5]{}((\Varid{f}\mathbin{\mathbin{>}\hspace{-5.2pt}\mathrel{=}\hspace{-5.5pt}\mathbin{>}}\Varid{pure})\;\Varid{x}){}\<[35]%
\>[35]{}=\hspace{-3pt}\{\; \Conid{Refl}\;\}\hspace{-3pt}={}\<[E]%
\\
\>[3]{}\hsindent{2}{}\<[5]%
\>[5]{}((\Varid{f}\mathbin{\mathbin{>}\hspace{-5.2pt}\mathrel{=}\hspace{-5.5pt}\mathbin{>}}\Varid{flowMonR}\;\Varid{f}\;\Conid{Z})\;\Varid{x})\;{}\<[35]%
\>[35]{}\Conid{QED}{}\<[E]%
\ColumnHook
\end{hscode}\resethooks
As we will see in subsection \ref{def:pureLeftIdKleisli},
\ensuremath{\Varid{pureLeftIdKleisli}} and \ensuremath{\Varid{pureRightIdKleisli}} are either ADT axioms or
theorems, depending of the formulation of the monad ADT.
The induction step of \ensuremath{\Varid{flowMonRLem}} relies on preservation of
extensional equality and on associativity of Kleisli composition:
\begin{hscode}\SaveRestoreHook
\column{B}{@{}>{\hspre}l<{\hspost}@{}}%
\column{3}{@{}>{\hspre}l<{\hspost}@{}}%
\column{4}{@{}>{\hspre}l<{\hspost}@{}}%
\column{5}{@{}>{\hspre}l<{\hspost}@{}}%
\column{18}{@{}>{\hspre}c<{\hspost}@{}}%
\column{18E}{@{}l@{}}%
\column{21}{@{}>{\hspre}l<{\hspost}@{}}%
\column{41}{@{}>{\hspre}l<{\hspost}@{}}%
\column{59}{@{}>{\hspre}l<{\hspost}@{}}%
\column{74}{@{}>{\hspre}l<{\hspost}@{}}%
\column{98}{@{}>{\hspre}l<{\hspost}@{}}%
\column{E}{@{}>{\hspre}l<{\hspost}@{}}%
\>[3]{}\Varid{flowMonRLem}\;\Varid{f}\;{}\<[18]%
\>[18]{}({}\<[18E]%
\>[21]{}\Conid{S}\;\Varid{n})\;\Varid{x}\mathrel{=}{}\<[E]%
\\
\>[3]{}\hsindent{1}{}\<[4]%
\>[4]{}\mathbf{let}\;\Varid{rest}\mathrel{=}\Varid{flowMonR}\;\Varid{f}\;\Varid{n}\;\mathbf{in}{}\<[E]%
\\
\>[4]{}\hsindent{1}{}\<[5]%
\>[5]{}((\Varid{flowMonR}\;\Varid{f}\;(\Conid{S}\;\Varid{n})\mathbin{\mathbin{>}\hspace{-5.2pt}\mathrel{=}\hspace{-5.5pt}\mathbin{>}}\Varid{f})\;\Varid{x}){}\<[41]%
\>[41]{}=\hspace{-3pt}\{\; \Conid{Refl}\;\}\hspace{-3pt}={}\<[E]%
\\
\>[4]{}\hsindent{1}{}\<[5]%
\>[5]{}(((\Varid{f}\mathbin{\mathbin{>}\hspace{-5.2pt}\mathrel{=}\hspace{-5.5pt}\mathbin{>}}\Varid{rest})\mathbin{\mathbin{>}\hspace{-5.2pt}\mathrel{=}\hspace{-5.5pt}\mathbin{>}}\Varid{f})\;\Varid{x}){}\<[41]%
\>[41]{}=\hspace{-3pt}\{\; \Varid{kleisliAssoc}\;\Varid{f}\;\Varid{rest}\;\Varid{f}\;\Varid{x}\;\}\hspace{-3pt}={}\<[E]%
\\
\>[4]{}\hsindent{1}{}\<[5]%
\>[5]{}((\Varid{f}\mathbin{\mathbin{>}\hspace{-5.2pt}\mathrel{=}\hspace{-5.5pt}\mathbin{>}}(\Varid{rest}\mathbin{\mathbin{>}\hspace{-5.2pt}\mathrel{=}\hspace{-5.5pt}\mathbin{>}}\Varid{f}))\;\Varid{x}){}\<[41]%
\>[41]{}=\hspace{-3pt}\{\; \Varid{kleisliPresEE}\;{}\<[59]%
\>[59]{}\Varid{f}\;\Varid{f}\;{}\<[74]%
\>[74]{}(\Varid{rest}\mathbin{\mathbin{>}\hspace{-5.2pt}\mathrel{=}\hspace{-5.5pt}\mathbin{>}}\Varid{f})\;(\Varid{f}\mathbin{\mathbin{>}\hspace{-5.2pt}\mathrel{=}\hspace{-5.5pt}\mathbin{>}}\Varid{rest})\;{}\<[E]%
\\
\>[59]{}(\lambda \Varid{v}\Rightarrow \Conid{Refl})\;{}\<[74]%
\>[74]{}(\Varid{flowMonRLem}\;\Varid{f}\;\Varid{n})\;\Varid{x}{}\<[98]%
\>[98]{}\;\}\hspace{-3pt}={}\<[E]%
\\
\>[4]{}\hsindent{1}{}\<[5]%
\>[5]{}((\Varid{f}\mathbin{\mathbin{>}\hspace{-5.2pt}\mathrel{=}\hspace{-5.5pt}\mathbin{>}}(\Varid{f}\mathbin{\mathbin{>}\hspace{-5.2pt}\mathrel{=}\hspace{-5.5pt}\mathbin{>}}\Varid{rest}))\;\Varid{x}){}\<[41]%
\>[41]{}=\hspace{-3pt}\{\; \Conid{Refl}\;\}\hspace{-3pt}={}\<[E]%
\\
\>[4]{}\hsindent{1}{}\<[5]%
\>[5]{}((\Varid{f}\mathbin{\mathbin{>}\hspace{-5.2pt}\mathrel{=}\hspace{-5.5pt}\mathbin{>}}\Varid{flowMonR}\;\Varid{f}\;(\Conid{S}\;\Varid{n}))\;\Varid{x})\;{}\<[41]%
\>[41]{}\Conid{QED}{}\<[E]%
\ColumnHook
\end{hscode}\resethooks
Finally, the extensional equality of \ensuremath{\Varid{flowMonL}} and \ensuremath{\Varid{flowMonR}}
\begin{hscode}\SaveRestoreHook
\column{B}{@{}>{\hspre}l<{\hspost}@{}}%
\column{3}{@{}>{\hspre}l<{\hspost}@{}}%
\column{4}{@{}>{\hspre}l<{\hspost}@{}}%
\column{5}{@{}>{\hspre}l<{\hspost}@{}}%
\column{9}{@{}>{\hspre}l<{\hspost}@{}}%
\column{14}{@{}>{\hspre}l<{\hspost}@{}}%
\column{17}{@{}>{\hspre}c<{\hspost}@{}}%
\column{17E}{@{}l@{}}%
\column{20}{@{}>{\hspre}l<{\hspost}@{}}%
\column{26}{@{}>{\hspre}l<{\hspost}@{}}%
\column{31}{@{}>{\hspre}l<{\hspost}@{}}%
\column{34}{@{}>{\hspre}l<{\hspost}@{}}%
\column{52}{@{}>{\hspre}l<{\hspost}@{}}%
\column{72}{@{}>{\hspre}l<{\hspost}@{}}%
\column{E}{@{}>{\hspre}l<{\hspost}@{}}%
\>[3]{}\Varid{flowMonLemma}{}\<[17]%
\>[17]{}\ \mathop{:}\ {}\<[17E]%
\>[20]{}\{\mskip1.5mu \Conid{X}\ \mathop{:}\ \Conid{Type}\mskip1.5mu\}\,\to\,\{\mskip1.5mu \Conid{M}\ \mathop{:}\ \Conid{Type}\,\to\,\Conid{Type}\mskip1.5mu\}\,\to\,\Conid{Monad}\;\Conid{M}\Rightarrow {}\<[E]%
\\
\>[20]{}(\Varid{f}\ \mathop{:}\ \Conid{X}\,\to\,\Conid{M}\;\Conid{X})\,\to\,(\Varid{n}\ \mathop{:}\ \mathbb{N})\,\to\,\Varid{flowMonL}\;\Varid{f}\;\Varid{n}\doteq\Varid{flowMonR}\;\Varid{f}\;\Varid{n}{}\<[E]%
\\[\blanklineskip]%
\>[3]{}\Varid{flowMonLemma}\;\Varid{f}\;\Conid{Z}\;\Varid{x}\mathrel{=}\Conid{Refl}{}\<[E]%
\\[\blanklineskip]%
\>[3]{}\Varid{flowMonLemma}\;\Varid{f}\;(\Conid{S}\;\Varid{n})\;\Varid{x}\mathrel{=}{}\<[E]%
\\
\>[3]{}\hsindent{1}{}\<[4]%
\>[4]{}\mathbf{let}\;{}\<[9]%
\>[9]{}\Varid{fLn}{}\<[14]%
\>[14]{}\mathrel{=}\Varid{flowMonL}\;{}\<[26]%
\>[26]{}\Varid{f}\;\Varid{n}{}\<[E]%
\\
\>[9]{}\Varid{fRn}{}\<[14]%
\>[14]{}\mathrel{=}\Varid{flowMonR}\;{}\<[26]%
\>[26]{}\Varid{f}\;\Varid{n}\;{}\<[31]%
\>[31]{}\mathbf{in}{}\<[E]%
\\
\>[4]{}\hsindent{1}{}\<[5]%
\>[5]{}(\Varid{flowMonL}\;\Varid{f}\;(\Conid{S}\;\Varid{n})\;\Varid{x}){}\<[34]%
\>[34]{}=\hspace{-3pt}\{\; \Conid{Refl}\;\}\hspace{-3pt}={}\<[E]%
\\
\>[4]{}\hsindent{1}{}\<[5]%
\>[5]{}((\Varid{fLn}\mathbin{\mathbin{>}\hspace{-5.2pt}\mathrel{=}\hspace{-5.5pt}\mathbin{>}}\Varid{f})\;\Varid{x}){}\<[34]%
\>[34]{}=\hspace{-3pt}\{\; \Varid{kleisliPresEE}\;{}\<[52]%
\>[52]{}\Varid{fLn}\;\Varid{fRn}\;{}\<[72]%
\>[72]{}\Varid{f}\;\Varid{f}\;{}\<[E]%
\\
\>[52]{}(\Varid{flowMonLemma}\;\Varid{f}\;\Varid{n})\;{}\<[72]%
\>[72]{}(\lambda \Varid{v}\Rightarrow \Conid{Refl})\;\Varid{x}\;\}\hspace{-3pt}={}\<[E]%
\\
\>[4]{}\hsindent{1}{}\<[5]%
\>[5]{}((\Varid{fRn}\mathbin{\mathbin{>}\hspace{-5.2pt}\mathrel{=}\hspace{-5.5pt}\mathbin{>}}\Varid{f})\;\Varid{x}){}\<[34]%
\>[34]{}=\hspace{-3pt}\{\; \Varid{flowMonRLem}\;\Varid{f}\;\Varid{n}\;\Varid{x}\;\}\hspace{-3pt}={}\<[E]%
\\
\>[4]{}\hsindent{1}{}\<[5]%
\>[5]{}((\Varid{f}\mathbin{\mathbin{>}\hspace{-5.2pt}\mathrel{=}\hspace{-5.5pt}\mathbin{>}}\Varid{fRn})\;\Varid{x}){}\<[34]%
\>[34]{}=\hspace{-3pt}\{\; \Conid{Refl}\;\}\hspace{-3pt}={}\<[E]%
\\
\>[4]{}\hsindent{1}{}\<[5]%
\>[5]{}(\Varid{flowMonR}\;\Varid{f}\;(\Conid{S}\;\Varid{n})\;\Varid{x})\;{}\<[34]%
\>[34]{}\Conid{QED}{}\<[E]%
\ColumnHook
\end{hscode}\resethooks
follows from \ensuremath{\Varid{flowMonRLem}} and, again, preservation of extensional equality.

\paragraph*{Discussion.} Before we turn to the next section, let us discuss one
objection to what we have just done.
Why have we not tried to prove that \ensuremath{\Varid{flowMonL}\;\Varid{f}\;\Varid{n}} and \ensuremath{\Varid{flowMonR}\;\Varid{f}\;\Varid{n}}
are \emph{intensionally} equal as we did for the deterministic flows?
If we managed to show the intensional equality of the two flow
computations, their extensional equality would follow.

The problem with that approach is that it would require much stronger
assumptions: \ensuremath{\Varid{pureLeftIdKleisli}}, \ensuremath{\Varid{pureRightIdKleisli}}, \ensuremath{\Varid{kleisliAssoc}}
and \ensuremath{\Varid{kleisliSpec}} would need to hold intensionally.
For example, it would require \ensuremath{\Varid{f}\mathbin{\mathbin{>}\hspace{-5.2pt}\mathrel{=}\hspace{-5.5pt}\mathbin{>}}\Varid{g}} to be intensionally equal to
\ensuremath{\Varid{join}\mathbin{\circ}\Varid{map}\;\Varid{g}\mathbin{\circ}\Varid{f}}.
In section \ref{section:monads} we will see that, in some abstract
data types for monads this is indeed the case, but to require all of
these would make our monad interface impossible (or at least very
hard) to implement.
In general, we cannot rely on \ensuremath{\Varid{f}\mathbin{\mathbin{>}\hspace{-5.2pt}\mathrel{=}\hspace{-5.5pt}\mathbin{>}}\Varid{g}} to be intensionally equal to
\ensuremath{\Varid{join}\mathbin{\circ}\Varid{map}\;\Varid{g}\mathbin{\circ}\Varid{f}}.

In designing ADTs and formulating generic results, we have to be careful
not to impose too strong proof obligations on implementors.
%
%
Requiring the monad operations to fulfil intensional equalities would
perhaps not be as bad as pretending that function extensionality holds
in general, but would still imply unnecessary restrictions.
By contrast, requiring proper functors to preserve the extensional
equality of arrows is a natural, minimal invasive specification:
it allows one to leverage on what \ensuremath{\Conid{List}}, \ensuremath{\Conid{Maybe}}, \ensuremath{\Conid{Dist}}
\citep{10.1017/S0956796805005721}, \ensuremath{\Conid{SimpleProb}}
\citep{2017_Botta_Jansson_Ionescu} and many other monads that are
relevant for applications are known to fulfil, derive generic verified
implementations, avoid boilerplate code and improve the
understandability of proofs.

\section{Functors and extensional equality preservation}
\label{section:functors}

In category theory, a functor \ensuremath{\Conid{F}} is a structure-preserving mapping
between two categories $\mathcal{C}$ and $\mathcal{D}$.
A functor is both
a total function from the objects of $\mathcal{C}$ to the objects of
$\mathcal{D}$ and
a total function from the arrows of $\mathcal{C}$ to the arrows of
$\mathcal{D}$ (often both denoted by \ensuremath{\Conid{F}}) such that for each arrow \ensuremath{\Varid{f}\ \mathop{:}\ \Conid{A}\,\to\,\Conid{B}} in $\mathcal{C}$ there is an arrow \ensuremath{\Conid{F}\;\Varid{f}\ \mathop{:}\ \Conid{F}\;\Conid{A}\,\to\,\Conid{F}\;\Conid{B}} in
$\mathcal{D}$.
For an introduction to category theory, see \citet{pierce_basic_1991}.
The arrow map preserves identity arrows and arrow composition.
In formulas:
\begin{hscode}\SaveRestoreHook
\column{B}{@{}>{\hspre}l<{\hspost}@{}}%
\column{3}{@{}>{\hspre}l<{\hspost}@{}}%
\column{14}{@{}>{\hspre}l<{\hspost}@{}}%
\column{E}{@{}>{\hspre}l<{\hspost}@{}}%
\>[3]{}\Conid{F}\;\Varid{id}_{\Conid{A}}{}\<[14]%
\>[14]{}\mathrel{=}\Varid{id}_{\Conid{F}\;\Conid{A}}{}\<[E]%
\\[\blanklineskip]%
\>[3]{}\Conid{F}\;(\Varid{g}\mathbin{\circ}\Varid{f}){}\<[14]%
\>[14]{}\mathrel{=}\Conid{F}\;\Varid{g}\mathbin{\circ}\Conid{F}\;\Varid{f}{}\<[E]%
\ColumnHook
\end{hscode}\resethooks
Here \ensuremath{\Conid{A}} denotes an object of $\mathcal{C}$, \ensuremath{\Conid{F}\;\Conid{A}} the corresponding
object of $\mathcal{D}$ under \ensuremath{\Conid{F}}, \ensuremath{\Varid{id}_{\Conid{A}}} and \ensuremath{\Varid{id}_{\Conid{F}\;\Conid{A}}} denote the
identity arrows of \ensuremath{\Conid{A}} and \ensuremath{\Conid{F}\;\Conid{A}} in $\mathcal{C}$ and $\mathcal{D}$,
respectively and $g$ and $f$ denote arrows between suitable objects in
$\mathcal{C}$.
In $\mathcal{D}$, \ensuremath{\Conid{F}\;\Varid{id}_{\Conid{A}}}, \ensuremath{\Conid{F}\;(\Varid{g}\mathbin{\circ}\Varid{f})}, \ensuremath{\Conid{F}\;\Varid{g}} and \ensuremath{\Conid{F}\;\Varid{f}} denote the
arrows corresponding to the $\mathcal{C}$-arrows \ensuremath{\Varid{id}_{\Conid{A}}}, \ensuremath{\Varid{g}\mathbin{\circ}\Varid{f}}, \ensuremath{\Varid{g}}
and \ensuremath{\Varid{f}}.

\paragraph*{Level of formalization.}
When considering ADT specifications of functor (and of natural
transformation, monad, etc.) in dependently typed languages, one has to
distinguish between two related but different situations.

One in which the specification is put forward in the context of
formalizations of category theory, see for example \citep{UniMath,
CoqProofAssistant}. In this situation, one has to expect the notion of
category to be in place and that of functor to be predicated on that of
its source and target categories. A functor ADT in this situation is an
answer to the question ``What shall the notion of functor look like in
dependently typed formalizations of category theory?''

A different situation is the one in which, in a dependently typed
language, we consider the category whose objects are types (in Idris,
values of type \ensuremath{\Conid{Type}}), arrows are functions, and functors are
of type \ensuremath{\Conid{Type}\,\to\,\Conid{Type}}.
In this case, a functor ADT is an answer to the question ``What does
it mean for a value of type \ensuremath{\Conid{Type}\,\to\,\Conid{Type}} to be a functor?''  and
category theory plays the role of a meta-theory that we use to
motivate the specification.

The latter situation is the one considered in this paper.
More specifically, we consider the ADTs encoded in the Haskell type
classes \ensuremath{\Conid{Functor}} and \ensuremath{\Conid{Monad}} and ask ourselves what are meaningful
specifications for these ADTs in dependently typed languages.

\paragraph*{Towards an ADT for functors.}
In Idris, the notion of a functor that preserves identity and
composition can be specified as follows (but this is not our final version):
\begin{hscode}\SaveRestoreHook
\column{B}{@{}>{\hspre}l<{\hspost}@{}}%
\column{3}{@{}>{\hspre}l<{\hspost}@{}}%
\column{5}{@{}>{\hspre}l<{\hspost}@{}}%
\column{18}{@{}>{\hspre}c<{\hspost}@{}}%
\column{18E}{@{}l@{}}%
\column{21}{@{}>{\hspre}l<{\hspost}@{}}%
\column{39}{@{}>{\hspre}c<{\hspost}@{}}%
\column{39E}{@{}l@{}}%
\column{43}{@{}>{\hspre}l<{\hspost}@{}}%
\column{60}{@{}>{\hspre}l<{\hspost}@{}}%
\column{E}{@{}>{\hspre}l<{\hspost}@{}}%
\>[3]{}\bf{interface}\;\Conid{Functor}\;(\Conid{F}\ \mathop{:}\ \Conid{Type}\,\to\,\Conid{Type})\;\mathbf{where}{}\<[E]%
\\
\>[3]{}\hsindent{2}{}\<[5]%
\>[5]{}\Varid{map}{}\<[18]%
\>[18]{}\ \mathop{:}\ {}\<[18E]%
\>[21]{}\{\mskip1.5mu \Conid{A},\Conid{B}\ \mathop{:}\ \Conid{Type}\mskip1.5mu\}{}\<[39]%
\>[39]{}\,\to\,{}\<[39E]%
\>[43]{}(\Conid{A}\,\to\,\Conid{B})\,\to\,\Conid{F}\;\Conid{A}\,\to\,\Conid{F}\;\Conid{B}{}\<[E]%
\\[\blanklineskip]%
\>[3]{}\hsindent{2}{}\<[5]%
\>[5]{}\Varid{mapPresId}{}\<[18]%
\>[18]{}\ \mathop{:}\ {}\<[18E]%
\>[21]{}\{\mskip1.5mu \Conid{A}\ \mathop{:}\ \Conid{Type}\mskip1.5mu\}{}\<[39]%
\>[39]{}\,\to\,{}\<[39E]%
\>[43]{}\Varid{map}\;\Varid{id}\doteq\Varid{id}\;\{\mskip1.5mu \Varid{a}\mathrel{=}\Conid{F}\;\Conid{A}\mskip1.5mu\}{}\<[E]%
\\
\>[3]{}\hsindent{2}{}\<[5]%
\>[5]{}\Varid{mapPresComp}{}\<[18]%
\>[18]{}\ \mathop{:}\ {}\<[18E]%
\>[21]{}\{\mskip1.5mu \Conid{A},\Conid{B},\Conid{C}\ \mathop{:}\ \Conid{Type}\mskip1.5mu\}{}\<[39]%
\>[39]{}\,\to\,{}\<[39E]%
\>[43]{}(\Varid{g}\ \mathop{:}\ \Conid{B}\,\to\,\Conid{C})\,\to\,(\Varid{f}\ \mathop{:}\ \Conid{A}\,\to\,\Conid{B})\,\to\,{}\<[E]%
\\
\>[43]{}\Varid{map}\;(\Varid{g}\mathbin{\circ}\Varid{f})\doteq{}\<[60]%
\>[60]{}\Varid{map}\;\Varid{g}\mathbin{\circ}\Varid{map}\;\Varid{f}{}\<[E]%
\ColumnHook
\end{hscode}\resethooks
In \ensuremath{\Varid{mapPresId}} we have to help the type checker a little bit and give
the domain of the two functions that are posited to be extensionally
equal explicitly.
%

Notice that the function \ensuremath{\Varid{map}} is required to preserve identity and
composition \emph{extensionally}.\label{text:MathEqExtEq}
In other words, \ensuremath{\Conid{Functor}} does not require \ensuremath{\Varid{map}\;\Varid{id}} and \ensuremath{\Varid{id}} (\ensuremath{\Varid{map}\;(\Varid{g}\mathbin{\circ}\Varid{f})} and \ensuremath{\Varid{map}\;\Varid{g}\mathbin{\circ}\Varid{map}\;\Varid{f}}) to be intensionally equal but only to be
equal extensionally.
This is for very good reasons!
If functors were required to preserve identity and composition
intensionally, the interface would be hardly implementable.
By contrast,
\ensuremath{\Conid{Identity}}, \ensuremath{\Conid{List}}, \ensuremath{\Conid{Maybe}},
\ensuremath{\Conid{Vect}\;\Varid{n}} and many other important type constructors are functors in the
sense specified by this \ensuremath{\Conid{Functor}} interface.

Does this \ensuremath{\Conid{Functor}} interface represent a suitable Idris implementation
of the notion of functor in dependently typed languages?
We argue that this is not the case and that beside requiring from \ensuremath{\Varid{map}}
preservation of identity and of composition, one should
additionally require preservation of extensional equality.
In other words, we argue that the above specification of \ensuremath{\Conid{Functor}} is
incomplete.
A more complete specification could look like
\begin{hscode}\SaveRestoreHook
\column{B}{@{}>{\hspre}l<{\hspost}@{}}%
\column{3}{@{}>{\hspre}l<{\hspost}@{}}%
\column{5}{@{}>{\hspre}l<{\hspost}@{}}%
\column{18}{@{}>{\hspre}c<{\hspost}@{}}%
\column{18E}{@{}l@{}}%
\column{21}{@{}>{\hspre}l<{\hspost}@{}}%
\column{39}{@{}>{\hspre}c<{\hspost}@{}}%
\column{39E}{@{}l@{}}%
\column{43}{@{}>{\hspre}l<{\hspost}@{}}%
\column{60}{@{}>{\hspre}l<{\hspost}@{}}%
\column{63}{@{}>{\hspre}l<{\hspost}@{}}%
\column{E}{@{}>{\hspre}l<{\hspost}@{}}%
\>[3]{}\bf{interface}\;\Conid{Functor}\;(\Conid{F}\ \mathop{:}\ \Conid{Type}\,\to\,\Conid{Type})\;\mathbf{where}{}\<[E]%
\\
\>[3]{}\hsindent{2}{}\<[5]%
\>[5]{}\Varid{map}{}\<[18]%
\>[18]{}\ \mathop{:}\ {}\<[18E]%
\>[21]{}\{\mskip1.5mu \Conid{A},\Conid{B}\ \mathop{:}\ \Conid{Type}\mskip1.5mu\}{}\<[39]%
\>[39]{}\,\to\,{}\<[39E]%
\>[43]{}(\Conid{A}\,\to\,\Conid{B})\,\to\,\Conid{F}\;\Conid{A}\,\to\,\Conid{F}\;\Conid{B}{}\<[E]%
\\[\blanklineskip]%
\>[3]{}\hsindent{2}{}\<[5]%
\>[5]{}\Varid{mapPresEE}{}\<[18]%
\>[18]{}\ \mathop{:}\ {}\<[18E]%
\>[21]{}\{\mskip1.5mu \Conid{A},\Conid{B}\ \mathop{:}\ \Conid{Type}\mskip1.5mu\}{}\<[39]%
\>[39]{}\,\to\,{}\<[39E]%
\>[43]{}(\Varid{f},\Varid{g}\ \mathop{:}\ \Conid{A}\,\to\,\Conid{B})\,\to\,{}\<[63]%
\>[63]{}\Varid{f}\doteq\Varid{g}\,\to\,\Varid{map}\;\Varid{f}\doteq\Varid{map}\;\Varid{g}\mbox{\onelinecomment  New!}{}\<[E]%
\\[\blanklineskip]%
\>[3]{}\hsindent{2}{}\<[5]%
\>[5]{}\Varid{mapPresId}{}\<[18]%
\>[18]{}\ \mathop{:}\ {}\<[18E]%
\>[21]{}\{\mskip1.5mu \Conid{A}\ \mathop{:}\ \Conid{Type}\mskip1.5mu\}{}\<[39]%
\>[39]{}\,\to\,{}\<[39E]%
\>[43]{}\Varid{map}\;\Varid{id}\doteq\Varid{id}\;\{\mskip1.5mu \Varid{a}\mathrel{=}\Conid{F}\;\Conid{A}\mskip1.5mu\}{}\<[E]%
\\
\>[3]{}\hsindent{2}{}\<[5]%
\>[5]{}\Varid{mapPresComp}{}\<[18]%
\>[18]{}\ \mathop{:}\ {}\<[18E]%
\>[21]{}\{\mskip1.5mu \Conid{A},\Conid{B},\Conid{C}\ \mathop{:}\ \Conid{Type}\mskip1.5mu\}{}\<[39]%
\>[39]{}\,\to\,{}\<[39E]%
\>[43]{}(\Varid{g}\ \mathop{:}\ \Conid{B}\,\to\,\Conid{C})\,\to\,(\Varid{f}\ \mathop{:}\ \Conid{A}\,\to\,\Conid{B})\,\to\,{}\<[E]%
\\
\>[43]{}\Varid{map}\;(\Varid{g}\mathbin{\circ}\Varid{f})\doteq{}\<[60]%
\>[60]{}\Varid{map}\;\Varid{g}\mathbin{\circ}\Varid{map}\;\Varid{f}{}\<[E]%
\ColumnHook
\end{hscode}\resethooks
The \ensuremath{\Conid{Identity}} functor, \ensuremath{\Conid{List}}, \ensuremath{\Conid{Maybe}}, \ensuremath{\Conid{Vect}\;\Varid{n}} and, more generally,
container-like functors built from algebraic datatypes, fulfil the
complete specification and the proofs for \ensuremath{\Varid{mapPresEE}} do not add
significant work.
But other prominent functors such as \ensuremath{\Conid{Reader}} do not fulfil the above
specification as we will explain below.

Note that it is quite possible to continue on the road towards full
generality (supporting a larger class of functors) by parameterising
over the equalities used, but this leads to quite a bit of book-keeping
(basically a setoid-based framework).
We instead stop at this point and show that it is a pragmatic
compromise between generality and convenient usage.

\paragraph*{Equality preservation examples.}
Let's first have a look at \ensuremath{\Varid{map}} and a proof of \ensuremath{\Varid{mapPresEE}} for
\ensuremath{\Conid{List}}, one of the functors that fulfil the above specification:
\begin{hscode}\SaveRestoreHook
\column{B}{@{}>{\hspre}l<{\hspost}@{}}%
\column{3}{@{}>{\hspre}l<{\hspost}@{}}%
\column{24}{@{}>{\hspre}l<{\hspost}@{}}%
\column{31}{@{}>{\hspre}c<{\hspost}@{}}%
\column{31E}{@{}l@{}}%
\column{35}{@{}>{\hspre}l<{\hspost}@{}}%
\column{E}{@{}>{\hspre}l<{\hspost}@{}}%
\>[3]{}\Varid{mapList}\ \mathop{:}\ \{\mskip1.5mu \Conid{A},\Conid{B}\ \mathop{:}\ \Conid{Type}\mskip1.5mu\}\,\to\,(\Conid{A}\,\to\,\Conid{B})\,\to\,(\Conid{List}\;\Conid{A}\,\to\,\Conid{List}\;\Conid{B}){}\<[E]%
\\
\>[3]{}\Varid{mapList}\;\Varid{f}\;[\mskip1.5mu \mskip1.5mu]{}\<[24]%
\>[24]{}\mathrel{=}[\mskip1.5mu \mskip1.5mu]{}\<[E]%
\\
\>[3]{}\Varid{mapList}\;\Varid{f}\;(\Varid{a}\mathbin{::}\Varid{as}){}\<[24]%
\>[24]{}\mathrel{=}\Varid{f}\;\Varid{a}{}\<[31]%
\>[31]{}\mathbin{::}{}\<[31E]%
\>[35]{}\Varid{mapList}\;\Varid{f}\;\Varid{as}{}\<[E]%
\ColumnHook
\end{hscode}\resethooks
Written out in equational reasoning style, the preservation of EE
proof looks as follows:
\begin{hscode}\SaveRestoreHook
\column{B}{@{}>{\hspre}l<{\hspost}@{}}%
\column{3}{@{}>{\hspre}l<{\hspost}@{}}%
\column{5}{@{}>{\hspre}l<{\hspost}@{}}%
\column{20}{@{}>{\hspre}l<{\hspost}@{}}%
\column{37}{@{}>{\hspre}c<{\hspost}@{}}%
\column{37E}{@{}l@{}}%
\column{40}{@{}>{\hspre}l<{\hspost}@{}}%
\column{44}{@{}>{\hspre}l<{\hspost}@{}}%
\column{E}{@{}>{\hspre}l<{\hspost}@{}}%
\>[3]{}\Varid{mapListPresEE}\ \mathop{:}\ {}\<[20]%
\>[20]{}\{\mskip1.5mu \Conid{A},\Conid{B}\ \mathop{:}\ \Conid{Type}\mskip1.5mu\}\,\to\,(\Varid{f},\Varid{g}\ \mathop{:}\ \Conid{A}\,\to\,\Conid{B})\,\to\,\Varid{f}\doteq\Varid{g}\,\to\,\Varid{mapList}\;\Varid{f}\doteq\Varid{mapList}\;\Varid{g}{}\<[E]%
\\
\>[3]{}\Varid{mapListPresEE}\;\Varid{f}\;\Varid{g}\;\Varid{fEEg}\;[\mskip1.5mu \mskip1.5mu]{}\<[37]%
\>[37]{}\mathrel{=}{}\<[37E]%
\>[44]{}\Conid{Refl}{}\<[E]%
\\
\>[3]{}\Varid{mapListPresEE}\;\Varid{f}\;\Varid{g}\;\Varid{fEEg}\;(\Varid{a}\mathbin{::}\Varid{as}){}\<[37]%
\>[37]{}\mathrel{=}{}\<[37E]%
\>[40]{}\,{}\<[E]%
\\
\>[3]{}\hsindent{2}{}\<[5]%
\>[5]{}(\Varid{mapList}\;\Varid{f}\;(\Varid{a}\mathbin{::}\Varid{as})){}\<[40]%
\>[40]{}=\hspace{-3pt}\{\; {}\<[44]%
\>[44]{}\Conid{Refl}\;\}\hspace{-3pt}={}\<[E]%
\\
\>[3]{}\hsindent{2}{}\<[5]%
\>[5]{}(\Varid{f}\;\Varid{a}\mathbin{::}\Varid{mapList}\;\Varid{f}\;\Varid{as}){}\<[40]%
\>[40]{}=\hspace{-3pt}\{\; {}\<[44]%
\>[44]{}\Varid{cong}\;\{\mskip1.5mu \Varid{f}\mathrel{=}(\mathbin{::}~\Varid{mapList}\;\Varid{f}\;\Varid{as})\mskip1.5mu\}\;(\Varid{fEEg}\;\Varid{a})\;\}\hspace{-3pt}={}\<[E]%
\\
\>[3]{}\hsindent{2}{}\<[5]%
\>[5]{}(\Varid{g}\;\Varid{a}\mathbin{::}\Varid{mapList}\;\Varid{f}\;\Varid{as}){}\<[40]%
\>[40]{}=\hspace{-3pt}\{\; {}\<[44]%
\>[44]{}\Varid{cong}\;(\Varid{mapListPresEE}\;\Varid{f}\;\Varid{g}\;\Varid{fEEg}\;\Varid{as})\;\}\hspace{-3pt}={}\<[E]%
\\
\>[3]{}\hsindent{2}{}\<[5]%
\>[5]{}(\Varid{g}\;\Varid{a}\mathbin{::}\Varid{mapList}\;\Varid{g}\;\Varid{as}){}\<[40]%
\>[40]{}=\hspace{-3pt}\{\; {}\<[44]%
\>[44]{}\Conid{Refl}\;\}\hspace{-3pt}={}\<[E]%
\\
\>[3]{}\hsindent{2}{}\<[5]%
\>[5]{}(\Varid{mapList}\;\Varid{g}\;(\Varid{a}\mathbin{::}\Varid{as}))\;{}\<[40]%
\>[40]{}\Conid{QED}{}\<[E]%
\ColumnHook
\end{hscode}\resethooks
In general the proofs have a very simple structure: they use the \ensuremath{\Varid{f}\doteq\Varid{g}} arguments at the ``leaves'', and otherwise only use the
induction hypotheses.
%
%
With a suitable universe of codes for types, or a library for
parametricity proofs, these proofs can be automated using
datatype-generic programming.

Let's now turn to a type constructor that is not an instance of
our \ensuremath{\Conid{Functor}}, namely \ensuremath{\Conid{Reader}\;\Conid{E}} for some environment \ensuremath{\Conid{E}\ \mathop{:}\ \Conid{Type}}.
\begin{hscode}\SaveRestoreHook
\column{B}{@{}>{\hspre}l<{\hspost}@{}}%
\column{3}{@{}>{\hspre}l<{\hspost}@{}}%
\column{E}{@{}>{\hspre}l<{\hspost}@{}}%
\>[3]{}\Conid{Reader}\ \mathop{:}\ \Conid{Type}\,\to\,\Conid{Type}\,\to\,\Conid{Type}{}\<[E]%
\\
\>[3]{}\Conid{Reader}\;\Conid{E}\;\Conid{A}\mathrel{=}\Conid{E}\,\to\,\Conid{A}{}\<[E]%
\\[\blanklineskip]%
\>[3]{}\Varid{mapR}\ \mathop{:}\ \{\mskip1.5mu \Conid{A},\Conid{B},\Conid{E}\ \mathop{:}\ \Conid{Type}\mskip1.5mu\}\,\to\,(\Conid{A}\,\to\,\Conid{B})\,\to\,(\Conid{Reader}\;\Conid{E}\;\Conid{A}\,\to\,\Conid{Reader}\;\Conid{E}\;\Conid{B}){}\<[E]%
\\
\>[3]{}\Varid{mapR}\;\Varid{f}\;\Varid{r}\mathrel{=}\Varid{f}\mathbin{\circ}\Varid{r}{}\<[E]%
\ColumnHook
\end{hscode}\resethooks
If we try to implement preservation of extensional equality we end up with

\begin{hscode}\SaveRestoreHook
\column{B}{@{}>{\hspre}l<{\hspost}@{}}%
\column{3}{@{}>{\hspre}l<{\hspost}@{}}%
\column{5}{@{}>{\hspre}l<{\hspost}@{}}%
\column{17}{@{}>{\hspre}l<{\hspost}@{}}%
\column{26}{@{}>{\hspre}c<{\hspost}@{}}%
\column{26E}{@{}l@{}}%
\column{29}{@{}>{\hspre}l<{\hspost}@{}}%
\column{50}{@{}>{\hspre}l<{\hspost}@{}}%
\column{E}{@{}>{\hspre}l<{\hspost}@{}}%
\>[3]{}\Varid{mapRPresEE}\ \mathop{:}\ {}\<[17]%
\>[17]{}\{\mskip1.5mu \Conid{A},\Conid{B}\ \mathop{:}\ \Conid{Type}\mskip1.5mu\}\,\to\,(\Varid{f},\Varid{g}\ \mathop{:}\ \Conid{A}\,\to\,\Conid{B})\,\to\,\Varid{f}\doteq\Varid{g}\,\to\,\Varid{mapR}\;\Varid{f}\doteq\Varid{mapR}\;\Varid{g}{}\<[E]%
\\
\>[3]{}\Varid{mapRPresEE}\;\Varid{f}\;\Varid{g}\;\Varid{fEEg}\;\Varid{r}{}\<[26]%
\>[26]{}\mathrel{=}{}\<[26E]%
\\
\>[3]{}\hsindent{2}{}\<[5]%
\>[5]{}(\Varid{mapR}\;\Varid{f}\;\Varid{r}){}\<[29]%
\>[29]{}=\hspace{-3pt}\{\; \Conid{Refl}\;\}\hspace{-3pt}={}\<[E]%
\\
\>[3]{}\hsindent{2}{}\<[5]%
\>[5]{}(\Varid{f}\mathbin{\circ}\Varid{r}){}\<[29]%
\>[29]{}=\hspace{-3pt}\{\; \bf{?whatnow}\;\}\hspace{-3pt}={}\<[50]%
\>[50]{}\mbox{\onelinecomment  here we need \ensuremath{\Varid{f}\mathrel{=}\Varid{g}} to proceed}{}\<[E]%
\\
\>[3]{}\hsindent{2}{}\<[5]%
\>[5]{}(\Varid{g}\mathbin{\circ}\Varid{r}){}\<[29]%
\>[29]{}=\hspace{-3pt}\{\; \Conid{Refl}\;\}\hspace{-3pt}={}\<[E]%
\\
\>[3]{}\hsindent{2}{}\<[5]%
\>[5]{}(\Varid{mapR}\;\Varid{g}\;\Varid{r})\;{}\<[29]%
\>[29]{}\Conid{QED}{}\<[E]%
\ColumnHook
\end{hscode}\resethooks
Notice the question mark in front of \ensuremath{\bf{whatnow}}. This introduces an
unresolved proof step and allows us to ask Idris to help us implementing
this step, see ``Elaborator Reflection -- Holes'' in \citep{idrisdocs}.
Among other things, we can ask about the
type of \ensuremath{\bf{whatnow}}. Perhaps not surprisingly, this turns out to be \ensuremath{\Varid{f}\mathbin{\circ}\Varid{r}\mathrel{=}\Varid{g}\mathbin{\circ}\Varid{r}}.

The problem is that, although we know that \ensuremath{(\Varid{f}\mathbin{\circ}\Varid{r})\;\Varid{e}\mathrel{=}(\Varid{g}\mathbin{\circ}\Varid{r})\;\Varid{e}} for
all \ensuremath{\Varid{e}\ \mathop{:}\ \Conid{E}}, we cannot deduce \ensuremath{\Varid{f}\mathbin{\circ}\Varid{r}\mathrel{=}\Varid{g}\mathbin{\circ}\Varid{r}} without extensionality.
Thus \ensuremath{\Conid{Reader}\;\Conid{E}} does not implement the \ensuremath{\Conid{Functor}} interface, but it
is ``very close''.
Using the 2-argument version of function extensionality \ensuremath{(\stackrel{..}{=})\mathrel{=}\Varid{extify}\;(\doteq)} it is easy to show
\begin{hscode}\SaveRestoreHook
\column{B}{@{}>{\hspre}l<{\hspost}@{}}%
\column{3}{@{}>{\hspre}l<{\hspost}@{}}%
\column{18}{@{}>{\hspre}l<{\hspost}@{}}%
\column{66}{@{}>{\hspre}c<{\hspost}@{}}%
\column{66E}{@{}l@{}}%
\column{70}{@{}>{\hspre}l<{\hspost}@{}}%
\column{E}{@{}>{\hspre}l<{\hspost}@{}}%
\>[3]{}\Varid{mapRPresEE2}\ \mathop{:}\ {}\<[18]%
\>[18]{}\{\mskip1.5mu \Conid{E},\Conid{A},\Conid{B}\ \mathop{:}\ \Conid{Type}\mskip1.5mu\}\,\to\,(\Varid{f},\Varid{g}\ \mathop{:}\ \Conid{A}\,\to\,\Conid{B})\,\to\,\Varid{f}\doteq\Varid{g}{}\<[66]%
\>[66]{}\,\to\,{}\<[66E]%
\>[70]{}\Varid{mapR}\;\Varid{f}\stackrel{..}{=}\Varid{mapR}\;\Varid{g}{}\<[E]%
\\
\>[3]{}\Varid{mapRPresEE2}\;\Varid{f}\;\Varid{g}\;\Varid{fEEg}\;\Varid{r}\;\Varid{x}\mathrel{=}\Varid{fEEg}\;(\Varid{r}\;\Varid{x}){}\<[E]%
\ColumnHook
\end{hscode}\resethooks
Thus, \ensuremath{\Conid{Reader}\;\Conid{E}} is an example of a functor which does not preserve,
but rather \emph{transforms} the notion of equality.
By a similar argument, \ensuremath{\Varid{mapPresEE}} does not hold for the continuation
monad.

As we mentioned earlier, it is tempting to start adding equalities to
the interface (towards a setoid-based framework), but this is not a
path we take here.
As a small hint of the problems the setoid path leads to, consider
that we already have four different objects (\ensuremath{\Conid{A}}, \ensuremath{\Conid{B}}, \ensuremath{\Conid{F}\;\Conid{A}}, \ensuremath{\Conid{F}\;\Conid{B}})
and two arrow types (\ensuremath{\Conid{A}\,\to\,\Conid{B}}, \ensuremath{\Conid{F}\;\Conid{A}\,\to\,\Conid{F}\;\Conid{B}}), all of which could be
allowed ``their own'' notion of equality.


%

\paragraph*{Wrapping up.}
As stated in section \ref{section:about}, we argue that, for verified
generic programming, it is useful to distinguish between type
constructors whose \ensuremath{\Varid{map}} can be shown to preserve extensional equality
and type constructors for which this is not the case.
A discussion of what are appropriate names for the respective ADTs is
beyond the scope of this paper.
%
%
In the next section we explore how functors with \ensuremath{\Varid{mapPresEE}} affect
the monad ADT design.
%

\section{Verified monad interfaces}
\label{section:monads}

In this section, we review two standard notions of monads.
We discuss their mathematical equivalence and consider ``thin'' and
``fat'' monad ADT formulations.
We discuss the role of extensional equality preservation for deriving
monad laws and for verifying the equivalence between the different ADT
formulations.
There are many possible ways of formulating Monad axioms. Here, we do not
argue for or against a specific formulation but rather discuss the most
popular ones and their trade-offs.

\subsection{The traditional view}\label{sec:MonadTradView}

In category theory, a monad is
an \emph{endofunctor} \ensuremath{\Conid{M}} on a category $\mathcal{C}$ together with
two \emph{natural transformations}
\ensuremath{\eta\ \mathop{:}\ \Conid{Id}\xrightarrow{.}\Conid{M}} (the \emph{unit}) and
\ensuremath{\mu\ \mathop{:}\ \Conid{M}\mathbin{\circ}\Conid{M}\xrightarrow{.}\Conid{M}} (the \emph{multiplication})
such that, for any object \ensuremath{\Conid{A}} of $\mathcal{C}$, the following diagrams commute:

\begin{minipage}{0.4\textwidth}
\beginpgfgraphicnamed{ExtEqPres-f2}
\begin{tikzcd}[row sep=large, column sep=large]
\ensuremath{\Conid{M}\;\Conid{A}} \arrow[r, "\ensuremath{\eta_{\Conid{M}\;\Conid{A}}}"] \arrow[dr, "\ensuremath{\Varid{id}_{\Conid{M}\;\Conid{A}}}"'] & \ensuremath{\Conid{M}\;(\Conid{M}\;\Conid{A})} \arrow[d, "\ensuremath{\mu_{\Conid{A}}}"] & \ensuremath{\Conid{M}\;\Conid{A}} \arrow[l, "\ensuremath{\Conid{M}\;\eta_{\Conid{A}}}"'] \arrow[dl, "\ensuremath{\Varid{id}_{\Conid{M}\;\Conid{A}}}"] \\
                                                      & \ensuremath{\Conid{M}\;\Conid{A}}                             &
\end{tikzcd}
\endpgfgraphicnamed
\end{minipage}
\hfill
\begin{minipage}{0.4\textwidth}
\beginpgfgraphicnamed{ExtEqPres-f3}
\begin{tikzcd}[row sep=large, column sep=large]
\ensuremath{\Conid{M}\;(\Conid{M}\;(\Conid{M}\;\Conid{A}))} \arrow[r, "\ensuremath{\Conid{M}\;\mu_{\Conid{A}}}"] \arrow[d, "\ensuremath{\mu_{\Conid{M}\;\Conid{A}}}"'] & \ensuremath{\Conid{M}\;(\Conid{M}\;\Conid{A})} \arrow[d, "\ensuremath{\mu_{\Conid{A}}}"] \\
      \ensuremath{\Conid{M}\;(\Conid{M}\;\Conid{A})} \arrow[r, "\ensuremath{\mu_{\Conid{A}}}"']                                     & \ensuremath{\Conid{M}\;\Conid{A}}
\end{tikzcd}
\endpgfgraphicnamed
\end{minipage}

\noindent
The transformations \ensuremath{\eta} and \ensuremath{\mu} are families of arrows, one for
each object \ensuremath{\Conid{A}}, with types \ensuremath{\eta_{\Conid{A}}\ \mathop{:}\ \Conid{A}\,\to\,\Conid{M}\;\Conid{A}} and \ensuremath{\mu_{\Conid{A}}\ \mathop{:}\ \Conid{M}\;(\Conid{M}\;\Conid{A})\,\to\,\Conid{M}\;\Conid{A}}.
That they are \emph{natural transformations} means that the following
diagrams commute for any arrow \ensuremath{\Varid{f}\ \mathop{:}\ \Conid{A}\,\to\,\Conid{B}} in $\mathcal{C}$:

\hfill
\begin{minipage}{0.3\textwidth}
\beginpgfgraphicnamed{ExtEqPres-f4}
\begin{tikzcd}[row sep=large, column sep=large]
    \ensuremath{\Conid{A}} \arrow[r, "\ensuremath{\Varid{f}}"] \arrow[d, "\ensuremath{\eta_{\Conid{A}}}"'] & \ensuremath{\Conid{B}} \arrow[d, "\ensuremath{\eta_{\Conid{B}}}"] \\
\ensuremath{\Conid{M}\;\Conid{A}}  \arrow[r, "\ensuremath{\Conid{M}\;\Varid{f}}"']                    & \ensuremath{\Conid{M}\;\Conid{B}}
\end{tikzcd}
\endpgfgraphicnamed
\end{minipage}
\hfill
\begin{minipage}{0.4\textwidth}
\beginpgfgraphicnamed{ExtEqPres-f5}
\begin{tikzcd}[row sep=large, column sep=large]
\ensuremath{\Conid{M}\;(\Conid{M}\;\Conid{A})} \arrow[r, "\ensuremath{\Conid{M}\;(\Conid{M}\;\Varid{f})}"] \arrow[d, "\ensuremath{\mu_{\Conid{A}}}"'] & \ensuremath{\Conid{M}\;(\Conid{M}\;\Conid{B})} \arrow[d, "\ensuremath{\mu_{\Conid{B}}}"] \\
   \ensuremath{\Conid{M}\;\Conid{A}}  \arrow[r, "\ensuremath{\Conid{M}\;\Varid{f}}"']                         & \ensuremath{\Conid{M}\;\Conid{B}}
\end{tikzcd}
\endpgfgraphicnamed
\end{minipage}

\noindent
From this perspective, a monad is a functor with additional structure,
namely families of maps \ensuremath{\eta} and \ensuremath{\mu}, satisfying, for any 
arrow \ensuremath{\Varid{f}\ \mathop{:}\ \Conid{A}\,\to\,\Conid{B}}, the five properties:
\newcommand{\myeta}{\ensuremath{\eta}}
\newcommand{\mymu} {\ensuremath{\mu}}
\begin{hscode}\SaveRestoreHook
\column{B}{@{}>{\hspre}l<{\hspost}@{}}%
\column{3}{@{}>{\hspre}l<{\hspost}@{}}%
\column{49}{@{}>{\hspre}l<{\hspost}@{}}%
\column{67}{@{}>{\hspre}l<{\hspost}@{}}%
\column{E}{@{}>{\hspre}l<{\hspost}@{}}%
\>[3]{}\text{T1. Triangle left:             }{}\<[49]%
\>[49]{}\mu_{\Conid{A}}\mathbin{\circ}\eta_{\Conid{M}\;\Conid{A}}{}\<[67]%
\>[67]{}\mathrel{=}\Varid{id}_{\Conid{M}\;\Conid{A}}{}\<[E]%
\\
\>[3]{}\text{T2. Triangle right:            }{}\<[49]%
\>[49]{}\mu_{\Conid{A}}\mathbin{\circ}\Conid{M}\;\eta_{\Conid{A}}{}\<[67]%
\>[67]{}\mathrel{=}\Varid{id}_{\Conid{M}\;\Conid{A}}{}\<[E]%
\\
\>[3]{}\text{T3. Square:                    }{}\<[49]%
\>[49]{}\mu_{\Conid{A}}\mathbin{\circ}\mu_{\Conid{M}\;\Conid{A}}{}\<[67]%
\>[67]{}\mathrel{=}\mu_{\Conid{A}}\mathbin{\circ}\Conid{M}\;\mu_{\Conid{A}}{}\<[E]%
\\
\>[3]{}\text{T4. Naturality of \myeta:\quad }{}\<[49]%
\>[49]{}\Conid{M}\;\Varid{f}\mathbin{\circ}\eta_{\Conid{A}}{}\<[67]%
\>[67]{}\mathrel{=}\eta_{\Conid{B}}\mathbin{\circ}\Varid{f}{}\<[E]%
\\
\>[3]{}\text{T5. Naturality of \mymu:       }{}\<[49]%
\>[49]{}\Conid{M}\;\Varid{f}\mathbin{\circ}\mu_{\Conid{A}}{}\<[67]%
\>[67]{}\mathrel{=}\mu_{\Conid{B}}\mathbin{\circ}\Conid{M}\;(\Conid{M}\;\Varid{f}){}\<[E]%
\ColumnHook
\end{hscode}\resethooks
In functional programming, \ensuremath{\eta} is traditionally denoted by \ensuremath{\Varid{return}} or by
\ensuremath{\Varid{pure}} and \ensuremath{\mu} is traditionally called \ensuremath{\Varid{join}}.
Idris provides language support for interface \emph{refinement}.
Thus, we can leverage on the functor ADT \ensuremath{\Conid{Functor}} from section
\ref{section:functors} and define a monad to be a functor with
additional methods \ensuremath{\Varid{pure}} and \ensuremath{\Varid{join}} that satisfy the requirements
T1--T5:

\begin{hscode}\SaveRestoreHook
\column{B}{@{}>{\hspre}l<{\hspost}@{}}%
\column{5}{@{}>{\hspre}l<{\hspost}@{}}%
\column{7}{@{}>{\hspre}l<{\hspost}@{}}%
\column{27}{@{}>{\hspre}c<{\hspost}@{}}%
\column{27E}{@{}l@{}}%
\column{30}{@{}>{\hspre}l<{\hspost}@{}}%
\column{48}{@{}>{\hspre}l<{\hspost}@{}}%
\column{68}{@{}>{\hspre}l<{\hspost}@{}}%
\column{70}{@{}>{\hspre}l<{\hspost}@{}}%
\column{84}{@{}>{\hspre}c<{\hspost}@{}}%
\column{84E}{@{}l@{}}%
\column{89}{@{}>{\hspre}l<{\hspost}@{}}%
\column{E}{@{}>{\hspre}l<{\hspost}@{}}%
\>[5]{}\bf{interface}\;\Conid{Functor}\;\Conid{M}\Rightarrow \Conid{Monad}_{\!\mathrm{1}}\;(\Conid{M}\ \mathop{:}\ \Conid{Type}\,\to\,\Conid{Type})\;\mathbf{where}{}\<[E]%
\\
\>[5]{}\hsindent{2}{}\<[7]%
\>[7]{}\Varid{pure}{}\<[27]%
\>[27]{}\ \mathop{:}\ {}\<[27E]%
\>[30]{}\{\mskip1.5mu \Conid{A}\ \mathop{:}\ \Conid{Type}\mskip1.5mu\}{}\<[48]%
\>[48]{}\,\to\,\Conid{A}\,\to\,\Conid{M}\;\Conid{A}{}\<[E]%
\\
\>[5]{}\hsindent{2}{}\<[7]%
\>[7]{}\Varid{join}{}\<[27]%
\>[27]{}\ \mathop{:}\ {}\<[27E]%
\>[30]{}\{\mskip1.5mu \Conid{A}\ \mathop{:}\ \Conid{Type}\mskip1.5mu\}{}\<[48]%
\>[48]{}\,\to\,\Conid{M}\;(\Conid{M}\;\Conid{A})\,\to\,\Conid{M}\;\Conid{A}{}\<[E]%
\\[\blanklineskip]%
\>[5]{}\hsindent{2}{}\<[7]%
\>[7]{}\Varid{triangleLeft}{}\<[27]%
\>[27]{}\ \mathop{:}\ {}\<[27E]%
\>[30]{}\{\mskip1.5mu \Conid{A}\ \mathop{:}\ \Conid{Type}\mskip1.5mu\}{}\<[48]%
\>[48]{}\,\to\,\Varid{join}\mathbin{\circ}\Varid{pure}{}\<[68]%
\>[68]{}\doteq\Varid{id}\;\{\mskip1.5mu \Varid{a}\mathrel{=}\Conid{M}\;\Conid{A}\mskip1.5mu\}{}\<[E]%
\\
\>[5]{}\hsindent{2}{}\<[7]%
\>[7]{}\Varid{triangleRight}{}\<[27]%
\>[27]{}\ \mathop{:}\ {}\<[27E]%
\>[30]{}\{\mskip1.5mu \Conid{A}\ \mathop{:}\ \Conid{Type}\mskip1.5mu\}{}\<[48]%
\>[48]{}\,\to\,\Varid{join}\mathbin{\circ}\Varid{map}\;\Varid{pure}{}\<[68]%
\>[68]{}\doteq\Varid{id}\;\{\mskip1.5mu \Varid{a}\mathrel{=}\Conid{M}\;\Conid{A}\mskip1.5mu\}{}\<[E]%
\\
\>[5]{}\hsindent{2}{}\<[7]%
\>[7]{}\Varid{square}{}\<[27]%
\>[27]{}\ \mathop{:}\ {}\<[27E]%
\>[30]{}\{\mskip1.5mu \Conid{A}\ \mathop{:}\ \Conid{Type}\mskip1.5mu\}{}\<[48]%
\>[48]{}\,\to\,\Varid{join}\mathbin{\circ}\Varid{join}{}\<[68]%
\>[68]{}\doteq\Varid{join}\mathbin{\circ}\Varid{map}\;\{\mskip1.5mu \Conid{A}\mathrel{=}\Conid{M}\;(\Conid{M}\;\Conid{A})\mskip1.5mu\}\;\Varid{join}{}\<[E]%
\\[\blanklineskip]%
\>[5]{}\hsindent{2}{}\<[7]%
\>[7]{}\Varid{pureNatTrans}{}\<[27]%
\>[27]{}\ \mathop{:}\ {}\<[27E]%
\>[30]{}\{\mskip1.5mu \Conid{A},\Conid{B}\ \mathop{:}\ \Conid{Type}\mskip1.5mu\}{}\<[48]%
\>[48]{}\,\to\,(\Varid{f}\ \mathop{:}\ \Conid{A}\,\to\,\Conid{B})\,\to\,{}\<[70]%
\>[70]{}\Varid{map}\;\Varid{f}\mathbin{\circ}\Varid{pure}{}\<[84]%
\>[84]{}\doteq{}\<[84E]%
\>[89]{}\Varid{pure}\mathbin{\circ}\Varid{f}{}\<[E]%
\\
\>[5]{}\hsindent{2}{}\<[7]%
\>[7]{}\Varid{joinNatTrans}{}\<[27]%
\>[27]{}\ \mathop{:}\ {}\<[27E]%
\>[30]{}\{\mskip1.5mu \Conid{A},\Conid{B}\ \mathop{:}\ \Conid{Type}\mskip1.5mu\}{}\<[48]%
\>[48]{}\,\to\,(\Varid{f}\ \mathop{:}\ \Conid{A}\,\to\,\Conid{B})\,\to\,{}\<[70]%
\>[70]{}\Varid{map}\;\Varid{f}\mathbin{\circ}\Varid{join}{}\<[84]%
\>[84]{}\doteq{}\<[84E]%
\>[89]{}\Varid{join}\mathbin{\circ}\Varid{map}\;(\Varid{map}\;\Varid{f}){}\<[E]%
\ColumnHook
\end{hscode}\resethooks
\paragraph*{Kleisli composition in the traditional view.}
In section \ref{section:example}, we have seen that monads are equipped
with a (Kleisli) composition \ensuremath{(\mathbin{\mathbin{>}\hspace{-5.2pt}\mathrel{=}\hspace{-5.5pt}\mathbin{>}})} that we required to fulfil \ensuremath{\Varid{kleisliSpec}}:
\begin{hscode}\SaveRestoreHook
\column{B}{@{}>{\hspre}l<{\hspost}@{}}%
\column{5}{@{}>{\hspre}l<{\hspost}@{}}%
\column{18}{@{}>{\hspre}c<{\hspost}@{}}%
\column{18E}{@{}l@{}}%
\column{21}{@{}>{\hspre}l<{\hspost}@{}}%
\column{E}{@{}>{\hspre}l<{\hspost}@{}}%
\>[5]{}(\mathbin{\mathbin{>}\hspace{-5.2pt}\mathrel{=}\hspace{-5.5pt}\mathbin{>}}){}\<[18]%
\>[18]{}\ \mathop{:}\ {}\<[18E]%
\>[21]{}\{\mskip1.5mu \Conid{A},\Conid{B},\Conid{C}\ \mathop{:}\ \Conid{Type}\mskip1.5mu\}\,\to\,\{\mskip1.5mu \Conid{M}\ \mathop{:}\ \Conid{Type}\,\to\,\Conid{Type}\mskip1.5mu\}\,\to\,\Conid{Monad}_{\!\mathrm{1}}\;\Conid{M}\Rightarrow {}\<[E]%
\\
\>[21]{}(\Conid{A}\,\to\,\Conid{M}\;\Conid{B})\,\to\,(\Conid{B}\,\to\,\Conid{M}\;\Conid{C})\,\to\,(\Conid{A}\,\to\,\Conid{M}\;\Conid{C}){}\<[E]%
\\[\blanklineskip]%
\>[5]{}\Varid{kleisliSpec}{}\<[18]%
\>[18]{}\ \mathop{:}\ {}\<[18E]%
\>[21]{}\{\mskip1.5mu \Conid{A},\Conid{B},\Conid{C}\ \mathop{:}\ \Conid{Type}\mskip1.5mu\}\,\to\,\{\mskip1.5mu \Conid{M}\ \mathop{:}\ \Conid{Type}\,\to\,\Conid{Type}\mskip1.5mu\}\,\to\,\Conid{Monad}_{\!\mathrm{1}}\;\Conid{M}\Rightarrow {}\<[E]%
\\
\>[21]{}(\Varid{f}\ \mathop{:}\ \Conid{A}\,\to\,\Conid{M}\;\Conid{B})\,\to\,(\Varid{g}\ \mathop{:}\ \Conid{B}\,\to\,\Conid{M}\;\Conid{C})\,\to\,(\Varid{f}\mathbin{\mathbin{>}\hspace{-5.2pt}\mathrel{=}\hspace{-5.5pt}\mathbin{>}}\Varid{g})\doteq\Varid{join}\mathbin{\circ}\Varid{map}\;\Varid{g}\mathbin{\circ}\Varid{f}{}\<[E]%
\ColumnHook
\end{hscode}\resethooks
One way of implementing \ensuremath{(\mathbin{\mathbin{>}\hspace{-5.2pt}\mathrel{=}\hspace{-5.5pt}\mathbin{>}})} that satisfies \ensuremath{\Varid{kleisliSpec}} is to \emph{define}
\begin{hscode}\SaveRestoreHook
\column{B}{@{}>{\hspre}l<{\hspost}@{}}%
\column{5}{@{}>{\hspre}l<{\hspost}@{}}%
\column{E}{@{}>{\hspre}l<{\hspost}@{}}%
\>[5]{}\Varid{f}\mathbin{\mathbin{>}\hspace{-5.2pt}\mathrel{=}\hspace{-5.5pt}\mathbin{>}}\Varid{g}\mathrel{=}\Varid{join}\mathbin{\circ}\Varid{map}\;\Varid{g}\mathbin{\circ}\Varid{f}{}\<[E]%
\ColumnHook
\end{hscode}\resethooks
The extensional equality between \ensuremath{\Varid{f}\mathbin{\mathbin{>}\hspace{-5.2pt}\mathrel{=}\hspace{-5.5pt}\mathbin{>}}\Varid{g}} and \ensuremath{\Varid{join}\mathbin{\circ}\Varid{map}\;\Varid{g}\mathbin{\circ}\Varid{f}}
then follows directly:
\begin{hscode}\SaveRestoreHook
\column{B}{@{}>{\hspre}l<{\hspost}@{}}%
\column{5}{@{}>{\hspre}l<{\hspost}@{}}%
\column{E}{@{}>{\hspre}l<{\hspost}@{}}%
\>[5]{}\Varid{kleisliSpec}\;\Varid{f}\;\Varid{g}\mathrel{=}\lambda \Varid{x}\Rightarrow \Conid{Refl}{}\<[E]%
\ColumnHook
\end{hscode}\resethooks
The same approach can be followed for implementing bind, another monad
combinator similar to Kleisli composition:
\begin{hscode}\SaveRestoreHook
\column{B}{@{}>{\hspre}l<{\hspost}@{}}%
\column{5}{@{}>{\hspre}l<{\hspost}@{}}%
\column{12}{@{}>{\hspre}c<{\hspost}@{}}%
\column{12E}{@{}l@{}}%
\column{15}{@{}>{\hspre}l<{\hspost}@{}}%
\column{E}{@{}>{\hspre}l<{\hspost}@{}}%
\>[5]{}(\bind ){}\<[12]%
\>[12]{}\ \mathop{:}\ {}\<[12E]%
\>[15]{}\{\mskip1.5mu \Conid{A},\Conid{B}\ \mathop{:}\ \Conid{Type}\mskip1.5mu\}\,\to\,\{\mskip1.5mu \Conid{M}\ \mathop{:}\ \Conid{Type}\,\to\,\Conid{Type}\mskip1.5mu\}\,\to\,\Conid{Monad}_{\!\mathrm{1}}\;\Conid{M}\Rightarrow \Conid{M}\;\Conid{A}\,\to\,(\Conid{A}\,\to\,\Conid{M}\;\Conid{B})\,\to\,\Conid{M}\;\Conid{B}{}\<[E]%
\\
\>[5]{}\Varid{ma}\bind \Varid{f}\mathrel{=}\Varid{join}\;(\Varid{map}\;\Varid{f}\;\Varid{ma}){}\<[E]%
\ColumnHook
\end{hscode}\resethooks
Starting from a \emph{thin} monad ADT as in the example above and
adding monadic operators that fulfil a specification
\emph{by-construction}, is a viable approach.
It leads to a rich structure entailing monad laws that can be
implemented generically.
Thus, one can show that \ensuremath{\Varid{pure}} is a left and a right identity of
Kleisli composition (we show only one side here)
\label{def:pureLeftIdKleisli}\label{def:pureRightIdKleisli}
\begin{hscode}\SaveRestoreHook
\column{B}{@{}>{\hspre}l<{\hspost}@{}}%
\column{5}{@{}>{\hspre}l<{\hspost}@{}}%
\column{7}{@{}>{\hspre}l<{\hspost}@{}}%
\column{24}{@{}>{\hspre}c<{\hspost}@{}}%
\column{24E}{@{}l@{}}%
\column{28}{@{}>{\hspre}l<{\hspost}@{}}%
\column{34}{@{}>{\hspre}l<{\hspost}@{}}%
\column{47}{@{}>{\hspre}l<{\hspost}@{}}%
\column{61}{@{}>{\hspre}l<{\hspost}@{}}%
\column{E}{@{}>{\hspre}l<{\hspost}@{}}%
\>[5]{}\Varid{pureLeftIdKleisli}{}\<[24]%
\>[24]{}\ \mathop{:}\ {}\<[24E]%
\>[28]{}\{\mskip1.5mu \Conid{A},\Conid{B}\ \mathop{:}\ \Conid{Type}\mskip1.5mu\}\,\to\,\{\mskip1.5mu \Conid{M}\ \mathop{:}\ \Conid{Type}\,\to\,\Conid{Type}\mskip1.5mu\}\,\to\,\Conid{Monad}_{\!\mathrm{1}}\;\Conid{M}\Rightarrow {}\<[E]%
\\
\>[28]{}(\Varid{f}\ \mathop{:}\ \Conid{A}\,\to\,\Conid{M}\;\Conid{B})\,\to\,{}\<[47]%
\>[47]{}(\Varid{pure}\mathbin{\mathbin{>}\hspace{-5.2pt}\mathrel{=}\hspace{-5.5pt}\mathbin{>}}\Varid{f}){}\<[61]%
\>[61]{}\doteq\Varid{f}{}\<[E]%
\\
\>[5]{}\Varid{pureLeftIdKleisli}\;\Varid{f}\;\Varid{a}\mathrel{=}{}\<[E]%
\\
\>[5]{}\hsindent{2}{}\<[7]%
\>[7]{}((\Varid{pure}\mathbin{\mathbin{>}\hspace{-5.2pt}\mathrel{=}\hspace{-5.5pt}\mathbin{>}}\Varid{f})\;\Varid{a}){}\<[34]%
\>[34]{}=\hspace{-3pt}\{\; \Conid{Refl}\;\}\hspace{-3pt}={}\<[E]%
\\
\>[5]{}\hsindent{2}{}\<[7]%
\>[7]{}(\Varid{join}\;(\Varid{map}\;\Varid{f}\;(\Varid{pure}\;\Varid{a}))){}\<[34]%
\>[34]{}=\hspace{-3pt}\{\; \Varid{cong}\;\{\mskip1.5mu \Varid{f}\mathrel{=}\Varid{join}\mskip1.5mu\}\;(\Varid{pureNatTrans}\;\Varid{f}\;\Varid{a})\;\}\hspace{-3pt}={}\<[E]%
\\
\>[5]{}\hsindent{2}{}\<[7]%
\>[7]{}(\Varid{join}\;(\Varid{pure}\;(\Varid{f}\;\Varid{a}))){}\<[34]%
\>[34]{}=\hspace{-3pt}\{\; \Varid{triangleLeft}\;(\Varid{f}\;\Varid{a})\;\}\hspace{-3pt}={}\<[E]%
\\
\>[5]{}\hsindent{2}{}\<[7]%
\>[7]{}(\Varid{f}\;\Varid{a})\;{}\<[34]%
\>[34]{}\Conid{QED}{}\<[E]%
\ColumnHook
\end{hscode}\resethooks
and that Kleisli composition is associative as stated in section
\ref{section:example}, almost straightforwardly and without having to
invoke the \ensuremath{\Varid{mapPresEE}} axiom of the underlying functor.
\begin{hscode}\SaveRestoreHook
\column{B}{@{}>{\hspre}l<{\hspost}@{}}%
\column{5}{@{}>{\hspre}l<{\hspost}@{}}%
\column{7}{@{}>{\hspre}l<{\hspost}@{}}%
\column{19}{@{}>{\hspre}c<{\hspost}@{}}%
\column{19E}{@{}l@{}}%
\column{22}{@{}>{\hspre}l<{\hspost}@{}}%
\column{26}{@{}>{\hspre}l<{\hspost}@{}}%
\column{32}{@{}>{\hspre}l<{\hspost}@{}}%
\column{58}{@{}>{\hspre}l<{\hspost}@{}}%
\column{67}{@{}>{\hspre}l<{\hspost}@{}}%
\column{E}{@{}>{\hspre}l<{\hspost}@{}}%
\>[5]{}\Varid{kleisliAssoc}{}\<[19]%
\>[19]{}\ \mathop{:}\ {}\<[19E]%
\>[22]{}\{\mskip1.5mu \Conid{A},\Conid{B},\Conid{C},\Conid{D}\ \mathop{:}\ \Conid{Type}\mskip1.5mu\}\,\to\,\{\mskip1.5mu \Conid{M}\ \mathop{:}\ \Conid{Type}\,\to\,\Conid{Type}\mskip1.5mu\}\,\to\,\Conid{Monad}_{\!\mathrm{1}}\;\Conid{M}\Rightarrow {}\<[E]%
\\
\>[22]{}(\Varid{f}\ \mathop{:}\ \Conid{A}\,\to\,\Conid{M}\;\Conid{B})\,\to\,(\Varid{g}\ \mathop{:}\ \Conid{B}\,\to\,\Conid{M}\;\Conid{C})\,\to\,(\Varid{h}\ \mathop{:}\ \Conid{C}\,\to\,\Conid{M}\;\Conid{D})\,\to\,{}\<[E]%
\\
\>[22]{}((\Varid{f}\mathbin{\mathbin{>}\hspace{-5.2pt}\mathrel{=}\hspace{-5.5pt}\mathbin{>}}\Varid{g})\mathbin{\mathbin{>}\hspace{-5.2pt}\mathrel{=}\hspace{-5.5pt}\mathbin{>}}\Varid{h})\doteq(\Varid{f}\mathbin{\mathbin{>}\hspace{-5.2pt}\mathrel{=}\hspace{-5.5pt}\mathbin{>}}(\Varid{g}\mathbin{\mathbin{>}\hspace{-5.2pt}\mathrel{=}\hspace{-5.5pt}\mathbin{>}}\Varid{h})){}\<[E]%
\\
\>[5]{}\Varid{kleisliAssoc}\;\{\mskip1.5mu \Conid{M}\mskip1.5mu\}\;\{\mskip1.5mu \Conid{A}\mskip1.5mu\}\;\{\mskip1.5mu \Conid{C}\mskip1.5mu\}\;\Varid{f}\;\Varid{g}\;\Varid{h}\;\Varid{a}\mathrel{=}{}\<[E]%
\\
\>[5]{}\hsindent{2}{}\<[7]%
\>[7]{}(((\Varid{f}\mathbin{\mathbin{>}\hspace{-5.2pt}\mathrel{=}\hspace{-5.5pt}\mathbin{>}}\Varid{g})\mathbin{\mathbin{>}\hspace{-5.2pt}\mathrel{=}\hspace{-5.5pt}\mathbin{>}}\Varid{h})\;\Varid{a}){}\<[58]%
\>[58]{}=\hspace{-3pt}\{\; \Conid{Refl}\;\}\hspace{-3pt}={}\<[E]%
\\
\>[5]{}\hsindent{2}{}\<[7]%
\>[7]{}((\Varid{join}\mathbin{\circ}\Varid{map}\;\Varid{h}\mathbin{\circ}{}\<[26]%
\>[26]{}\Varid{join}{}\<[32]%
\>[32]{}\mathbin{\circ}\Varid{map}\;\Varid{g}\mathbin{\circ}\Varid{f})\;\Varid{a}){}\<[58]%
\>[58]{}=\hspace{-3pt}\{\; \Varid{cong}\;(\Varid{joinNatTrans}\;\Varid{h}\;(\Varid{map}\;\Varid{g}\;(\Varid{f}\;\Varid{a})))\;\}\hspace{-3pt}={}\<[E]%
\\
\>[5]{}\hsindent{2}{}\<[7]%
\>[7]{}((\Varid{join}\mathbin{\circ}\Varid{join}\mathbin{\circ}\Varid{map}\;(\Varid{map}\;\Varid{h})\mathbin{\circ}\Varid{map}\;\Varid{g}\mathbin{\circ}\Varid{f})\;\Varid{a}){}\<[58]%
\>[58]{}=\hspace{-3pt}\{\; \Varid{square}\;\anonymous \;\}\hspace{-3pt}={}\<[E]%
\\
\>[5]{}\hsindent{2}{}\<[7]%
\>[7]{}((\Varid{join}\mathbin{\circ}\Varid{map}\;\Varid{join}\mathbin{\circ}\Varid{map}\;(\Varid{map}\;\Varid{h})\mathbin{\circ}\Varid{map}\;\Varid{g}\mathbin{\circ}\Varid{f})\;\Varid{a}){}\<[58]%
\>[58]{}=\hspace{-3pt}\{\; \Varid{cong}\;{}\<[67]%
\>[67]{}\{\mskip1.5mu \Varid{f}\mathrel{=}\Varid{join}\mathbin{\circ}\Varid{map}\;\Varid{join}\mskip1.5mu\}\;{}\<[E]%
\\
\>[67]{}(\Varid{sym}\;(\Varid{mapPresComp}\;\anonymous \;\anonymous \;\anonymous ))\;\}\hspace{-3pt}={}\<[E]%
\\
\>[5]{}\hsindent{2}{}\<[7]%
\>[7]{}((\Varid{join}\mathbin{\circ}\Varid{map}\;\Varid{join}\mathbin{\circ}\Varid{map}\;(\Varid{map}\;\Varid{h}\mathbin{\circ}\Varid{g})\mathbin{\circ}\Varid{f})\;\Varid{a}){}\<[58]%
\>[58]{}=\hspace{-3pt}\{\; \Varid{cong}\;{}\<[67]%
\>[67]{}\{\mskip1.5mu \Varid{f}\mathrel{=}\Varid{join}\mskip1.5mu\}\;{}\<[E]%
\\
\>[67]{}(\Varid{sym}\;(\Varid{mapPresComp}\;\anonymous \;\anonymous \;\anonymous ))\;\}\hspace{-3pt}={}\<[E]%
\\
\>[5]{}\hsindent{2}{}\<[7]%
\>[7]{}((\Varid{join}\mathbin{\circ}\Varid{map}\;(\Varid{join}\mathbin{\circ}\Varid{map}\;\Varid{h}\mathbin{\circ}\Varid{g})\mathbin{\circ}\Varid{f})\;\Varid{a}){}\<[58]%
\>[58]{}=\hspace{-3pt}\{\; \Conid{Refl}\;\}\hspace{-3pt}={}\<[E]%
\\
\>[5]{}\hsindent{2}{}\<[7]%
\>[7]{}((\Varid{f}\mathbin{\mathbin{>}\hspace{-5.2pt}\mathrel{=}\hspace{-5.5pt}\mathbin{>}}(\Varid{g}\mathbin{\mathbin{>}\hspace{-5.2pt}\mathrel{=}\hspace{-5.5pt}\mathbin{>}}\Varid{h}))\;\Varid{a})\;{}\<[58]%
\>[58]{}\Conid{QED}{}\<[E]%
\ColumnHook
\end{hscode}\resethooks
Notice that in order to show that Kleisli composition preserves
extensional equality, one has to rely on the \ensuremath{\Varid{mapPresEE}} axiom of the
underlying functor, as one would expect.

\begin{joincode}%
\begin{hscode}\SaveRestoreHook
\column{B}{@{}>{\hspre}l<{\hspost}@{}}%
\column{5}{@{}>{\hspre}l<{\hspost}@{}}%
\column{20}{@{}>{\hspre}c<{\hspost}@{}}%
\column{20E}{@{}l@{}}%
\column{23}{@{}>{\hspre}l<{\hspost}@{}}%
\column{E}{@{}>{\hspre}l<{\hspost}@{}}%
\>[5]{}\Varid{kleisliPresEE}{}\<[20]%
\>[20]{}\ \mathop{:}\ {}\<[20E]%
\>[23]{}\{\mskip1.5mu \Conid{A},\Conid{B},\Conid{C}\ \mathop{:}\ \Conid{Type}\mskip1.5mu\}\,\to\,\{\mskip1.5mu \Conid{M}\ \mathop{:}\ \Conid{Type}\,\to\,\Conid{Type}\mskip1.5mu\}\,\to\,\Conid{Monad}_{\!\mathrm{1}}\;\Conid{M}\Rightarrow {}\<[E]%
\\
\>[23]{}(\Varid{f},\Varid{f'}\ \mathop{:}\ \Conid{A}\,\to\,\Conid{M}\;\Conid{B})\,\to\,(\Varid{g},\Varid{g'}\ \mathop{:}\ \Conid{B}\,\to\,\Conid{M}\;\Conid{C})\,\to\,{}\<[E]%
\\
\>[23]{}\Varid{f}\doteq\Varid{f'}\,\to\,\Varid{g}\doteq\Varid{g'}\,\to\,(\Varid{f}\mathbin{\mathbin{>}\hspace{-5.2pt}\mathrel{=}\hspace{-5.5pt}\mathbin{>}}\Varid{g})\doteq(\Varid{f'}\mathbin{\mathbin{>}\hspace{-5.2pt}\mathrel{=}\hspace{-5.5pt}\mathbin{>}}\Varid{g'}){}\<[E]%
\ColumnHook
\end{hscode}\resethooks
\begin{hscode}\SaveRestoreHook
\column{B}{@{}>{\hspre}l<{\hspost}@{}}%
\column{5}{@{}>{\hspre}l<{\hspost}@{}}%
\column{7}{@{}>{\hspre}l<{\hspost}@{}}%
\column{26}{@{}>{\hspre}l<{\hspost}@{}}%
\column{31}{@{}>{\hspre}l<{\hspost}@{}}%
\column{35}{@{}>{\hspre}l<{\hspost}@{}}%
\column{38}{@{}>{\hspre}l<{\hspost}@{}}%
\column{E}{@{}>{\hspre}l<{\hspost}@{}}%
\>[5]{}\Varid{kleisliPresEE}\;\Varid{f}\;\Varid{f'}\;\Varid{g}\;\Varid{g'}\;\Varid{fE}\;\Varid{gE}\;\Varid{a}\mathrel{=}{}\<[E]%
\\
\>[5]{}\hsindent{2}{}\<[7]%
\>[7]{}((\Varid{f}\mathbin{\mathbin{>}\hspace{-5.2pt}\mathrel{=}\hspace{-5.5pt}\mathbin{>}}\Varid{g})\;{}\<[31]%
\>[31]{}\Varid{a}{}\<[35]%
\>[35]{}){}\<[38]%
\>[38]{}=\hspace{-3pt}\{\; \Varid{kleisliLeapfrog}\;\Varid{f}\;\Varid{g}\;\Varid{a}\;\}\hspace{-3pt}={}\<[E]%
\\
\>[5]{}\hsindent{2}{}\<[7]%
\>[7]{}((\Varid{id}\mathbin{\mathbin{>}\hspace{-5.2pt}\mathrel{=}\hspace{-5.5pt}\mathbin{>}}\Varid{g})\;{}\<[26]%
\>[26]{}(\Varid{f}\;{}\<[31]%
\>[31]{}\Varid{a}){}\<[35]%
\>[35]{}){}\<[38]%
\>[38]{}=\hspace{-3pt}\{\; \Varid{cong}\;(\Varid{fE}\;\Varid{a})\;\}\hspace{-3pt}={}\<[E]%
\\
\>[5]{}\hsindent{2}{}\<[7]%
\>[7]{}((\Varid{id}\mathbin{\mathbin{>}\hspace{-5.2pt}\mathrel{=}\hspace{-5.5pt}\mathbin{>}}\Varid{g})\;{}\<[26]%
\>[26]{}(\Varid{f'}\;{}\<[31]%
\>[31]{}\Varid{a}){}\<[35]%
\>[35]{}){}\<[38]%
\>[38]{}=\hspace{-3pt}\{\; \Conid{Refl}\;\}\hspace{-3pt}={}\<[E]%
\\
\>[5]{}\hsindent{2}{}\<[7]%
\>[7]{}((\Varid{join}\mathbin{\circ}\Varid{map}\;\Varid{g})\;{}\<[26]%
\>[26]{}(\Varid{f'}\;{}\<[31]%
\>[31]{}\Varid{a}){}\<[35]%
\>[35]{}){}\<[38]%
\>[38]{}=\hspace{-3pt}\{\; \Varid{cong}\;(\Varid{mapPresEE}\;\Varid{g}\;\Varid{g'}\;\Varid{gE}\;(\Varid{f'}\;\Varid{a}))\;\}\hspace{-3pt}={}\<[E]%
\\
\>[5]{}\hsindent{2}{}\<[7]%
\>[7]{}((\Varid{join}\mathbin{\circ}\Varid{map}\;\Varid{g'})\;{}\<[26]%
\>[26]{}(\Varid{f'}\;{}\<[31]%
\>[31]{}\Varid{a}){}\<[35]%
\>[35]{}){}\<[38]%
\>[38]{}=\hspace{-3pt}\{\; \Conid{Refl}\;\}\hspace{-3pt}={}\<[E]%
\\
\>[5]{}\hsindent{2}{}\<[7]%
\>[7]{}((\Varid{f'}\mathbin{\mathbin{>}\hspace{-5.2pt}\mathrel{=}\hspace{-5.5pt}\mathbin{>}}\Varid{g'})\;{}\<[31]%
\>[31]{}\Varid{a}{}\<[35]%
\>[35]{})\;{}\<[38]%
\>[38]{}\Conid{QED}{}\<[E]%
\ColumnHook
\end{hscode}\resethooks
\end{joincode}

In the implementation of \ensuremath{\Varid{kleisliPresEE}}, we have
applied the ``leapfrogging'' rule (compare \citep{bird2014thinking},
p. 250):
\begin{hscode}\SaveRestoreHook
\column{B}{@{}>{\hspre}l<{\hspost}@{}}%
\column{5}{@{}>{\hspre}l<{\hspost}@{}}%
\column{22}{@{}>{\hspre}c<{\hspost}@{}}%
\column{22E}{@{}l@{}}%
\column{25}{@{}>{\hspre}l<{\hspost}@{}}%
\column{E}{@{}>{\hspre}l<{\hspost}@{}}%
\>[5]{}\Varid{kleisliLeapfrog}{}\<[22]%
\>[22]{}\ \mathop{:}\ {}\<[22E]%
\>[25]{}\{\mskip1.5mu \Conid{A},\Conid{B},\Conid{C}\ \mathop{:}\ \Conid{Type}\mskip1.5mu\}\,\to\,\{\mskip1.5mu \Conid{M}\ \mathop{:}\ \Conid{Type}\,\to\,\Conid{Type}\mskip1.5mu\}\,\to\,\Conid{Monad}_{\!\mathrm{1}}\;\Conid{M}\Rightarrow {}\<[E]%
\\
\>[25]{}(\Varid{f}\ \mathop{:}\ \Conid{A}\,\to\,\Conid{M}\;\Conid{B})\,\to\,(\Varid{g}\ \mathop{:}\ \Conid{B}\,\to\,\Conid{M}\;\Conid{C})\,\to\,(\Varid{f}\mathbin{\mathbin{>}\hspace{-5.2pt}\mathrel{=}\hspace{-5.5pt}\mathbin{>}}\Varid{g})\doteq(\Varid{id}\mathbin{\mathbin{>}\hspace{-5.2pt}\mathrel{=}\hspace{-5.5pt}\mathbin{>}}\Varid{g})\mathbin{\circ}\Varid{f}{}\<[E]%
\\
\>[5]{}\Varid{kleisliLeapfrog}\;\Varid{f}\;\Varid{g}\;\Varid{a}\mathrel{=}\Conid{Refl}{}\<[E]%
\ColumnHook
\end{hscode}\resethooks

\paragraph*{Fat ADTs.}

The main advantages of the approach outlined above -- a thin ADT and
explicit definitions of the monadic combinators -- are readability and
straightforwardness of proofs: thanks to the intensional equality
between \ensuremath{\Varid{f}\mathbin{\mathbin{>}\hspace{-5.2pt}\mathrel{=}\hspace{-5.5pt}\mathbin{>}}\Varid{g}} and \ensuremath{\Varid{join}\mathbin{\circ}\Varid{map}\;\Varid{g}\mathbin{\circ}\Varid{f}}, we were able to implement many
proof steps with just \ensuremath{\Conid{Refl}}.

The strength of thin ADT designs is also their weakness: in many
practical cases, one would like to be able to define \ensuremath{\Varid{join}} in terms
of bind and not the other way round.
In other words, one would like to weaken the requirements on, e.g.,
\ensuremath{\Varid{join}}, \ensuremath{\Varid{map}} and Kleisli composition and just require that \ensuremath{\Varid{f}\mathbin{\mathbin{>}\hspace{-5.2pt}\mathrel{=}\hspace{-5.5pt}\mathbin{>}}\Varid{g}}
and \ensuremath{\Varid{join}\mathbin{\circ}\Varid{map}\;\Varid{g}\mathbin{\circ}\Varid{f}} are extensionally equal.
If they happen to be intensionally equal for a specific instance, the
better.

This suggests that an alternative way of formalizing the traditional
notion of monads from category theory could be through a \emph{fat} ADT:
\pagebreak
\begin{hscode}\SaveRestoreHook
\column{B}{@{}>{\hspre}l<{\hspost}@{}}%
\column{5}{@{}>{\hspre}l<{\hspost}@{}}%
\column{7}{@{}>{\hspre}l<{\hspost}@{}}%
\column{27}{@{}>{\hspre}c<{\hspost}@{}}%
\column{27E}{@{}l@{}}%
\column{30}{@{}>{\hspre}l<{\hspost}@{}}%
\column{44}{@{}>{\hspre}c<{\hspost}@{}}%
\column{44E}{@{}l@{}}%
\column{46}{@{}>{\hspre}c<{\hspost}@{}}%
\column{46E}{@{}l@{}}%
\column{48}{@{}>{\hspre}l<{\hspost}@{}}%
\column{50}{@{}>{\hspre}l<{\hspost}@{}}%
\column{52}{@{}>{\hspre}l<{\hspost}@{}}%
\column{65}{@{}>{\hspre}l<{\hspost}@{}}%
\column{67}{@{}>{\hspre}l<{\hspost}@{}}%
\column{68}{@{}>{\hspre}c<{\hspost}@{}}%
\column{68E}{@{}l@{}}%
\column{72}{@{}>{\hspre}l<{\hspost}@{}}%
\column{81}{@{}>{\hspre}c<{\hspost}@{}}%
\column{81E}{@{}l@{}}%
\column{83}{@{}>{\hspre}c<{\hspost}@{}}%
\column{83E}{@{}l@{}}%
\column{86}{@{}>{\hspre}l<{\hspost}@{}}%
\column{88}{@{}>{\hspre}l<{\hspost}@{}}%
\column{E}{@{}>{\hspre}l<{\hspost}@{}}%
\>[5]{}\bf{interface}\;\Conid{Functor}\;\Conid{M}\Rightarrow \Conid{Monad}_{\mathrm{2}}\;(\Conid{M}\ \mathop{:}\ \Conid{Type}\,\to\,\Conid{Type})\;\mathbf{where}{}\<[E]%
\\[\blanklineskip]%
\>[5]{}\hsindent{2}{}\<[7]%
\>[7]{}\Varid{pure}{}\<[27]%
\>[27]{}\ \mathop{:}\ {}\<[27E]%
\>[30]{}\{\mskip1.5mu \Conid{A}\ \mathop{:}\ \Conid{Type}\mskip1.5mu\}{}\<[48]%
\>[48]{}\,\to\,{}\<[52]%
\>[52]{}\Conid{A}\,\to\,\Conid{M}\;\Conid{A}{}\<[E]%
\\
\>[5]{}\hsindent{2}{}\<[7]%
\>[7]{}\Varid{join}{}\<[27]%
\>[27]{}\ \mathop{:}\ {}\<[27E]%
\>[30]{}\{\mskip1.5mu \Conid{A}\ \mathop{:}\ \Conid{Type}\mskip1.5mu\}{}\<[48]%
\>[48]{}\,\to\,{}\<[52]%
\>[52]{}\Conid{M}\;(\Conid{M}\;\Conid{A})\,\to\,\Conid{M}\;\Conid{A}{}\<[E]%
\\
\>[5]{}\hsindent{2}{}\<[7]%
\>[7]{}(\bind ){}\<[27]%
\>[27]{}\ \mathop{:}\ {}\<[27E]%
\>[30]{}\{\mskip1.5mu \Conid{A},\Conid{B}\ \mathop{:}\ \Conid{Type}\mskip1.5mu\}{}\<[48]%
\>[48]{}\,\to\,{}\<[52]%
\>[52]{}\Conid{M}\;\Conid{A}\,\to\,(\Conid{A}\,\to\,\Conid{M}\;\Conid{B})\,\to\,\Conid{M}\;\Conid{B}{}\<[E]%
\\
\>[5]{}\hsindent{2}{}\<[7]%
\>[7]{}(\mathbin{\mathbin{>}\hspace{-5.2pt}\mathrel{=}\hspace{-5.5pt}\mathbin{>}}){}\<[27]%
\>[27]{}\ \mathop{:}\ {}\<[27E]%
\>[30]{}\{\mskip1.5mu \Conid{A},\Conid{B},\Conid{C}\ \mathop{:}\ \Conid{Type}\mskip1.5mu\}{}\<[48]%
\>[48]{}\,\to\,{}\<[52]%
\>[52]{}(\Conid{A}\,\to\,\Conid{M}\;\Conid{B})\,\to\,(\Conid{B}\,\to\,\Conid{M}\;\Conid{C})\,\to\,(\Conid{A}\,\to\,\Conid{M}\;\Conid{C}){}\<[E]%
\\[\blanklineskip]%
\>[5]{}\hsindent{2}{}\<[7]%
\>[7]{}\Varid{bindJoinMapSpec}{}\<[27]%
\>[27]{}\ \mathop{:}\ {}\<[27E]%
\>[30]{}\{\mskip1.5mu \Conid{A},\Conid{B}\ \mathop{:}\ \Conid{Type}\mskip1.5mu\}{}\<[48]%
\>[48]{}\,\to\,{}\<[52]%
\>[52]{}(\Varid{f}\ \mathop{:}\ \Conid{A}\,\to\,\Conid{M}\;\Conid{B}){}\<[68]%
\>[68]{}\,\to\,{}\<[68E]%
\>[72]{}(\bind \Varid{f}){}\<[83]%
\>[83]{}\doteq{}\<[83E]%
\>[88]{}\Varid{join}\mathbin{\circ}\Varid{map}\;\Varid{f}{}\<[E]%
\\
\>[5]{}\hsindent{2}{}\<[7]%
\>[7]{}\Varid{kleisliJoinMapSpec}{}\<[27]%
\>[27]{}\ \mathop{:}\ {}\<[27E]%
\>[30]{}\{\mskip1.5mu \Conid{A},\Conid{B},\Conid{C}\ \mathop{:}\ \Conid{Type}\mskip1.5mu\}{}\<[48]%
\>[48]{}\,\to\,{}\<[52]%
\>[52]{}(\Varid{f}\ \mathop{:}\ \Conid{A}\,\to\,\Conid{M}\;\Conid{B}){}\<[68]%
\>[68]{}\,\to\,{}\<[68E]%
\\
\>[52]{}(\Varid{g}\ \mathop{:}\ \Conid{B}\,\to\,\Conid{M}\;\Conid{C}){}\<[68]%
\>[68]{}\,\to\,{}\<[68E]%
\>[72]{}(\Varid{f}\mathbin{\mathbin{>}\hspace{-5.2pt}\mathrel{=}\hspace{-5.5pt}\mathbin{>}}\Varid{g}){}\<[83]%
\>[83]{}\doteq{}\<[83E]%
\>[88]{}\Varid{join}\mathbin{\circ}\Varid{map}\;\Varid{g}\mathbin{\circ}\Varid{f}{}\<[E]%
\\[\blanklineskip]%
\>[5]{}\hsindent{2}{}\<[7]%
\>[7]{}\Varid{triangleLeft}{}\<[27]%
\>[27]{}\ \mathop{:}\ {}\<[27E]%
\>[30]{}\{\mskip1.5mu \Conid{A}\ \mathop{:}\ \Conid{Type}\mskip1.5mu\}{}\<[44]%
\>[44]{}\,\to\,{}\<[44E]%
\>[48]{}\Varid{join}\mathbin{\circ}\Varid{pure}{}\<[65]%
\>[65]{}\doteq\Varid{id}\;\{\mskip1.5mu \Varid{a}\mathrel{=}\Conid{M}\;\Conid{A}\mskip1.5mu\}{}\<[E]%
\\
\>[5]{}\hsindent{2}{}\<[7]%
\>[7]{}\Varid{triangleRight}{}\<[27]%
\>[27]{}\ \mathop{:}\ {}\<[27E]%
\>[30]{}\{\mskip1.5mu \Conid{A}\ \mathop{:}\ \Conid{Type}\mskip1.5mu\}{}\<[44]%
\>[44]{}\,\to\,{}\<[44E]%
\>[48]{}\Varid{join}\mathbin{\circ}\Varid{map}\;\Varid{pure}{}\<[65]%
\>[65]{}\doteq\Varid{id}\;\{\mskip1.5mu \Varid{a}\mathrel{=}\Conid{M}\;\Conid{A}\mskip1.5mu\}{}\<[E]%
\\
\>[5]{}\hsindent{2}{}\<[7]%
\>[7]{}\Varid{square}{}\<[27]%
\>[27]{}\ \mathop{:}\ {}\<[27E]%
\>[30]{}\{\mskip1.5mu \Conid{A}\ \mathop{:}\ \Conid{Type}\mskip1.5mu\}{}\<[44]%
\>[44]{}\,\to\,{}\<[44E]%
\>[48]{}\Varid{join}\mathbin{\circ}\Varid{join}{}\<[65]%
\>[65]{}\doteq\Varid{join}\mathbin{\circ}\Varid{map}\;\{\mskip1.5mu \Conid{A}\mathrel{=}\Conid{M}\;(\Conid{M}\;\Conid{A})\mskip1.5mu\}\;\Varid{join}{}\<[E]%
\\[\blanklineskip]%
\>[5]{}\hsindent{2}{}\<[7]%
\>[7]{}\Varid{pureNatTrans}{}\<[27]%
\>[27]{}\ \mathop{:}\ {}\<[27E]%
\>[30]{}\{\mskip1.5mu \Conid{A},\Conid{B}\ \mathop{:}\ \Conid{Type}\mskip1.5mu\}{}\<[46]%
\>[46]{}\,\to\,{}\<[46E]%
\>[50]{}(\Varid{f}\ \mathop{:}\ \Conid{A}\,\to\,\Conid{B})\,\to\,{}\<[67]%
\>[67]{}\Varid{map}\;\Varid{f}\mathbin{\circ}\Varid{pure}{}\<[81]%
\>[81]{}\doteq{}\<[81E]%
\>[86]{}\Varid{pure}\mathbin{\circ}\Varid{f}{}\<[E]%
\\
\>[5]{}\hsindent{2}{}\<[7]%
\>[7]{}\Varid{joinNatTrans}{}\<[27]%
\>[27]{}\ \mathop{:}\ {}\<[27E]%
\>[30]{}\{\mskip1.5mu \Conid{A},\Conid{B}\ \mathop{:}\ \Conid{Type}\mskip1.5mu\}{}\<[46]%
\>[46]{}\,\to\,{}\<[46E]%
\>[50]{}(\Varid{f}\ \mathop{:}\ \Conid{A}\,\to\,\Conid{B})\,\to\,{}\<[67]%
\>[67]{}\Varid{map}\;\Varid{f}\mathbin{\circ}\Varid{join}{}\<[81]%
\>[81]{}\doteq{}\<[81E]%
\>[86]{}\Varid{join}\mathbin{\circ}\Varid{map}\;(\Varid{map}\;\Varid{f}){}\<[E]%
\ColumnHook
\end{hscode}\resethooks
One could go even further and add more combinators (and their axioms)
to the ADT, but for the purpose of this discussion the above example
will do.
(In any case, many of these methods could be filled in by defaults.)
What are the implications of having replaced definitions with
specifications?
A direct implication is that now, in implementing generic proofs of the monad
laws, we have to replace some \ensuremath{\Conid{Refl}} steps with suitable
specifications.
Thus, for instance \ensuremath{\Varid{pureLeftIdKleisli}} becomes
\begin{hscode}\SaveRestoreHook
\column{B}{@{}>{\hspre}l<{\hspost}@{}}%
\column{5}{@{}>{\hspre}l<{\hspost}@{}}%
\column{24}{@{}>{\hspre}c<{\hspost}@{}}%
\column{24E}{@{}l@{}}%
\column{27}{@{}>{\hspre}l<{\hspost}@{}}%
\column{30}{@{}>{\hspre}l<{\hspost}@{}}%
\column{57}{@{}>{\hspre}l<{\hspost}@{}}%
\column{E}{@{}>{\hspre}l<{\hspost}@{}}%
\>[5]{}\Varid{pureLeftIdKleisli}{}\<[24]%
\>[24]{}\ \mathop{:}\ {}\<[24E]%
\>[27]{}\{\mskip1.5mu \Conid{A},\Conid{B}\ \mathop{:}\ \Conid{Type}\mskip1.5mu\}\,\to\,\{\mskip1.5mu \Conid{M}\ \mathop{:}\ \Conid{Type}\,\to\,\Conid{Type}\mskip1.5mu\}\,\to\,\Conid{Monad}_{\mathrm{2}}\;\Conid{M}\Rightarrow {}\<[E]%
\\
\>[27]{}(\Varid{f}\ \mathop{:}\ \Conid{A}\,\to\,\Conid{M}\;\Conid{B})\,\to\,(\Varid{pure}\mathbin{\mathbin{>}\hspace{-5.2pt}\mathrel{=}\hspace{-5.5pt}\mathbin{>}}\Varid{f})\doteq\Varid{f}{}\<[E]%
\\
\>[5]{}\Varid{pureLeftIdKleisli}\;\Varid{f}\;\Varid{a}\mathrel{=}{}\<[30]%
\>[30]{}((\Varid{pure}\mathbin{\mathbin{>}\hspace{-5.2pt}\mathrel{=}\hspace{-5.5pt}\mathbin{>}}\Varid{f})\;\Varid{a}){}\<[57]%
\>[57]{}=\hspace{-3pt}\{\; \Varid{kleisliJoinMapSpec}\;\Varid{pure}\;\Varid{f}\;\Varid{a}\;\}\hspace{-3pt}={}\<[E]%
\\
\>[30]{}(\Varid{join}\;(\Varid{map}\;\Varid{f}\;(\Varid{pure}\;\Varid{a}))){}\<[57]%
\>[57]{}=\hspace{-3pt}\{\; \Varid{cong}\;\{\mskip1.5mu \Varid{f}\mathrel{=}\Varid{join}\mskip1.5mu\}\;(\Varid{pureNatTrans}\;\Varid{f}\;\Varid{a})\;\}\hspace{-3pt}={}\<[E]%
\\
\>[30]{}(\Varid{join}\;(\Varid{pure}\;(\Varid{f}\;\Varid{a}))){}\<[57]%
\>[57]{}=\hspace{-3pt}\{\; \Varid{triangleLeft}\;(\Varid{f}\;\Varid{a})\;\}\hspace{-3pt}={}\<[E]%
\\
\>[30]{}(\Varid{f}\;\Varid{a})\;{}\<[57]%
\>[57]{}\Conid{QED}{}\<[E]%
\ColumnHook
\end{hscode}\resethooks
where we have replaced the first proof step, \ensuremath{\Conid{Refl}}, with
\ensuremath{\Varid{kleisliJoinMapSpec}\;\Varid{pure}\;\Varid{f}\;\Varid{a}}.
Similar transformations have to be done for \ensuremath{\Varid{pureRightIdKleisli}},
\ensuremath{\Varid{kleisliAssoc}}, etc.
However, completing the proof of the monad laws for the fat interface
is not just a matter of replacing definitions with
specifications.
Consider associativity:

\begin{hscode}\SaveRestoreHook
\column{B}{@{}>{\hspre}l<{\hspost}@{}}%
\column{5}{@{}>{\hspre}l<{\hspost}@{}}%
\column{7}{@{}>{\hspre}l<{\hspost}@{}}%
\column{9}{@{}>{\hspre}l<{\hspost}@{}}%
\column{19}{@{}>{\hspre}c<{\hspost}@{}}%
\column{19E}{@{}l@{}}%
\column{22}{@{}>{\hspre}l<{\hspost}@{}}%
\column{31}{@{}>{\hspre}l<{\hspost}@{}}%
\column{39}{@{}>{\hspre}l<{\hspost}@{}}%
\column{47}{@{}>{\hspre}l<{\hspost}@{}}%
\column{56}{@{}>{\hspre}l<{\hspost}@{}}%
\column{65}{@{}>{\hspre}l<{\hspost}@{}}%
\column{85}{@{}>{\hspre}l<{\hspost}@{}}%
\column{E}{@{}>{\hspre}l<{\hspost}@{}}%
\>[5]{}\Varid{kleisliAssoc}{}\<[19]%
\>[19]{}\ \mathop{:}\ {}\<[19E]%
\>[22]{}\{\mskip1.5mu \Conid{A},\Conid{B},\Conid{C},\Conid{D}\ \mathop{:}\ \Conid{Type}\mskip1.5mu\}\,\to\,\{\mskip1.5mu \Conid{M}\ \mathop{:}\ \Conid{Type}\,\to\,\Conid{Type}\mskip1.5mu\}\,\to\,\Conid{Monad}_{\mathrm{2}}\;\Conid{M}\Rightarrow {}\<[E]%
\\
\>[22]{}(\Varid{f}\ \mathop{:}\ \Conid{A}\,\to\,\Conid{M}\;\Conid{B})\,\to\,(\Varid{g}\ \mathop{:}\ \Conid{B}\,\to\,\Conid{M}\;\Conid{C})\,\to\,(\Varid{h}\ \mathop{:}\ \Conid{C}\,\to\,\Conid{M}\;\Conid{D})\,\to\,{}\<[E]%
\\
\>[22]{}((\Varid{f}\mathbin{\mathbin{>}\hspace{-5.2pt}\mathrel{=}\hspace{-5.5pt}\mathbin{>}}\Varid{g})\mathbin{\mathbin{>}\hspace{-5.2pt}\mathrel{=}\hspace{-5.5pt}\mathbin{>}}\Varid{h})\doteq(\Varid{f}\mathbin{\mathbin{>}\hspace{-5.2pt}\mathrel{=}\hspace{-5.5pt}\mathbin{>}}(\Varid{g}\mathbin{\mathbin{>}\hspace{-5.2pt}\mathrel{=}\hspace{-5.5pt}\mathbin{>}}\Varid{h})){}\<[E]%
\\
\>[5]{}\Varid{kleisliAssoc}\;\{\mskip1.5mu \Conid{M}\mskip1.5mu\}\;\{\mskip1.5mu \Conid{A}\mskip1.5mu\}\;\{\mskip1.5mu \Conid{C}\mskip1.5mu\}\;\Varid{f}\;\Varid{g}\;\Varid{h}\;\Varid{a}\mathrel{=}{}\<[E]%
\\
\>[5]{}\hsindent{2}{}\<[7]%
\>[7]{}(((\Varid{f}\mathbin{\mathbin{>}\hspace{-5.2pt}\mathrel{=}\hspace{-5.5pt}\mathbin{>}}\Varid{g})\mathbin{\mathbin{>}\hspace{-5.2pt}\mathrel{=}\hspace{-5.5pt}\mathbin{>}}\Varid{h})\;\Varid{a}){}\<[56]%
\>[56]{}=\hspace{-3pt}\{\; \Varid{kleisliJoinMapSpec}\;(\Varid{f}\mathbin{\mathbin{>}\hspace{-5.2pt}\mathrel{=}\hspace{-5.5pt}\mathbin{>}}\Varid{g})\;\Varid{h}\;\Varid{a}\;\}\hspace{-3pt}={}\<[E]%
\\
\>[5]{}\hsindent{2}{}\<[7]%
\>[7]{}((\Varid{join}\mathbin{\circ}\Varid{map}\;\Varid{h}\mathbin{\circ}(\Varid{f}\mathbin{\mathbin{>}\hspace{-5.2pt}\mathrel{=}\hspace{-5.5pt}\mathbin{>}}\Varid{g}))\;\Varid{a}){}\<[56]%
\>[56]{}=\hspace{-3pt}\{\; \Varid{cong}\;{}\<[65]%
\>[65]{}\{\mskip1.5mu \Varid{f}\mathrel{=}\Varid{join}\mathbin{\circ}\Varid{map}\;\Varid{h}\mskip1.5mu\}\;{}\<[E]%
\\
\>[65]{}(\Varid{kleisliJoinMapSpec}\;\Varid{f}\;\Varid{g}\;\Varid{a})\;\}\hspace{-3pt}={}\<[E]%
\\
\>[5]{}\hsindent{2}{}\<[7]%
\>[7]{}((\Varid{join}\mathbin{\circ}\Varid{map}\;\Varid{h}\mathbin{\circ}\Varid{join}\mathbin{\circ}\Varid{map}\;\Varid{g}\mathbin{\circ}\Varid{f})\;\Varid{a}){}\<[56]%
\>[56]{}=\hspace{-3pt}\{\; \Varid{cong}\;(\Varid{joinNatTrans}\;\Varid{h}\;(\Varid{map}\;\Varid{g}\;(\Varid{f}\;\Varid{a})))\;\}\hspace{-3pt}={}\<[E]%
\\
\>[5]{}\hsindent{2}{}\<[7]%
\>[7]{}((\Varid{join}\mathbin{\circ}\Varid{join}\mathbin{\circ}\Varid{map}\;(\Varid{map}\;\Varid{h})\mathbin{\circ}\Varid{map}\;\Varid{g}\mathbin{\circ}\Varid{f})\;\Varid{a}){}\<[56]%
\>[56]{}=\hspace{-3pt}\{\; \Varid{square}\;\anonymous \;\}\hspace{-3pt}={}\<[E]%
\\
\>[5]{}\hsindent{2}{}\<[7]%
\>[7]{}((\Varid{join}\mathbin{\circ}\Varid{map}\;\Varid{join}\mathbin{\circ}\Varid{map}\;(\Varid{map}\;\Varid{h})\mathbin{\circ}\Varid{map}\;\Varid{g}\mathbin{\circ}\Varid{f})\;\Varid{a}){}\<[56]%
\>[56]{}=\hspace{-3pt}\{\; \Varid{cong}\;{}\<[65]%
\>[65]{}\{\mskip1.5mu \Varid{f}\mathrel{=}\Varid{join}\mathbin{\circ}\Varid{map}\;\Varid{join}\mskip1.5mu\}\;{}\<[E]%
\\
\>[65]{}(\Varid{sym}\;(\Varid{mapPresComp}\;\anonymous \;\anonymous \;\anonymous ))\;\}\hspace{-3pt}={}\<[E]%
\\
\>[5]{}\hsindent{2}{}\<[7]%
\>[7]{}((\Varid{join}\mathbin{\circ}\Varid{map}\;\Varid{join}\mathbin{\circ}\Varid{map}\;(\Varid{map}\;\Varid{h}\mathbin{\circ}\Varid{g})\mathbin{\circ}\Varid{f})\;\Varid{a}){}\<[56]%
\>[56]{}=\hspace{-3pt}\{\; \Varid{cong}\;{}\<[65]%
\>[65]{}\{\mskip1.5mu \Varid{f}\mathrel{=}\Varid{join}\mskip1.5mu\}\;{}\<[E]%
\\
\>[65]{}(\Varid{sym}\;(\Varid{mapPresComp}\;\anonymous \;\anonymous \;\anonymous ))\;\}\hspace{-3pt}={}\<[E]%
\\
\>[5]{}\hsindent{2}{}\<[7]%
\>[7]{}((\Varid{join}\mathbin{\circ}\Varid{map}\;(\Varid{join}\mathbin{\circ}\Varid{map}\;\Varid{h}\mathbin{\circ}\Varid{g})\mathbin{\circ}\Varid{f})\;\Varid{a})\quad=\hspace{-3pt}\{\; \Varid{sym}\;(\Varid{kleisliJoinMapSpec}\;\Varid{f}\;(\Varid{join}\mathbin{\circ}\Varid{map}\;\Varid{h}\mathbin{\circ}\Varid{g})\;\Varid{a})\;\}\hspace{-3pt}={}\<[E]%
\\
\>[5]{}\hsindent{2}{}\<[7]%
\>[7]{}((\Varid{f}\mathbin{\mathbin{>}\hspace{-5.2pt}\mathrel{=}\hspace{-5.5pt}\mathbin{>}}(\Varid{join}\mathbin{\circ}\Varid{map}\;\Varid{h}\mathbin{\circ}\Varid{g}))\;\Varid{a}){}\<[E]%
\\
\>[7]{}\hsindent{2}{}\<[9]%
\>[9]{}=\hspace{-3pt}\{\; \Varid{kleisliPresEE}\;{}\<[31]%
\>[31]{}\Varid{f}\;\Varid{f}\;{}\<[39]%
\>[39]{}(\Varid{join}\mathbin{\circ}\Varid{map}\;\Varid{h}\mathbin{\circ}\Varid{g})\;(\Varid{g}\mathbin{\mathbin{>}\hspace{-5.2pt}\mathrel{=}\hspace{-5.5pt}\mathbin{>}}\Varid{h})\;{}\<[E]%
\\
\>[31]{}\Varid{reflEE}\;{}\<[39]%
\>[39]{}(\Varid{symEE}\;{}\<[47]%
\>[47]{}\{\mskip1.5mu \Varid{f}\mathrel{=}\Varid{g}\mathbin{\mathbin{>}\hspace{-5.2pt}\mathrel{=}\hspace{-5.5pt}\mathbin{>}}\Varid{h}\mskip1.5mu\}\;\{\mskip1.5mu \Varid{g}\mathrel{=}\Varid{join}\mathbin{\circ}\Varid{map}\;\Varid{h}\mathbin{\circ}\Varid{g}\mskip1.5mu\}\;{}\<[E]%
\\
\>[47]{}(\Varid{kleisliJoinMapSpec}\;\Varid{g}\;\Varid{h}))\;{}\<[85]%
\>[85]{}\Varid{a}\;\}\hspace{-3pt}={}\<[E]%
\\
\>[5]{}\hsindent{2}{}\<[7]%
\>[7]{}((\Varid{f}\mathbin{\mathbin{>}\hspace{-5.2pt}\mathrel{=}\hspace{-5.5pt}\mathbin{>}}(\Varid{g}\mathbin{\mathbin{>}\hspace{-5.2pt}\mathrel{=}\hspace{-5.5pt}\mathbin{>}}\Varid{h}))\;\Varid{a})\;\Conid{QED}{}\<[E]%
\ColumnHook
\end{hscode}\resethooks
Here, in order to deduce \ensuremath{(\Varid{f}\mathbin{\mathbin{>}\hspace{-5.2pt}\mathrel{=}\hspace{-5.5pt}\mathbin{>}}(\Varid{g}\mathbin{\mathbin{>}\hspace{-5.2pt}\mathrel{=}\hspace{-5.5pt}\mathbin{>}}\Varid{h}))\;\Varid{a}} from \ensuremath{(\Varid{f}\mathbin{\mathbin{>}\hspace{-5.2pt}\mathrel{=}\hspace{-5.5pt}\mathbin{>}}(\Varid{join}\mathbin{\circ}\Varid{map}\;\Varid{h}\mathbin{\circ}\Varid{g}))\;\Varid{a}} in the last step of the proof, we had to apply
\begin{hscode}\SaveRestoreHook
\column{B}{@{}>{\hspre}l<{\hspost}@{}}%
\column{5}{@{}>{\hspre}l<{\hspost}@{}}%
\column{20}{@{}>{\hspre}c<{\hspost}@{}}%
\column{20E}{@{}l@{}}%
\column{23}{@{}>{\hspre}l<{\hspost}@{}}%
\column{E}{@{}>{\hspre}l<{\hspost}@{}}%
\>[5]{}\Varid{kleisliPresEE}{}\<[20]%
\>[20]{}\ \mathop{:}\ {}\<[20E]%
\>[23]{}\{\mskip1.5mu \Conid{A},\Conid{B},\Conid{C}\ \mathop{:}\ \Conid{Type}\mskip1.5mu\}\,\to\,\{\mskip1.5mu \Conid{M}\ \mathop{:}\ \Conid{Type}\,\to\,\Conid{Type}\mskip1.5mu\}\,\to\,\Conid{Monad}_{\mathrm{2}}\;\Conid{M}\Rightarrow {}\<[E]%
\\
\>[23]{}(\Varid{f},\Varid{f'}\ \mathop{:}\ \Conid{A}\,\to\,\Conid{M}\;\Conid{B})\,\to\,(\Varid{g},\Varid{g'}\ \mathop{:}\ \Conid{B}\,\to\,\Conid{M}\;\Conid{C})\,\to\,{}\<[E]%
\\
\>[23]{}\Varid{f}\doteq\Varid{f'}\,\to\,\Varid{g}\doteq\Varid{g'}\,\to\,(\Varid{f}\mathbin{\mathbin{>}\hspace{-5.2pt}\mathrel{=}\hspace{-5.5pt}\mathbin{>}}\Varid{g})\doteq(\Varid{f'}\mathbin{\mathbin{>}\hspace{-5.2pt}\mathrel{=}\hspace{-5.5pt}\mathbin{>}}\Varid{g'}){}\<[E]%
\ColumnHook
\end{hscode}\resethooks
instead of just \ensuremath{\Conid{Refl}} as in the case of thin interfaces.
In other words: the specification \ensuremath{\Varid{kleisliJoinMapSpec}} alone is not
strong enough to grant the last step.
It allows one to deduce that \ensuremath{\Varid{join}\mathbin{\circ}\Varid{map}\;\Varid{h}\mathbin{\circ}\Varid{g}} and \ensuremath{\Varid{g}\mathbin{\mathbin{>}\hspace{-5.2pt}\mathrel{=}\hspace{-5.5pt}\mathbin{>}}\Varid{h}} are
extensionally equal.
But this is not enough: we need a proof that Kleisli composition
preserves extensional equality.
This relies on the functorial \ensuremath{\Varid{map}} of \ensuremath{\Conid{M}} preserving extensional
equality, as in the case of thin ADTs.


The moral is that, when the relationships between the monadic operations
\ensuremath{\Varid{pure}}, \ensuremath{\Varid{join}} and \ensuremath{(\mathbin{\mathbin{>}\hspace{-5.2pt}\mathrel{=}\hspace{-5.5pt}\mathbin{>}})} are specified rather then defined, preservation
of extensional equality plays a crucial role even in proofs of
straightforward properties like associativity.
The same situation occurs if we specify bind in terms of \ensuremath{\Varid{pure}} and
\ensuremath{\Varid{join}}.

\subsection{The Wadler view}\label{sec:MonadWadlerView}

A different perspective on monads goes back to
\citep{manes1976algebraic} and has been popularized by P.\
\citet{wadler1992essence}:
a monad on a category $\mathcal{C}$ can be defined by giving an
endo\-function \ensuremath{\Conid{M}} on the objects of $\mathcal{C}$, a family of
arrows \ensuremath{\eta_{\Conid{A}}\ \mathop{:}\ \Conid{A}\,\to\,\Conid{M}\;\Conid{A}} (like \ensuremath{\eta} above, but not required to be
natural), and a ``lifting'' operation that maps any
arrow \ensuremath{\Varid{f}\ \mathop{:}\ \Conid{A}\,\to\,\Conid{M}\;\Conid{B}} to an arrow \ensuremath{\Varid{f}^*\ \mathop{:}\ \Conid{M}\;\Conid{A}\,\to\,\Conid{M}\;\Conid{B}}.
The lifting operation is required to satisfy W1--W3 for any objects
\ensuremath{\Conid{A},\Conid{B},\Conid{C}} and arrows \ensuremath{\Varid{f}\ \mathop{:}\ \Conid{A}\,\to\,\Conid{M}\;\Conid{B}} and \ensuremath{\Varid{g}\ \mathop{:}\ \Conid{B}\,\to\,\Conid{M}\;\Conid{C}}, see
\citep{streicher2003category}:
\begin{hscode}\SaveRestoreHook
\column{B}{@{}>{\hspre}l<{\hspost}@{}}%
\column{3}{@{}>{\hspre}l<{\hspost}@{}}%
\column{44}{@{}>{\hspre}l<{\hspost}@{}}%
\column{E}{@{}>{\hspre}l<{\hspost}@{}}%
\>[3]{}\text{W1.\quad}\Varid{f}^*\mathbin{\circ}\eta_{\Conid{A}}{}\<[44]%
\>[44]{}\mathrel{=}\Varid{f}{}\<[E]%
\\
\>[3]{}\text{W2.\quad}\eta_{\Conid{A}}^*{}\<[44]%
\>[44]{}\mathrel{=}\Varid{id}_{\Conid{M}\;\Conid{A}}{}\<[E]%
\\
\>[3]{}\text{W3.\quad}\Varid{g}^*\mathbin{\circ}\Varid{f}^*{}\<[44]%
\>[44]{}\mathrel{=}(\Varid{g}^*\mathbin{\circ}\Varid{f})^*{}\<[E]%
\ColumnHook
\end{hscode}\resethooks
\paragraph*{From tradition to Wadler and back.}
The two monad definitions can be seen as two views on the same mathematical
concept and we would like the corresponding ADT formulations to preserve
this relationship.

It turns out that, if (\ensuremath{\Conid{M}}, \ensuremath{\eta}, \ensuremath{\mu}) fulfil the properties T1--T5
of the traditional view, then the object part of \ensuremath{\Conid{M}}, \ensuremath{\eta}, and the
lifting operation defined by \ensuremath{\Varid{f}^*\mathrel{=}\mu_{\Conid{M}\;\Conid{B}}\mathbin{\circ}\Varid{map}\;\Varid{f}} satisfy W1--W3.

In turn, given \ensuremath{\Conid{M}}, \ensuremath{\eta} and \ensuremath{\cdot ^*} that satisfy W1--W3, one can define
\ensuremath{\Varid{map}\;\Varid{f}\mathrel{=}(\eta_{\Conid{B}}\mathbin{\circ}\Varid{f})^*} and \ensuremath{\mu_{\Conid{A}}\mathrel{=}\Varid{id}_{\Conid{M}\;\Conid{A}}^*} and prove that
\ensuremath{(\Conid{M},\Varid{map})} is a functor, and that T1--T5 are all satisfied.

This economical way to define a monad has become very popular in
functional programming, where \ensuremath{\Varid{lift}\;\Varid{f}\mathrel{=}\Varid{f}^*} is usually given in
infix form with flipped arguments and called bind:
\ensuremath{\Varid{ma}\bind \Varid{f}\mathrel{=}\Varid{f}^*\;\Varid{ma}}.
This suggests yet another ADT for monads:

\begin{hscode}\SaveRestoreHook
\column{B}{@{}>{\hspre}l<{\hspost}@{}}%
\column{5}{@{}>{\hspre}l<{\hspost}@{}}%
\column{7}{@{}>{\hspre}l<{\hspost}@{}}%
\column{24}{@{}>{\hspre}c<{\hspost}@{}}%
\column{24E}{@{}l@{}}%
\column{27}{@{}>{\hspre}l<{\hspost}@{}}%
\column{41}{@{}>{\hspre}l<{\hspost}@{}}%
\column{E}{@{}>{\hspre}l<{\hspost}@{}}%
\>[5]{}\bf{interface}\;\Conid{Monad}_{\mathrm{3}}\;(\Conid{M}\ \mathop{:}\ \Conid{Type}\,\to\,\Conid{Type})\;\mathbf{where}{}\<[E]%
\\
\>[5]{}\hsindent{2}{}\<[7]%
\>[7]{}\Varid{pure}{}\<[24]%
\>[24]{}\ \mathop{:}\ {}\<[24E]%
\>[27]{}\{\mskip1.5mu \Conid{A}\ \mathop{:}\ \Conid{Type}\mskip1.5mu\}{}\<[41]%
\>[41]{}\,\to\,\Conid{A}\,\to\,\Conid{M}\;\Conid{A}{}\<[E]%
\\
\>[5]{}\hsindent{2}{}\<[7]%
\>[7]{}(\bind ){}\<[24]%
\>[24]{}\ \mathop{:}\ {}\<[24E]%
\>[27]{}\{\mskip1.5mu \Conid{A},\Conid{B}\ \mathop{:}\ \Conid{Type}\mskip1.5mu\}\,\to\,\Conid{M}\;\Conid{A}\,\to\,(\Conid{A}\,\to\,\Conid{M}\;\Conid{B})\,\to\,\Conid{M}\;\Conid{B}{}\<[E]%
\\[\blanklineskip]%
\>[5]{}\hsindent{2}{}\<[7]%
\>[7]{}\Varid{pureLeftIdBind}{}\<[24]%
\>[24]{}\ \mathop{:}\ {}\<[24E]%
\>[27]{}\{\mskip1.5mu \Conid{A},\Conid{B}\ \mathop{:}\ \Conid{Type}\mskip1.5mu\}\,\to\,(\Varid{f}\ \mathop{:}\ \Conid{A}\,\to\,\Conid{M}\;\Conid{B})\,\to\,(\lambda \Varid{a}\Rightarrow \Varid{pure}\;\Varid{a}\bind \Varid{f})\doteq\Varid{f}{}\<[E]%
\\
\>[5]{}\hsindent{2}{}\<[7]%
\>[7]{}\Varid{pureRightIdBind}{}\<[24]%
\>[24]{}\ \mathop{:}\ {}\<[24E]%
\>[27]{}\{\mskip1.5mu \Conid{A}\ \mathop{:}\ \Conid{Type}\mskip1.5mu\}{}\<[41]%
\>[41]{}\,\to\,(\bind \Varid{pure})\doteq\Varid{id}\;\{\mskip1.5mu \Varid{a}\mathrel{=}\Conid{M}\;\Conid{A}\mskip1.5mu\}{}\<[E]%
\\
\>[5]{}\hsindent{2}{}\<[7]%
\>[7]{}\Varid{bindAssoc}{}\<[24]%
\>[24]{}\ \mathop{:}\ {}\<[24E]%
\>[27]{}\{\mskip1.5mu \Conid{A},\Conid{B},\Conid{C}\ \mathop{:}\ \Conid{Type}\mskip1.5mu\}\,\to\,(\Varid{f}\ \mathop{:}\ \Conid{A}\,\to\,\Conid{M}\;\Conid{B})\,\to\,(\Varid{g}\ \mathop{:}\ \Conid{B}\,\to\,\Conid{M}\;\Conid{C})\,\to\,{}\<[E]%
\\
\>[27]{}(\lambda \Varid{ma}\Rightarrow (\Varid{ma}\bind \Varid{f})\bind \Varid{g})\doteq(\lambda \Varid{ma}\Rightarrow \Varid{ma}\bind (\lambda \Varid{a}\Rightarrow \Varid{f}\;\Varid{a}\bind \Varid{g})){}\<[E]%
\ColumnHook
\end{hscode}\resethooks
%
%
The three axioms are formulations of the properties of \ensuremath{\Varid{lift}} W1--W3 in
terms of bind.
We can now \emph{define} \ensuremath{\Varid{map}}, \ensuremath{\Varid{join}} and Kleisli composition in
terms of bind and \ensuremath{\Varid{pure}}:
\begin{hscode}\SaveRestoreHook
\column{B}{@{}>{\hspre}l<{\hspost}@{}}%
\column{5}{@{}>{\hspre}l<{\hspost}@{}}%
\column{12}{@{}>{\hspre}l<{\hspost}@{}}%
\column{13}{@{}>{\hspre}l<{\hspost}@{}}%
\column{14}{@{}>{\hspre}l<{\hspost}@{}}%
\column{E}{@{}>{\hspre}l<{\hspost}@{}}%
\>[5]{}\Varid{map}\ \mathop{:}\ {}\<[12]%
\>[12]{}\{\mskip1.5mu \Conid{A},\Conid{B}\ \mathop{:}\ \Conid{Type}\mskip1.5mu\}\,\to\,\{\mskip1.5mu \Conid{M}\ \mathop{:}\ \Conid{Type}\,\to\,\Conid{Type}\mskip1.5mu\}\,\to\,\Conid{Monad}_{\mathrm{3}}\;\Conid{M}\Rightarrow (\Conid{A}\,\to\,\Conid{B})\,\to\,(\Conid{M}\;\Conid{A}\,\to\,\Conid{M}\;\Conid{B}){}\<[E]%
\\
\>[5]{}\Varid{map}\;\Varid{f}\;\Varid{ma}\mathrel{=}\Varid{ma}\bind (\Varid{pure}\mathbin{\circ}\Varid{f}){}\<[E]%
\\[\blanklineskip]%
\>[5]{}\Varid{join}\ \mathop{:}\ {}\<[13]%
\>[13]{}\{\mskip1.5mu \Conid{A}\ \mathop{:}\ \Conid{Type}\mskip1.5mu\}\,\to\,\{\mskip1.5mu \Conid{M}\ \mathop{:}\ \Conid{Type}\,\to\,\Conid{Type}\mskip1.5mu\}\,\to\,\Conid{Monad}_{\mathrm{3}}\;\Conid{M}\Rightarrow \Conid{M}\;(\Conid{M}\;\Conid{A})\,\to\,\Conid{M}\;\Conid{A}{}\<[E]%
\\
\>[5]{}\Varid{join}\;\Varid{mma}\mathrel{=}\Varid{mma}\bind \Varid{id}{}\<[E]%
\\[\blanklineskip]%
\>[5]{}(\mathbin{\mathbin{>}\hspace{-5.2pt}\mathrel{=}\hspace{-5.5pt}\mathbin{>}})\ \mathop{:}\ {}\<[14]%
\>[14]{}\{\mskip1.5mu \Conid{A},\Conid{B},\Conid{C}\ \mathop{:}\ \Conid{Type}\mskip1.5mu\}\,\to\,\{\mskip1.5mu \Conid{M}\ \mathop{:}\ \Conid{Type}\,\to\,\Conid{Type}\mskip1.5mu\}\,\to\,\Conid{Monad}_{\mathrm{3}}\;\Conid{M}\Rightarrow {}\<[E]%
\\
\>[14]{}(\Conid{A}\,\to\,\Conid{M}\;\Conid{B})\,\to\,(\Conid{B}\,\to\,\Conid{M}\;\Conid{C})\,\to\,(\Conid{A}\,\to\,\Conid{M}\;\Conid{C}){}\<[E]%
\\
\>[5]{}\Varid{f}\mathbin{\mathbin{>}\hspace{-5.2pt}\mathrel{=}\hspace{-5.5pt}\mathbin{>}}\Varid{g}\mathrel{=}\lambda \Varid{a}\Rightarrow \Varid{f}\;\Varid{a}\bind \Varid{g}{}\<[E]%
\ColumnHook
\end{hscode}\resethooks
The obligation is now to prove that \ensuremath{\Varid{pure}} and \ensuremath{\Varid{join}} fulfil the
properties T1--T5, for instance, that \ensuremath{\Varid{pure}} is a natural transformation.
In much the same way as for formalizations of the traditional view, some
of these proofs can be implemented straightforwardly.
But in some cases, one runs into trouble.
Consider the proof of T2. Triangle right: \ensuremath{\mu_{\Conid{A}}\mathbin{\circ}\Conid{M}\;\eta_{\Conid{A}}\mathrel{=}\Varid{id}_{\Conid{M}\;\Conid{A}}}

\begin{hscode}\SaveRestoreHook
\column{B}{@{}>{\hspre}l<{\hspost}@{}}%
\column{5}{@{}>{\hspre}l<{\hspost}@{}}%
\column{12}{@{}>{\hspre}l<{\hspost}@{}}%
\column{28}{@{}>{\hspre}c<{\hspost}@{}}%
\column{28E}{@{}l@{}}%
\column{31}{@{}>{\hspre}l<{\hspost}@{}}%
\column{57}{@{}>{\hspre}l<{\hspost}@{}}%
\column{E}{@{}>{\hspre}l<{\hspost}@{}}%
\>[5]{}\Varid{triangleRightFromBind}{}\<[28]%
\>[28]{}\ \mathop{:}\ {}\<[28E]%
\>[31]{}\{\mskip1.5mu \Conid{A}\ \mathop{:}\ \Conid{Type}\mskip1.5mu\}\,\to\,\{\mskip1.5mu \Conid{M}\ \mathop{:}\ \Conid{Type}\,\to\,\Conid{Type}\mskip1.5mu\}\,\to\,\Conid{Monad}_{\mathrm{3}}\;\Conid{M}\Rightarrow {}\<[E]%
\\
\>[31]{}\Varid{join}\mathbin{\circ}\Varid{map}\;\Varid{pure}\doteq\Varid{id}\;\{\mskip1.5mu \Varid{a}\mathrel{=}\Conid{M}\;\Conid{A}\mskip1.5mu\}{}\<[E]%
\\
\>[5]{}\Varid{triangleRightFromBind}\;\{\mskip1.5mu \Conid{A}\mskip1.5mu\}\;\{\mskip1.5mu \Conid{M}\mskip1.5mu\}\;\Varid{ma}\mathrel{=}{}\<[E]%
\\
\>[5]{}\hsindent{7}{}\<[12]%
\>[12]{}(\Varid{join}\;(\Varid{map}\;\Varid{pure}\;\Varid{ma})){}\<[57]%
\>[57]{}=\hspace{-3pt}\{\; \Conid{Refl}\;\}\hspace{-3pt}={}\<[E]%
\\
\>[5]{}\hsindent{7}{}\<[12]%
\>[12]{}((\Varid{ma}\bind (\Varid{pure}\mathbin{\circ}\Varid{pure}))\bind \Varid{id}){}\<[57]%
\>[57]{}=\hspace{-3pt}\{\; \Varid{bindAssoc}\;(\Varid{pure}\mathbin{\circ}\Varid{pure})\;\Varid{id}\;\Varid{ma}\;\}\hspace{-3pt}={}\<[E]%
\\
\>[5]{}\hsindent{7}{}\<[12]%
\>[12]{}(\Varid{ma}\bind (\lambda \Varid{a}\Rightarrow \Varid{pure}\;(\Varid{pure}\;\Varid{a})\bind \Varid{id})){}\<[57]%
\>[57]{}=\hspace{-3pt}\{\; \Conid{Refl}\;\}\hspace{-3pt}={}\<[E]%
\\
\>[5]{}\hsindent{7}{}\<[12]%
\>[12]{}(\Varid{ma}\bind ((\lambda \Varid{a}\Rightarrow \Varid{pure}\;\Varid{a}\bind \Varid{id})\mathbin{\circ}\Varid{pure})){}\<[57]%
\>[57]{}=\hspace{-3pt}\{\; \bf{?whatnow}\;\}\hspace{-3pt}={}\<[E]%
\\
\>[5]{}\hsindent{7}{}\<[12]%
\>[12]{}(\Varid{ma}\bind \Varid{id}\mathbin{\circ}\Varid{pure}){}\<[57]%
\>[57]{}=\hspace{-3pt}\{\; \Conid{Refl}\;\}\hspace{-3pt}={}\<[E]%
\\
\>[5]{}\hsindent{7}{}\<[12]%
\>[12]{}(\Varid{ma}\bind \Varid{pure}){}\<[57]%
\>[57]{}=\hspace{-3pt}\{\; \Varid{pureRightIdBind}\;\Varid{ma}\;\}\hspace{-3pt}={}\<[E]%
\\
\>[5]{}\hsindent{7}{}\<[12]%
\>[12]{}(\Varid{ma})\;{}\<[57]%
\>[57]{}\Conid{QED}{}\<[E]%
\ColumnHook
\end{hscode}\resethooks
\noindent
We know that \ensuremath{\lambda \Varid{a}\Rightarrow \Varid{pure}\;\Varid{a}\bind \Varid{id}} and \ensuremath{\Varid{id}} are \emph{extensionally}
equal by \ensuremath{\Varid{pureLeftIdBind}\;\Varid{id}}.
If this equality would hold \emph{intensionally}, we could fill the
hole by congruence.
But we cannot strengthen \ensuremath{(\doteq)} to \ensuremath{(\mathrel{=})} in \ensuremath{\Varid{pureLeftIdBind}}.
Instead, we extend our ADT with
%
\begin{hscode}\SaveRestoreHook
\column{B}{@{}>{\hspre}l<{\hspost}@{}}%
\column{7}{@{}>{\hspre}l<{\hspost}@{}}%
\column{23}{@{}>{\hspre}c<{\hspost}@{}}%
\column{23E}{@{}l@{}}%
\column{26}{@{}>{\hspre}l<{\hspost}@{}}%
\column{E}{@{}>{\hspre}l<{\hspost}@{}}%
\>[7]{}\Varid{liftPresEE}{}\<[23]%
\>[23]{}\ \mathop{:}\ {}\<[23E]%
\>[26]{}\{\mskip1.5mu \Conid{A},\Conid{B}\ \mathop{:}\ \Conid{Type}\mskip1.5mu\}\,\to\,(\Varid{f},\Varid{g}\ \mathop{:}\ \Conid{A}\,\to\,\Conid{M}\;\Conid{B})\,\to\,\Varid{f}\doteq\Varid{g}\,\to\,(\bind \Varid{f})\doteq(\bind \Varid{g}){}\<[E]%
\ColumnHook
\end{hscode}\resethooks
and complete the proof by filling in \ensuremath{\bf{?whatnow}} by:
\begin{hscode}\SaveRestoreHook
\column{B}{@{}>{\hspre}l<{\hspost}@{}}%
\column{14}{@{}>{\hspre}l<{\hspost}@{}}%
\column{26}{@{}>{\hspre}l<{\hspost}@{}}%
\column{E}{@{}>{\hspre}l<{\hspost}@{}}%
\>[14]{}\Varid{liftPresEE}\;{}\<[26]%
\>[26]{}((\lambda \Varid{a}\Rightarrow \Varid{pure}\;\Varid{a}\bind \Varid{id})\mathbin{\circ}\Varid{pure})\;(\Varid{id}\mathbin{\circ}\Varid{pure})\;(\lambda \Varid{a}\Rightarrow \Varid{pureLeftIdBind}\;\Varid{id}\;(\Varid{pure}\;\Varid{a}))\;\Varid{ma}{}\<[E]%
\ColumnHook
\end{hscode}\resethooks
Notice that, in this approach, \ensuremath{\Varid{map}} is defined in terms of \ensuremath{\Varid{pure}} and
bind. Thus, we do not have at our disposal the axioms of the functor ADT
and thus we cannot leverage \ensuremath{\Varid{mapPresEE}} to \emph{derive} \ensuremath{\Varid{liftPresEE}}
as we have done in the traditional formulation for Kleisli
composition.

The moral is that, even if we adopt the Wadler view on monads and a more
economical specification, we have to require \ensuremath{\Varid{lift}} to preserve
extensional equality if we want our specification to be consistent with the
traditional one.

This completes the discussion on different notions of monads and on the
role of extensional equality preservation in generic proofs of monad
laws.
For the rest of the paper, we apply the traditional view on monads and the fat
monad interface.

\subsection{More monad properties}\label{sec:MonadMoreResults}

As we have seen in section \ref{section:example} for \ensuremath{\Varid{flowMonLemma}},
extensional equality preservation is crucially needed in inductive
proofs.
We discuss more examples of applications of the principle in the
context of DSLs for dynamical systems theory in section \ref{section:applications}.
In the rest of this section, we prepare by deriving two intermediate
properties. 
The first is the extensional equality between \ensuremath{\Varid{map}\;\Varid{f}\mathbin{\circ}\Varid{join}\mathbin{\circ}\Varid{map}\;\Varid{g}} and \ensuremath{\Varid{join}\mathbin{\circ}\Varid{map}\;(\Varid{map}\;\Varid{f}\mathbin{\circ}\Varid{g})}:
\begin{hscode}\SaveRestoreHook
\column{B}{@{}>{\hspre}l<{\hspost}@{}}%
\column{5}{@{}>{\hspre}l<{\hspost}@{}}%
\column{7}{@{}>{\hspre}l<{\hspost}@{}}%
\column{19}{@{}>{\hspre}c<{\hspost}@{}}%
\column{19E}{@{}l@{}}%
\column{22}{@{}>{\hspre}l<{\hspost}@{}}%
\column{47}{@{}>{\hspre}l<{\hspost}@{}}%
\column{E}{@{}>{\hspre}l<{\hspost}@{}}%
\>[5]{}\Varid{mapJoinLemma}{}\<[19]%
\>[19]{}\ \mathop{:}\ {}\<[19E]%
\>[22]{}\{\mskip1.5mu \Conid{M}\ \mathop{:}\ \Conid{Type}\,\to\,\Conid{Type}\mskip1.5mu\}\,\to\,\{\mskip1.5mu \Conid{A},\Conid{B},\Conid{C}\ \mathop{:}\ \Conid{Type}\mskip1.5mu\}\,\to\,\Conid{Monad}\;\Conid{M}\Rightarrow {}\<[E]%
\\
\>[22]{}(\Varid{f}\ \mathop{:}\ \Conid{B}\,\to\,\Conid{C})\,\to\,(\Varid{g}\ \mathop{:}\ \Conid{A}\,\to\,\Conid{M}\;\Conid{B})\,\to\,{}\<[E]%
\\
\>[22]{}(\doteq)\;\{\mskip1.5mu \Conid{A}\mathrel{=}\Conid{M}\;\Conid{A}\mskip1.5mu\}\;\{\mskip1.5mu \Conid{B}\mathrel{=}\Conid{M}\;\Conid{C}\mskip1.5mu\}\;(\Varid{map}\;\Varid{f}\mathbin{\circ}\Varid{join}\mathbin{\circ}\Varid{map}\;\Varid{g})\;(\Varid{join}\mathbin{\circ}\Varid{map}\;(\Varid{map}\;\Varid{f}\mathbin{\circ}\Varid{g})){}\<[E]%
\\
\>[5]{}\Varid{mapJoinLemma}\;\Varid{f}\;\Varid{g}\;\Varid{ma}\mathrel{=}{}\<[E]%
\\
\>[5]{}\hsindent{2}{}\<[7]%
\>[7]{}(\Varid{map}\;\Varid{f}\;(\Varid{join}\;(\Varid{map}\;\Varid{g}\;\Varid{ma}))){}\<[47]%
\>[47]{}=\hspace{-3pt}\{\; \Varid{joinNatTrans}\;\Varid{f}\;(\Varid{map}\;\Varid{g}\;\Varid{ma})\;\}\hspace{-3pt}={}\<[E]%
\\
\>[5]{}\hsindent{2}{}\<[7]%
\>[7]{}(\Varid{join}\;(\Varid{map}\;(\Varid{map}\;\Varid{f})\;(\Varid{map}\;\Varid{g}\;\Varid{ma}))){}\<[47]%
\>[47]{}=\hspace{-3pt}\{\; \Conid{Refl}\;\}\hspace{-3pt}={}\<[E]%
\\
\>[5]{}\hsindent{2}{}\<[7]%
\>[7]{}(\Varid{join}\;((\Varid{map}\;(\Varid{map}\;\Varid{f})\mathbin{\circ}(\Varid{map}\;\Varid{g}))\;\Varid{ma})){}\<[47]%
\>[47]{}=\hspace{-3pt}\{\; \Varid{cong}\;(\Varid{sym}\;(\Varid{mapPresComp}\;(\Varid{map}\;\Varid{f})\;\Varid{g}\;\Varid{ma}))\;\}\hspace{-3pt}={}\<[E]%
\\
\>[5]{}\hsindent{2}{}\<[7]%
\>[7]{}(\Varid{join}\;(\Varid{map}\;(\Varid{map}\;\Varid{f}\mathbin{\circ}\Varid{g})\;\Varid{ma}))\;{}\<[47]%
\>[47]{}\Conid{QED}{}\<[E]%
\ColumnHook
\end{hscode}\resethooks
The second,
\begin{hscode}\SaveRestoreHook
\column{B}{@{}>{\hspre}l<{\hspost}@{}}%
\column{5}{@{}>{\hspre}l<{\hspost}@{}}%
\column{7}{@{}>{\hspre}l<{\hspost}@{}}%
\column{22}{@{}>{\hspre}c<{\hspost}@{}}%
\column{22E}{@{}l@{}}%
\column{25}{@{}>{\hspre}l<{\hspost}@{}}%
\column{42}{@{}>{\hspre}l<{\hspost}@{}}%
\column{E}{@{}>{\hspre}l<{\hspost}@{}}%
\>[5]{}\Varid{mapKleisliLemma}{}\<[22]%
\>[22]{}\ \mathop{:}\ {}\<[22E]%
\>[25]{}\{\mskip1.5mu \Conid{A},\Conid{B},\Conid{C},\Conid{D}\ \mathop{:}\ \Conid{Type}\mskip1.5mu\}\,\to\,\{\mskip1.5mu \Conid{M}\ \mathop{:}\ \Conid{Type}\,\to\,\Conid{Type}\mskip1.5mu\}\,\to\,\Conid{Monad}\;\Conid{M}\Rightarrow {}\<[E]%
\\
\>[25]{}(\Varid{f}\ \mathop{:}\ \Conid{A}\,\to\,\Conid{M}\;\Conid{B})\,\to\,(\Varid{g}\ \mathop{:}\ \Conid{B}\,\to\,\Conid{M}\;\Conid{C})\,\to\,(\Varid{h}\ \mathop{:}\ \Conid{C}\,\to\,\Conid{D})\,\to\,{}\<[E]%
\\
\>[25]{}(\Varid{map}\;\Varid{h}\mathbin{\circ}(\Varid{f}\mathbin{\mathbin{>}\hspace{-5.2pt}\mathrel{=}\hspace{-5.5pt}\mathbin{>}}\Varid{g}))\doteq(\Varid{f}\mathbin{\mathbin{>}\hspace{-5.2pt}\mathrel{=}\hspace{-5.5pt}\mathbin{>}}\Varid{map}\;\Varid{h}\mathbin{\circ}\Varid{g}){}\<[E]%
\\
\>[5]{}\Varid{mapKleisliLemma}\;\Varid{f}\;\Varid{g}\;\Varid{h}\;\Varid{ma}\mathrel{=}{}\<[E]%
\\
\>[5]{}\hsindent{2}{}\<[7]%
\>[7]{}((\Varid{map}\;\Varid{h}\mathbin{\circ}(\Varid{f}\mathbin{\mathbin{>}\hspace{-5.2pt}\mathrel{=}\hspace{-5.5pt}\mathbin{>}}\Varid{g}))\;\Varid{ma}){}\<[42]%
\>[42]{}=\hspace{-3pt}\{\; \Varid{cong}\;(\Varid{kleisliJoinMapSpec}\;\Varid{f}\;\Varid{g}\;\Varid{ma})\;\}\hspace{-3pt}={}\<[E]%
\\
\>[5]{}\hsindent{2}{}\<[7]%
\>[7]{}(\Varid{map}\;\Varid{h}\;(\Varid{join}\;(\Varid{map}\;\Varid{g}\;(\Varid{f}\;\Varid{ma})))){}\<[42]%
\>[42]{}=\hspace{-3pt}\{\; \Varid{mapJoinLemma}\;\Varid{h}\;\Varid{g}\;(\Varid{f}\;\Varid{ma})\;\}\hspace{-3pt}={}\<[E]%
\\
\>[5]{}\hsindent{2}{}\<[7]%
\>[7]{}(\Varid{join}\;(\Varid{map}\;(\Varid{map}\;\Varid{h}\mathbin{\circ}\Varid{g})\;(\Varid{f}\;\Varid{ma}))){}\<[42]%
\>[42]{}=\hspace{-3pt}\{\; \Varid{sym}\;(\Varid{kleisliJoinMapSpec}\;\Varid{f}\;(\Varid{map}\;\Varid{h}\mathbin{\circ}\Varid{g})\;\Varid{ma})\;\}\hspace{-3pt}={}\<[E]%
\\
\>[5]{}\hsindent{2}{}\<[7]%
\>[7]{}((\Varid{f}\mathbin{\mathbin{>}\hspace{-5.2pt}\mathrel{=}\hspace{-5.5pt}\mathbin{>}}\Varid{map}\;\Varid{h}\mathbin{\circ}\Varid{g})\;\Varid{ma})\;{}\<[42]%
\>[42]{}\Conid{QED}{}\<[E]%
\ColumnHook
\end{hscode}\resethooks
can be seen as an associativity law by rewriting it in terms of
\ensuremath{(\mathbin{\mathbin{<}\hspace{-5.5pt}\mathrel{=}\hspace{-5.2pt}\mathbin{<}})\mathrel{=}\Varid{flip}\;(\mathbin{\mathbin{>}\hspace{-5.2pt}\mathrel{=}\hspace{-5.5pt}\mathbin{>}})}:

\begin{hscode}\SaveRestoreHook
\column{B}{@{}>{\hspre}l<{\hspost}@{}}%
\column{5}{@{}>{\hspre}l<{\hspost}@{}}%
\column{23}{@{}>{\hspre}c<{\hspost}@{}}%
\column{23E}{@{}l@{}}%
\column{26}{@{}>{\hspre}l<{\hspost}@{}}%
\column{E}{@{}>{\hspre}l<{\hspost}@{}}%
\>[5]{}\Varid{mapKleisliLemma}{}\<[23]%
\>[23]{}\ \mathop{:}\ {}\<[23E]%
\>[26]{}\{\mskip1.5mu \Conid{A},\Conid{B},\Conid{C},\Conid{D}\ \mathop{:}\ \Conid{Type}\mskip1.5mu\}\,\to\,\{\mskip1.5mu \Conid{M}\ \mathop{:}\ \Conid{Type}\,\to\,\Conid{Type}\mskip1.5mu\}\,\to\,\Conid{Monad}\;\Conid{M}\Rightarrow {}\<[E]%
\\
\>[26]{}(\Varid{f}\ \mathop{:}\ \Conid{A}\,\to\,\Conid{M}\;\Conid{B})\,\to\,(\Varid{g}\ \mathop{:}\ \Conid{B}\,\to\,\Conid{M}\;\Conid{C})\,\to\,(\Varid{h}\ \mathop{:}\ \Conid{C}\,\to\,\Conid{D})\,\to\,{}\<[E]%
\\
\>[26]{}(\Varid{map}\;\Varid{h}\mathbin{\circ}(\Varid{g}\mathbin{\mathbin{<}\hspace{-5.5pt}\mathrel{=}\hspace{-5.2pt}\mathbin{<}}\Varid{f}))\doteq((\Varid{map}\;\Varid{h}\mathbin{\circ}\Varid{g})\mathbin{\mathbin{<}\hspace{-5.5pt}\mathrel{=}\hspace{-5.2pt}\mathbin{<}}\Varid{f}){}\<[E]%
\ColumnHook
\end{hscode}\resethooks
%

\section{Applications in dynamical systems and control theory}
\label{section:applications}

In this section we discuss applications of the principle of preservation
of extensional equality to dynamical systems and control theory.
We have seen in section \ref{section:example} that time discrete
deterministic dynamical systems on a set \ensuremath{\Conid{X}} are functions of type \ensuremath{\Conid{X}\,\to\,\Conid{X}}
\begin{hscode}\SaveRestoreHook
\column{B}{@{}>{\hspre}l<{\hspost}@{}}%
\column{3}{@{}>{\hspre}l<{\hspost}@{}}%
\column{E}{@{}>{\hspre}l<{\hspost}@{}}%
\>[3]{}\Conid{DetSys}\ \mathop{:}\ \Conid{Type}\,\to\,\Conid{Type}{}\<[E]%
\\
\>[3]{}\Conid{DetSys}\;\Conid{X}\mathrel{=}\Conid{X}\,\to\,\Conid{X}{}\<[E]%
\ColumnHook
\end{hscode}\resethooks
and that generalizing this notion to systems with uncertainties leads to
monadic systems
\begin{hscode}\SaveRestoreHook
\column{B}{@{}>{\hspre}l<{\hspost}@{}}%
\column{3}{@{}>{\hspre}l<{\hspost}@{}}%
\column{E}{@{}>{\hspre}l<{\hspost}@{}}%
\>[3]{}\Conid{MonSys}\ \mathop{:}\ (\Conid{Type}\,\to\,\Conid{Type})\,\to\,\Conid{Type}\,\to\,\Conid{Type}{}\<[E]%
\\
\>[3]{}\Conid{MonSys}\;\Conid{M}\;\Conid{X}\mathrel{=}\Conid{X}\,\to\,\Conid{M}\;\Conid{X}{}\<[E]%
\ColumnHook
\end{hscode}\resethooks
where \ensuremath{\Conid{M}} is an \emph{uncertainty} monad: \ensuremath{\Conid{List}}, \ensuremath{\Conid{Maybe}}, \ensuremath{\Conid{Dist}}
\citep{10.1017/S0956796805005721}, \ensuremath{\Conid{SimpleProb}}
\citep{ionescu2009, 2017_Botta_Jansson_Ionescu}, etc.
For monadic systems, one can derive a number of general results.
One is that every deterministic system can be embedded in a monadic
system:

\begin{hscode}\SaveRestoreHook
\column{B}{@{}>{\hspre}l<{\hspost}@{}}%
\column{3}{@{}>{\hspre}l<{\hspost}@{}}%
\column{E}{@{}>{\hspre}l<{\hspost}@{}}%
\>[3]{}\Varid{embed}\ \mathop{:}\ \{\mskip1.5mu \Conid{X}\ \mathop{:}\ \Conid{Type}\mskip1.5mu\}\,\to\,\{\mskip1.5mu \Conid{M}\ \mathop{:}\ \Conid{Type}\,\to\,\Conid{Type}\mskip1.5mu\}\,\to\,\Conid{Monad}\;\Conid{M}\Rightarrow \Conid{DetSys}\;\Conid{X}\,\to\,\Conid{MonSys}\;\Conid{M}\;\Conid{X}{}\<[E]%
\\
\>[3]{}\Varid{embed}\;\Varid{f}\mathrel{=}\Varid{pure}\mathbin{\circ}\Varid{f}{}\<[E]%
\ColumnHook
\end{hscode}\resethooks
A more interesting result is that the flow of a monadic system is a monoid morphism from \ensuremath{(\mathbb{N},(\mathbin{+}),\mathrm{0})} to \ensuremath{(\Conid{MonSys}\;\Conid{M}\;\Conid{X},(\mathbin{\mathbin{>}\hspace{-5.2pt}\mathrel{=}\hspace{-5.5pt}\mathbin{>}}),\Varid{pure})}.
As discussed in section~\ref{section:example}, \ensuremath{\Varid{flowMonL}\doteq\Varid{flowMonR}} and here we write just \ensuremath{\Varid{flow}}.
The two parts of the monoid morphism proof are
\begin{hscode}\SaveRestoreHook
\column{B}{@{}>{\hspre}l<{\hspost}@{}}%
\column{3}{@{}>{\hspre}l<{\hspost}@{}}%
\column{15}{@{}>{\hspre}c<{\hspost}@{}}%
\column{15E}{@{}l@{}}%
\column{18}{@{}>{\hspre}l<{\hspost}@{}}%
\column{E}{@{}>{\hspre}l<{\hspost}@{}}%
\>[3]{}\Varid{flowLemma1}{}\<[15]%
\>[15]{}\ \mathop{:}\ {}\<[15E]%
\>[18]{}\{\mskip1.5mu \Conid{X}\ \mathop{:}\ \Conid{Type}\mskip1.5mu\}\,\to\,\{\mskip1.5mu \Conid{M}\ \mathop{:}\ \Conid{Type}\,\to\,\Conid{Type}\mskip1.5mu\}\,\to\,\Conid{Monad}\;\Conid{M}\Rightarrow {}\<[E]%
\\
\>[18]{}(\Varid{f}\ \mathop{:}\ \Conid{MonSys}\;\Conid{M}\;\Conid{X})\,\to\,\Varid{flow}\;\Varid{f}\;\Conid{Z}\doteq\Varid{pure}{}\<[E]%
\\[\blanklineskip]%
\>[3]{}\Varid{flowLemma2}{}\<[15]%
\>[15]{}\ \mathop{:}\ {}\<[15E]%
\>[18]{}\{\mskip1.5mu \Conid{X}\ \mathop{:}\ \Conid{Type}\mskip1.5mu\}\,\to\,\{\mskip1.5mu \Conid{M}\ \mathop{:}\ \Conid{Type}\,\to\,\Conid{Type}\mskip1.5mu\}\,\to\,\Conid{Monad}\;\Conid{M}\Rightarrow \{\mskip1.5mu \Varid{m},\Varid{n}\ \mathop{:}\ \mathbb{N}\mskip1.5mu\}\,\to\,{}\<[E]%
\\
\>[18]{}(\Varid{f}\ \mathop{:}\ \Conid{MonSys}\;\Conid{M}\;\Conid{X})\,\to\,\Varid{flow}\;\Varid{f}\;(\Varid{m}\mathbin{+}\Varid{n})\doteq(\Varid{flow}\;\Varid{f}\;\Varid{m}\mathbin{\mathbin{>}\hspace{-5.2pt}\mathrel{=}\hspace{-5.5pt}\mathbin{>}}\Varid{flow}\;\Varid{f}\;\Varid{n}){}\<[E]%
\ColumnHook
\end{hscode}\resethooks
Proving \ensuremath{\Varid{flowLemma1}} is immediate (because \ensuremath{\Varid{flow}\;\Varid{f}\;\Conid{Z}\mathrel{=}\Varid{pure}}):
\begin{hscode}\SaveRestoreHook
\column{B}{@{}>{\hspre}l<{\hspost}@{}}%
\column{3}{@{}>{\hspre}l<{\hspost}@{}}%
\column{E}{@{}>{\hspre}l<{\hspost}@{}}%
\>[3]{}\Varid{flowLemma1}\;\Varid{f}\mathrel{=}\Varid{reflEE}{}\<[E]%
\ColumnHook
\end{hscode}\resethooks
We prove \ensuremath{\Varid{flowLemma2}} by induction on \ensuremath{\Varid{m}} using the properties from
section \ref{section:monads}: \ensuremath{\Varid{pure}} is a left and right identity of
Kleisli composition and Kleisli composition is associative.
The base case is straightforward
\begin{hscode}\SaveRestoreHook
\column{B}{@{}>{\hspre}l<{\hspost}@{}}%
\column{3}{@{}>{\hspre}l<{\hspost}@{}}%
\column{5}{@{}>{\hspre}l<{\hspost}@{}}%
\column{37}{@{}>{\hspre}l<{\hspost}@{}}%
\column{E}{@{}>{\hspre}l<{\hspost}@{}}%
\>[3]{}\Varid{flowLemma2}\;\{\mskip1.5mu \Varid{m}\mathrel{=}\Conid{Z}\mskip1.5mu\}\;\{\mskip1.5mu \Varid{n}\mskip1.5mu\}\;\Varid{f}\;\Varid{x}\mathrel{=}{}\<[E]%
\\
\>[3]{}\hsindent{2}{}\<[5]%
\>[5]{}(\Varid{flow}\;\Varid{f}\;(\Conid{Z}\mathbin{+}\Varid{n})\;\Varid{x}){}\<[37]%
\>[37]{}=\hspace{-3pt}\{\; \Conid{Refl}\;\}\hspace{-3pt}={}\<[E]%
\\
\>[3]{}\hsindent{2}{}\<[5]%
\>[5]{}(\Varid{flow}\;\Varid{f}\;\Varid{n}\;\Varid{x}){}\<[37]%
\>[37]{}=\hspace{-3pt}\{\; \Varid{sym}\;(\Varid{pureLeftIdKleisli}\;(\Varid{flow}\;\Varid{f}\;\Varid{n})\;\Varid{x})\;\}\hspace{-3pt}={}\<[E]%
\\
\>[3]{}\hsindent{2}{}\<[5]%
\>[5]{}((\Varid{pure}\mathbin{\mathbin{>}\hspace{-5.2pt}\mathrel{=}\hspace{-5.5pt}\mathbin{>}}\Varid{flow}\;\Varid{f}\;\Varid{n})\;\Varid{x}){}\<[37]%
\>[37]{}=\hspace{-3pt}\{\; \Conid{Refl}\;\}\hspace{-3pt}={}\<[E]%
\\
\>[3]{}\hsindent{2}{}\<[5]%
\>[5]{}((\Varid{flow}\;\Varid{f}\;\Conid{Z}\mathbin{\mathbin{>}\hspace{-5.2pt}\mathrel{=}\hspace{-5.5pt}\mathbin{>}}\Varid{flow}\;\Varid{f}\;\Varid{n})\;\Varid{x})\;{}\<[37]%
\>[37]{}\Conid{QED}{}\<[E]%
\ColumnHook
\end{hscode}\resethooks
but the induction step again relies on Kleisli composition preserving extensional equality.
\begin{hscode}\SaveRestoreHook
\column{B}{@{}>{\hspre}l<{\hspost}@{}}%
\column{3}{@{}>{\hspre}l<{\hspost}@{}}%
\column{5}{@{}>{\hspre}l<{\hspost}@{}}%
\column{44}{@{}>{\hspre}l<{\hspost}@{}}%
\column{62}{@{}>{\hspre}l<{\hspost}@{}}%
\column{70}{@{}>{\hspre}l<{\hspost}@{}}%
\column{E}{@{}>{\hspre}l<{\hspost}@{}}%
\>[3]{}\Varid{flowLemma2}\;\Varid{f}\;\{\mskip1.5mu \Varid{m}\mathrel{=}\Conid{S}\;\Varid{l}\mskip1.5mu\}\;\{\mskip1.5mu \Varid{n}\mskip1.5mu\}\;\Varid{x}\mathrel{=}{}\<[E]%
\\
\>[3]{}\hsindent{2}{}\<[5]%
\>[5]{}(\Varid{flow}\;\Varid{f}\;(\Conid{S}\;\Varid{l}\mathbin{+}\Varid{n})\;\Varid{x}){}\<[44]%
\>[44]{}=\hspace{-3pt}\{\; \Conid{Refl}\;\}\hspace{-3pt}={}\<[E]%
\\
\>[3]{}\hsindent{2}{}\<[5]%
\>[5]{}((\Varid{f}\mathbin{\mathbin{>}\hspace{-5.2pt}\mathrel{=}\hspace{-5.5pt}\mathbin{>}}\Varid{flow}\;\Varid{f}\;(\Varid{l}\mathbin{+}\Varid{n}))\;\Varid{x}){}\<[44]%
\>[44]{}=\hspace{-3pt}\{\; \Varid{kleisliPresEE}\;{}\<[62]%
\>[62]{}\Varid{f}\;\Varid{f}\;{}\<[70]%
\>[70]{}(\Varid{flow}\;\Varid{f}\;(\Varid{l}\mathbin{+}\Varid{n}))\;(\Varid{flow}\;\Varid{f}\;\Varid{l}\mathbin{\mathbin{>}\hspace{-5.2pt}\mathrel{=}\hspace{-5.5pt}\mathbin{>}}\Varid{flow}\;\Varid{f}\;\Varid{n})\;{}\<[E]%
\\
\>[62]{}\Varid{reflEE}\;{}\<[70]%
\>[70]{}(\Varid{flowLemma2}\;\Varid{f})\;\Varid{x}\;\}\hspace{-3pt}={}\<[E]%
\\
\>[3]{}\hsindent{2}{}\<[5]%
\>[5]{}((\Varid{f}\mathbin{\mathbin{>}\hspace{-5.2pt}\mathrel{=}\hspace{-5.5pt}\mathbin{>}}(\Varid{flow}\;\Varid{f}\;\Varid{l}\mathbin{\mathbin{>}\hspace{-5.2pt}\mathrel{=}\hspace{-5.5pt}\mathbin{>}}\Varid{flow}\;\Varid{f}\;\Varid{n}))\;\Varid{x}){}\<[44]%
\>[44]{}=\hspace{-3pt}\{\; \Varid{sym}\;(\Varid{kleisliAssoc}\;\Varid{f}\;(\Varid{flow}\;\Varid{f}\;\Varid{l})\;(\Varid{flow}\;\Varid{f}\;\Varid{n})\;\Varid{x})\;\}\hspace{-3pt}={}\<[E]%
\\
\>[3]{}\hsindent{2}{}\<[5]%
\>[5]{}(((\Varid{f}\mathbin{\mathbin{>}\hspace{-5.2pt}\mathrel{=}\hspace{-5.5pt}\mathbin{>}}\Varid{flow}\;\Varid{f}\;\Varid{l})\mathbin{\mathbin{>}\hspace{-5.2pt}\mathrel{=}\hspace{-5.5pt}\mathbin{>}}\Varid{flow}\;\Varid{f}\;\Varid{n})\;\Varid{x}){}\<[44]%
\>[44]{}=\hspace{-3pt}\{\; \Conid{Refl}\;\}\hspace{-3pt}={}\<[E]%
\\
\>[3]{}\hsindent{2}{}\<[5]%
\>[5]{}((\Varid{flow}\;\Varid{f}\;(\Conid{S}\;\Varid{l})\mathbin{\mathbin{>}\hspace{-5.2pt}\mathrel{=}\hspace{-5.5pt}\mathbin{>}}\Varid{flow}\;\Varid{f}\;\Varid{n})\;\Varid{x})\;{}\<[44]%
\>[44]{}\Conid{QED}{}\<[E]%
\ColumnHook
\end{hscode}\resethooks
As seen in section \ref{section:monads}, this follows directly from
the monad ADT and from the preservation of extensional equality for
functors.

\paragraph*{A representation theorem.}
Another important result for monadic systems is a representation
theorem: any monadic system \ensuremath{\Varid{f}\ \mathop{:}\ \Conid{MonSys}\;\Conid{M}\;\Conid{X}} can be represented by
a deterministic system on \ensuremath{\Conid{M}\;\Conid{X}}. With
\begin{hscode}\SaveRestoreHook
\column{B}{@{}>{\hspre}l<{\hspost}@{}}%
\column{3}{@{}>{\hspre}l<{\hspost}@{}}%
\column{E}{@{}>{\hspre}l<{\hspost}@{}}%
\>[3]{}\Varid{repr}\ \mathop{:}\ \{\mskip1.5mu \Conid{X}\ \mathop{:}\ \Conid{Type}\mskip1.5mu\}\,\to\,\{\mskip1.5mu \Conid{M}\ \mathop{:}\ \Conid{Type}\,\to\,\Conid{Type}\mskip1.5mu\}\,\to\,\Conid{Monad}\;\Conid{M}\Rightarrow \Conid{MonSys}\;\Conid{M}\;\Conid{X}\,\to\,\Conid{DetSys}\;(\Conid{M}\;\Conid{X}){}\<[E]%
\\
\>[3]{}\Varid{repr}\;\Varid{f}\mathrel{=}\Varid{id}\mathbin{\mathbin{>}\hspace{-5.2pt}\mathrel{=}\hspace{-5.5pt}\mathbin{>}}\Varid{f}{}\<[E]%
\ColumnHook
\end{hscode}\resethooks
%
and for an arbitrary monadic system \ensuremath{\Varid{f}}, \ensuremath{\Varid{repr}\;\Varid{f}} is equivalent to \ensuremath{\Varid{f}}
in the sense that
\begin{hscode}\SaveRestoreHook
\column{B}{@{}>{\hspre}l<{\hspost}@{}}%
\column{3}{@{}>{\hspre}l<{\hspost}@{}}%
\column{14}{@{}>{\hspre}c<{\hspost}@{}}%
\column{14E}{@{}l@{}}%
\column{17}{@{}>{\hspre}l<{\hspost}@{}}%
\column{E}{@{}>{\hspre}l<{\hspost}@{}}%
\>[3]{}\Varid{reprLemma}{}\<[14]%
\>[14]{}\ \mathop{:}\ {}\<[14E]%
\>[17]{}\{\mskip1.5mu \Conid{X}\ \mathop{:}\ \Conid{Type}\mskip1.5mu\}\,\to\,\{\mskip1.5mu \Conid{M}\ \mathop{:}\ \Conid{Type}\,\to\,\Conid{Type}\mskip1.5mu\}\,\to\,\Conid{Monad}\;\Conid{M}\Rightarrow {}\<[E]%
\\
\>[17]{}(\Varid{f}\ \mathop{:}\ \Conid{MonSys}\;\Conid{M}\;\Conid{X})\,\to\,(\Varid{n}\ \mathop{:}\ \mathbb{N})\,\to\,\Varid{repr}\;(\Varid{flow}\;\Varid{f}\;\Varid{n})\doteq\Varid{flowDet}\;(\Varid{repr}\;\Varid{f})\;\Varid{n}{}\<[E]%
\ColumnHook
\end{hscode}\resethooks
where \ensuremath{\Varid{flowDet}} is the flow of a deterministic system
\begin{hscode}\SaveRestoreHook
\column{B}{@{}>{\hspre}l<{\hspost}@{}}%
\column{3}{@{}>{\hspre}l<{\hspost}@{}}%
\column{14}{@{}>{\hspre}c<{\hspost}@{}}%
\column{14E}{@{}l@{}}%
\column{17}{@{}>{\hspre}l<{\hspost}@{}}%
\column{23}{@{}>{\hspre}c<{\hspost}@{}}%
\column{23E}{@{}l@{}}%
\column{26}{@{}>{\hspre}l<{\hspost}@{}}%
\column{E}{@{}>{\hspre}l<{\hspost}@{}}%
\>[3]{}\Varid{flowDet}\ \mathop{:}\ \{\mskip1.5mu \Conid{X}\ \mathop{:}\ \Conid{Type}\mskip1.5mu\}\,\to\,\Conid{DetSys}\;\Conid{X}\,\to\,\mathbb{N}\,\to\,\Conid{DetSys}\;\Conid{X}{}\<[E]%
\\
\>[3]{}\Varid{flowDet}\;\Varid{f}\;{}\<[17]%
\>[17]{}\Conid{Z}{}\<[23]%
\>[23]{}\mathrel{=}{}\<[23E]%
\>[26]{}\Varid{id}{}\<[E]%
\\
\>[3]{}\Varid{flowDet}\;\Varid{f}\;{}\<[14]%
\>[14]{}({}\<[14E]%
\>[17]{}\Conid{S}\;\Varid{n}){}\<[23]%
\>[23]{}\mathrel{=}{}\<[23E]%
\>[26]{}\Varid{flowDet}\;\Varid{f}\;\Varid{n}\mathbin{\circ}\Varid{f}{}\<[E]%
\ColumnHook
\end{hscode}\resethooks
As for \ensuremath{\Varid{flowLemma2}}, proving the representation lemma is straightforward
but crucially relies on associativity of Kleisli composition and thus, as
seen in section \ref{section:monads}, on preservation of extensional
equality:
\pagebreak

\begin{hscode}\SaveRestoreHook
\column{B}{@{}>{\hspre}l<{\hspost}@{}}%
\column{3}{@{}>{\hspre}l<{\hspost}@{}}%
\column{5}{@{}>{\hspre}l<{\hspost}@{}}%
\column{22}{@{}>{\hspre}l<{\hspost}@{}}%
\column{46}{@{}>{\hspre}l<{\hspost}@{}}%
\column{50}{@{}>{\hspre}l<{\hspost}@{}}%
\column{E}{@{}>{\hspre}l<{\hspost}@{}}%
\>[3]{}\Varid{reprLemma}\;\Varid{f}\;\Conid{Z}\;{}\<[22]%
\>[22]{}\Varid{mx}\mathrel{=}\Varid{pureRightIdKleisli}\;\Varid{id}\;\Varid{mx}{}\<[E]%
\\[\blanklineskip]%
\>[3]{}\Varid{reprLemma}\;\Varid{f}\;(\Conid{S}\;\Varid{m})\;{}\<[22]%
\>[22]{}\Varid{mx}\mathrel{=}{}\<[E]%
\\
\>[3]{}\hsindent{2}{}\<[5]%
\>[5]{}(\Varid{repr}\;(\Varid{flow}\;\Varid{f}\;(\Conid{S}\;\Varid{m}))\;\Varid{mx}){}\<[46]%
\>[46]{}=\hspace{-3pt}\{\; \Conid{Refl}\;\}\hspace{-3pt}={}\<[E]%
\\
\>[3]{}\hsindent{2}{}\<[5]%
\>[5]{}((\Varid{id}\mathbin{\mathbin{>}\hspace{-5.2pt}\mathrel{=}\hspace{-5.5pt}\mathbin{>}}\Varid{flow}\;\Varid{f}\;(\Conid{S}\;\Varid{m}))\;\Varid{mx}){}\<[46]%
\>[46]{}=\hspace{-3pt}\{\; \Conid{Refl}\;\}\hspace{-3pt}={}\<[E]%
\\
\>[3]{}\hsindent{2}{}\<[5]%
\>[5]{}((\Varid{id}\mathbin{\mathbin{>}\hspace{-5.2pt}\mathrel{=}\hspace{-5.5pt}\mathbin{>}}(\Varid{f}\mathbin{\mathbin{>}\hspace{-5.2pt}\mathrel{=}\hspace{-5.5pt}\mathbin{>}}\Varid{flow}\;\Varid{f}\;\Varid{m}))\;\Varid{mx}){}\<[46]%
\>[46]{}=\hspace{-3pt}\{\; \Varid{sym}\;(\Varid{kleisliAssoc}\;\Varid{id}\;\Varid{f}\;(\Varid{flow}\;\Varid{f}\;\Varid{m})\;\Varid{mx})\;\}\hspace{-3pt}={}\<[E]%
\\
\>[3]{}\hsindent{2}{}\<[5]%
\>[5]{}(((\Varid{id}\mathbin{\mathbin{>}\hspace{-5.2pt}\mathrel{=}\hspace{-5.5pt}\mathbin{>}}\Varid{f})\mathbin{\mathbin{>}\hspace{-5.2pt}\mathrel{=}\hspace{-5.5pt}\mathbin{>}}\Varid{flow}\;\Varid{f}\;\Varid{m})\;\Varid{mx}){}\<[46]%
\>[46]{}=\hspace{-3pt}\{\; \Varid{kleisliLeapfrog}\;(\Varid{id}\mathbin{\mathbin{>}\hspace{-5.2pt}\mathrel{=}\hspace{-5.5pt}\mathbin{>}}\Varid{f})\;(\Varid{flow}\;\Varid{f}\;\Varid{m})\;\Varid{mx}\;\}\hspace{-3pt}={}\<[E]%
\\
\>[3]{}\hsindent{2}{}\<[5]%
\>[5]{}((\Varid{id}\mathbin{\mathbin{>}\hspace{-5.2pt}\mathrel{=}\hspace{-5.5pt}\mathbin{>}}\Varid{flow}\;\Varid{f}\;\Varid{m})\;((\Varid{id}\mathbin{\mathbin{>}\hspace{-5.2pt}\mathrel{=}\hspace{-5.5pt}\mathbin{>}}\Varid{f})\;\Varid{mx})){}\<[46]%
\>[46]{}=\hspace{-3pt}\{\; \Conid{Refl}\;\}\hspace{-3pt}={}\<[E]%
\\
\>[3]{}\hsindent{2}{}\<[5]%
\>[5]{}(\Varid{repr}\;(\Varid{flow}\;\Varid{f}\;\Varid{m})\;((\Varid{id}\mathbin{\mathbin{>}\hspace{-5.2pt}\mathrel{=}\hspace{-5.5pt}\mathbin{>}}\Varid{f})\;\Varid{mx})){}\<[46]%
\>[46]{}=\hspace{-3pt}\{\; \Varid{reprLemma}\;\Varid{f}\;\Varid{m}\;((\Varid{id}\mathbin{\mathbin{>}\hspace{-5.2pt}\mathrel{=}\hspace{-5.5pt}\mathbin{>}}\Varid{f})\;\Varid{mx})\;\}\hspace{-3pt}={}\<[E]%
\\
\>[3]{}\hsindent{2}{}\<[5]%
\>[5]{}(\Varid{flowDet}\;(\Varid{repr}\;\Varid{f})\;\Varid{m}\;((\Varid{id}\mathbin{\mathbin{>}\hspace{-5.2pt}\mathrel{=}\hspace{-5.5pt}\mathbin{>}}\Varid{f})\;\Varid{mx})){}\<[46]%
\>[46]{}=\hspace{-3pt}\{\; \Conid{Refl}\;\}\hspace{-3pt}={}\<[E]%
\\
\>[3]{}\hsindent{2}{}\<[5]%
\>[5]{}(\Varid{flowDet}\;(\Varid{repr}\;\Varid{f})\;\Varid{m}\;(\Varid{repr}\;\Varid{f}\;\Varid{mx})){}\<[46]%
\>[46]{}=\hspace{-3pt}\{\; {}\<[50]%
\>[50]{}\Conid{Refl}\;\}\hspace{-3pt}={}\<[E]%
\\
\>[3]{}\hsindent{2}{}\<[5]%
\>[5]{}(\Varid{flowDet}\;(\Varid{repr}\;\Varid{f})\;(\Conid{S}\;\Varid{m})\;\Varid{mx})\;{}\<[46]%
\>[46]{}\Conid{QED}{}\<[E]%
\ColumnHook
\end{hscode}\resethooks
Notice also the application of \ensuremath{\Varid{kleisliLeapfrog}} to deduce \ensuremath{(\Varid{id}\mathbin{\mathbin{>}\hspace{-5.2pt}\mathrel{=}\hspace{-5.5pt}\mathbin{>}}\Varid{flow}\;\Varid{f}\;\Varid{m})\;((\Varid{id}\mathbin{\mathbin{>}\hspace{-5.2pt}\mathrel{=}\hspace{-5.5pt}\mathbin{>}}\Varid{f})\;\Varid{mx})} from \ensuremath{((\Varid{id}\mathbin{\mathbin{>}\hspace{-5.2pt}\mathrel{=}\hspace{-5.5pt}\mathbin{>}}\Varid{f})\mathbin{\mathbin{>}\hspace{-5.2pt}\mathrel{=}\hspace{-5.5pt}\mathbin{>}}\Varid{flow}\;\Varid{f}\;\Varid{m})\;\Varid{mx}}.
If we had formulated the theory in terms of bind instead of Kleisli
composition, the two expressions would have been intensionally equal.

\paragraph*{Flows and trajectories.}
Our last application of preservation of extensional
equality in the context of dynamical systems theory is a result about
flows and trajectories.
For a monadic system \ensuremath{\Varid{f}}, the trajectories of length \ensuremath{\Varid{n}\mathbin{+}\mathrm{1}} starting at
state \ensuremath{\Varid{x}\ \mathop{:}\ \Conid{X}} are
\begin{hscode}\SaveRestoreHook
\column{B}{@{}>{\hspre}l<{\hspost}@{}}%
\column{3}{@{}>{\hspre}l<{\hspost}@{}}%
\column{8}{@{}>{\hspre}c<{\hspost}@{}}%
\column{8E}{@{}l@{}}%
\column{10}{@{}>{\hspre}c<{\hspost}@{}}%
\column{10E}{@{}l@{}}%
\column{11}{@{}>{\hspre}l<{\hspost}@{}}%
\column{13}{@{}>{\hspre}l<{\hspost}@{}}%
\column{19}{@{}>{\hspre}l<{\hspost}@{}}%
\column{22}{@{}>{\hspre}c<{\hspost}@{}}%
\column{22E}{@{}l@{}}%
\column{25}{@{}>{\hspre}l<{\hspost}@{}}%
\column{39}{@{}>{\hspre}l<{\hspost}@{}}%
\column{E}{@{}>{\hspre}l<{\hspost}@{}}%
\>[3]{}\Varid{trj}{}\<[8]%
\>[8]{}\ \mathop{:}\ {}\<[8E]%
\>[11]{}\{\mskip1.5mu \Conid{M}\ \mathop{:}\ \Conid{Type}\,\to\,\Conid{Type}\mskip1.5mu\}\,\to\,\{\mskip1.5mu \Conid{X}\ \mathop{:}\ \Conid{Type}\mskip1.5mu\}\,\to\,\Conid{Monad}\;\Conid{M}\Rightarrow {}\<[E]%
\\
\>[11]{}\Conid{MonSys}\;\Conid{M}\;\Conid{X}\,\to\,(\Varid{n}\ \mathop{:}\ \mathbb{N})\,\to\,{}\<[39]%
\>[39]{}\Conid{X}\,\to\,\Conid{M}\;(\Conid{Vect}\;(\Conid{S}\;\Varid{n})\;\Conid{X}){}\<[E]%
\\
\>[3]{}\Varid{trj}\;\Varid{f}\;{}\<[13]%
\>[13]{}\Conid{Z}\;{}\<[19]%
\>[19]{}\Varid{x}{}\<[22]%
\>[22]{}\mathrel{=}{}\<[22E]%
\>[25]{}\Varid{map}\;(\Varid{x}\mathbin{::})\;(\Varid{pure}\;\Conid{Nil}){}\<[E]%
\\
\>[3]{}\Varid{trj}\;\Varid{f}\;{}\<[10]%
\>[10]{}({}\<[10E]%
\>[13]{}\Conid{S}\;\Varid{n})\;{}\<[19]%
\>[19]{}\Varid{x}{}\<[22]%
\>[22]{}\mathrel{=}{}\<[22E]%
\>[25]{}\Varid{map}\;(\Varid{x}\mathbin{::})\;((\Varid{f}\mathbin{\mathbin{>}\hspace{-5.2pt}\mathrel{=}\hspace{-5.5pt}\mathbin{>}}\Varid{trj}\;\Varid{f}\;\Varid{n})\;\Varid{x}){}\<[E]%
\ColumnHook
\end{hscode}\resethooks
In words, the trajectory obtained by making zero steps starting at \ensuremath{\Varid{x}}
is an \ensuremath{\Conid{M}}-structure containing just \ensuremath{[\mskip1.5mu \Varid{x}\mskip1.5mu]}.
To compute the trajectories for \ensuremath{\Conid{S}\;\Varid{n}} steps, we first bind the outcome
of a single step \ensuremath{\Varid{f}\;\Varid{x}\ \mathop{:}\ \Conid{M}\;\Conid{X}} into \ensuremath{\Varid{trj}\;\Varid{f}\;\Varid{n}}.
This results in an \ensuremath{\Conid{M}}-structure of vectors of length \ensuremath{\Varid{n}}.
Finally, we prepend these possible trajectories with the initial state
\ensuremath{\Varid{x}}.

Since \ensuremath{\Varid{trj}\;\Varid{f}\;\Varid{n}\;\Varid{x}} is an \ensuremath{\Conid{M}}-structure of vectors of states, we can
compute the last state of each trajectory.
It turns out that this is point-wise equal to \ensuremath{\Varid{flow}\;\Varid{f}\;\Varid{n}}:
\begin{hscode}\SaveRestoreHook
\column{B}{@{}>{\hspre}l<{\hspost}@{}}%
\column{3}{@{}>{\hspre}l<{\hspost}@{}}%
\column{17}{@{}>{\hspre}c<{\hspost}@{}}%
\column{17E}{@{}l@{}}%
\column{20}{@{}>{\hspre}l<{\hspost}@{}}%
\column{E}{@{}>{\hspre}l<{\hspost}@{}}%
\>[3]{}\Varid{flowTrjLemma}{}\<[17]%
\>[17]{}\ \mathop{:}\ {}\<[17E]%
\>[20]{}\{\mskip1.5mu \Conid{X}\ \mathop{:}\ \Conid{Type}\mskip1.5mu\}\,\to\,\{\mskip1.5mu \Conid{M}\ \mathop{:}\ \Conid{Type}\,\to\,\Conid{Type}\mskip1.5mu\}\,\to\,\Conid{Monad}\;\Conid{M}\Rightarrow {}\<[E]%
\\
\>[20]{}(\Varid{f}\ \mathop{:}\ \Conid{MonSys}\;\Conid{M}\;\Conid{X})\,\to\,(\Varid{n}\ \mathop{:}\ \mathbb{N})\,\to\,{}\<[E]%
\\
\>[20]{}\Varid{flow}\;\Varid{f}\;\Varid{n}\doteq\Varid{map}\;\{\mskip1.5mu \Conid{A}\mathrel{=}\Conid{Vect}\;(\Conid{S}\;\Varid{n})\;\Conid{X}\mskip1.5mu\}\;\Varid{last}\mathbin{\circ}\Varid{trj}\;\Varid{f}\;\Varid{n}{}\<[E]%
\ColumnHook
\end{hscode}\resethooks
To prove this result, we first derive the auxiliary lemma
\begin{hscode}\SaveRestoreHook
\column{B}{@{}>{\hspre}l<{\hspost}@{}}%
\column{3}{@{}>{\hspre}l<{\hspost}@{}}%
\column{5}{@{}>{\hspre}l<{\hspost}@{}}%
\column{7}{@{}>{\hspre}l<{\hspost}@{}}%
\column{17}{@{}>{\hspre}c<{\hspost}@{}}%
\column{17E}{@{}l@{}}%
\column{20}{@{}>{\hspre}l<{\hspost}@{}}%
\column{31}{@{}>{\hspre}c<{\hspost}@{}}%
\column{31E}{@{}l@{}}%
\column{E}{@{}>{\hspre}l<{\hspost}@{}}%
\>[3]{}\Varid{mapLastLemma}{}\<[17]%
\>[17]{}\ \mathop{:}\ {}\<[17E]%
\>[20]{}\{\mskip1.5mu \Conid{F}\ \mathop{:}\ \Conid{Type}\,\to\,\Conid{Type}\mskip1.5mu\}\,\to\,\{\mskip1.5mu \Conid{X}\ \mathop{:}\ \Conid{Type}\mskip1.5mu\}\,\to\,\{\mskip1.5mu \Varid{n}\ \mathop{:}\ \mathbb{N}\mskip1.5mu\}\,\to\,\Conid{Functor}\;\Conid{F}\Rightarrow {}\<[E]%
\\
\>[20]{}(\Varid{x}\ \mathop{:}\ \Conid{X})\,\to\,(\Varid{mvx}\ \mathop{:}\ \Conid{F}\;(\Conid{Vect}\;(\Conid{S}\;\Varid{n})\;\Conid{X}))\,\to\,{}\<[E]%
\\
\>[20]{}(\Varid{map}\;\Varid{last}\mathbin{\circ}\Varid{map}\;(\Varid{x}\mathbin{::}))\;\Varid{mvx}\mathrel{=}\Varid{map}\;\Varid{last}\;\Varid{mvx}{}\<[E]%
\\
\>[3]{}\Varid{mapLastLemma}\;\{\mskip1.5mu \Conid{X}\mskip1.5mu\}\;\{\mskip1.5mu \Varid{n}\mskip1.5mu\}\;\Varid{x}\;\Varid{mvx}{}\<[31]%
\>[31]{}\mathrel{=}{}\<[31E]%
\\
\>[3]{}\hsindent{2}{}\<[5]%
\>[5]{}(\Varid{map}\;\{\mskip1.5mu \Conid{A}\mathrel{=}\Conid{Vect}\;(\Conid{S}\;(\Conid{S}\;\Varid{n}))\;\Conid{X}\mskip1.5mu\}\;\Varid{last}\;(\Varid{map}\;(\Varid{x}\mathbin{::})\;\Varid{mvx})){}\<[E]%
\\
\>[5]{}\hsindent{2}{}\<[7]%
\>[7]{}=\hspace{-3pt}\{\; \Varid{sym}\;(\Varid{mapPresComp}\;\{\mskip1.5mu \Conid{A}\mathrel{=}\Conid{Vect}\;(\Conid{S}\;\Varid{n})\;\Conid{X}\mskip1.5mu\}\;\Varid{last}\;(\Varid{x}\mathbin{::})\;\Varid{mvx})\;\}\hspace{-3pt}={}\<[E]%
\\
\>[3]{}\hsindent{2}{}\<[5]%
\>[5]{}(\Varid{map}\;(\Varid{last}\mathbin{\circ}(\Varid{x}\mathbin{::}))\;\Varid{mvx}){}\<[E]%
\\
\>[5]{}\hsindent{2}{}\<[7]%
\>[7]{}=\hspace{-3pt}\{\; \Varid{mapPresEE}\;(\Varid{last}\mathbin{\circ}(\Varid{x}\mathbin{::}))\;\Varid{last}\;(\Varid{lastLemma}\;\Varid{x})\;\Varid{mvx}\;\}\hspace{-3pt}={}\<[E]%
\\
\>[3]{}\hsindent{2}{}\<[5]%
\>[5]{}(\Varid{map}\;\Varid{last}\;\Varid{mvx})\;\Conid{QED}{}\<[E]%
\ColumnHook
\end{hscode}\resethooks
where \ensuremath{\Varid{lastLemma}\;\Varid{x}\ \mathop{:}\ \Varid{last}\mathbin{\circ}(\Varid{x}\mathbin{::})\doteq\Varid{last}}.

In the implementation of \ensuremath{\Varid{mapLastLemma}} we have applied both
preservation of composition and preservation of extensional
equality.
With \ensuremath{\Varid{mapLastLemma}} in place, \ensuremath{\Varid{flowTrjLemma}} is readily implemented by
induction on the number of steps
\begin{hscode}\SaveRestoreHook
\column{B}{@{}>{\hspre}l<{\hspost}@{}}%
\column{3}{@{}>{\hspre}l<{\hspost}@{}}%
\column{5}{@{}>{\hspre}l<{\hspost}@{}}%
\column{21}{@{}>{\hspre}l<{\hspost}@{}}%
\column{29}{@{}>{\hspre}l<{\hspost}@{}}%
\column{43}{@{}>{\hspre}l<{\hspost}@{}}%
\column{47}{@{}>{\hspre}l<{\hspost}@{}}%
\column{52}{@{}>{\hspre}l<{\hspost}@{}}%
\column{55}{@{}>{\hspre}l<{\hspost}@{}}%
\column{E}{@{}>{\hspre}l<{\hspost}@{}}%
\>[3]{}\Varid{flowTrjLemma}\;\{\mskip1.5mu \Conid{X}\mskip1.5mu\}\;\Varid{f}\;\Conid{Z}\;\Varid{x}\mathrel{=}{}\<[E]%
\\
\>[3]{}\hsindent{2}{}\<[5]%
\>[5]{}(\Varid{flow}\;\Varid{f}\;\Conid{Z}\;\Varid{x}){}\<[21]%
\>[21]{}=\hspace{-3pt}\{\; \Conid{Refl}\;\}\hspace{-3pt}=(\Varid{pure}\;\Varid{x}){}\<[43]%
\>[43]{}=\hspace{-3pt}\{\; \Conid{Refl}\;\}\hspace{-3pt}={}\<[E]%
\\
\>[3]{}\hsindent{2}{}\<[5]%
\>[5]{}(\Varid{pure}\;(\Varid{last}\;(\Varid{x}\mathbin{::}\Conid{Nil}))){}\<[43]%
\>[43]{}=\hspace{-3pt}\{\; \Varid{sym}\;(\Varid{pureNatTrans}\;\Varid{last}\;(\Varid{x}\mathbin{::}\Conid{Nil}))\;\}\hspace{-3pt}={}\<[E]%
\\
\>[3]{}\hsindent{2}{}\<[5]%
\>[5]{}(\Varid{map}\;\Varid{last}\;(\Varid{pure}\;(\Varid{x}\mathbin{::}\Conid{Nil}))){}\<[43]%
\>[43]{}=\hspace{-3pt}\{\; \Varid{cong}\;{}\<[52]%
\>[52]{}\{\mskip1.5mu \Varid{f}\mathrel{=}\Varid{map}\;\Varid{last}\mskip1.5mu\}\;{}\<[E]%
\\
\>[52]{}(\Varid{sym}\;(\Varid{pureNatTrans}\;\{\mskip1.5mu \Conid{A}\mathrel{=}\Conid{Vect}\;\Conid{Z}\;\Conid{X}\mskip1.5mu\}\;(\Varid{x}\mathbin{::})\;\Conid{Nil}))\;\}\hspace{-3pt}={}\<[E]%
\\
\>[3]{}\hsindent{2}{}\<[5]%
\>[5]{}(\Varid{map}\;\Varid{last}\;(\Varid{map}\;(\Varid{x}\mathbin{::})\;(\Varid{pure}\;\Conid{Nil}))){}\<[43]%
\>[43]{}=\hspace{-3pt}\{\; \Conid{Refl}\;\}\hspace{-3pt}={}\<[E]%
\\
\>[3]{}\hsindent{2}{}\<[5]%
\>[5]{}(\Varid{map}\;\Varid{last}\;(\Varid{trj}\;\Varid{f}\;\Conid{Z}\;\Varid{x}))\;{}\<[43]%
\>[43]{}\Conid{QED}{}\<[E]%
\\[\blanklineskip]%
\>[3]{}\Varid{flowTrjLemma}\;\Varid{f}\;(\Conid{S}\;\Varid{m})\;\Varid{x}\mathrel{=}{}\<[E]%
\\
\>[3]{}\hsindent{2}{}\<[5]%
\>[5]{}(\Varid{flow}\;\Varid{f}\;(\Conid{S}\;\Varid{m})\;\Varid{x}){}\<[52]%
\>[52]{}=\hspace{-3pt}\{\; \Conid{Refl}\;\}\hspace{-3pt}={}\<[E]%
\\
\>[3]{}\hsindent{2}{}\<[5]%
\>[5]{}((\Varid{f}\mathbin{\mathbin{>}\hspace{-5.2pt}\mathrel{=}\hspace{-5.5pt}\mathbin{>}}\Varid{flow}\;\Varid{f}\;\Varid{m})\;\Varid{x}){}\<[29]%
\>[29]{}=\hspace{-3pt}\{\; \Varid{kleisliPresEE}\;{}\<[47]%
\>[47]{}\Varid{f}\;\Varid{f}\;{}\<[55]%
\>[55]{}(\Varid{flow}\;\Varid{f}\;\Varid{m})\;(\Varid{map}\;(\Varid{last}\;\{\mskip1.5mu \Varid{len}\mathrel{=}\Varid{m}\mskip1.5mu\})\mathbin{\circ}\Varid{trj}\;\Varid{f}\;\Varid{m})\;{}\<[E]%
\\
\>[47]{}\Varid{reflEE}\;{}\<[55]%
\>[55]{}(\Varid{flowTrjLemma}\;\Varid{f}\;\Varid{m})\;\Varid{x}\;\}\hspace{-3pt}={}\<[E]%
\\[\blanklineskip]%
\>[3]{}\hsindent{2}{}\<[5]%
\>[5]{}((\Varid{f}\mathbin{\mathbin{>}\hspace{-5.2pt}\mathrel{=}\hspace{-5.5pt}\mathbin{>}}\Varid{map}\;(\Varid{last}\;\{\mskip1.5mu \Varid{len}\mathrel{=}\Varid{m}\mskip1.5mu\})\mathbin{\circ}\Varid{trj}\;\Varid{f}\;\Varid{m})\;\Varid{x}){}\<[52]%
\>[52]{}=\hspace{-3pt}\{\; \Varid{sym}\;(\Varid{mapKleisliLemma}\;\Varid{f}\;(\Varid{trj}\;\Varid{f}\;\Varid{m})\;\Varid{last}\;\Varid{x})\;\}\hspace{-3pt}={}\<[E]%
\\
\>[3]{}\hsindent{2}{}\<[5]%
\>[5]{}(\Varid{map}\;\Varid{last}\;((\Varid{f}\mathbin{\mathbin{>}\hspace{-5.2pt}\mathrel{=}\hspace{-5.5pt}\mathbin{>}}\Varid{trj}\;\Varid{f}\;\Varid{m})\;\Varid{x})){}\<[52]%
\>[52]{}=\hspace{-3pt}\{\; \Varid{sym}\;(\Varid{mapLastLemma}\;\Varid{x}\;((\Varid{f}\mathbin{\mathbin{>}\hspace{-5.2pt}\mathrel{=}\hspace{-5.5pt}\mathbin{>}}\Varid{trj}\;\Varid{f}\;\Varid{m})\;\Varid{x}))\;\}\hspace{-3pt}={}\<[E]%
\\
\>[3]{}\hsindent{2}{}\<[5]%
\>[5]{}(\Varid{map}\;\Varid{last}\;(\Varid{map}\;(\Varid{x}\mathbin{::})\;((\Varid{f}\mathbin{\mathbin{>}\hspace{-5.2pt}\mathrel{=}\hspace{-5.5pt}\mathbin{>}}\Varid{trj}\;\Varid{f}\;\Varid{m})\;\Varid{x}))){}\<[52]%
\>[52]{}=\hspace{-3pt}\{\; \Conid{Refl}\;\}\hspace{-3pt}={}\<[E]%
\\
\>[3]{}\hsindent{2}{}\<[5]%
\>[5]{}(\Varid{map}\;\Varid{last}\;(\Varid{trj}\;\Varid{f}\;(\Conid{S}\;\Varid{m})\;\Varid{x}))\;\Conid{QED}{}\<[E]%
\ColumnHook
\end{hscode}\resethooks
Again, preservation of extensional equality proves essential for the induction step.

\paragraph*{Dynamic programming (DP).} The relationship between the
flow and the trajectory of a monadic dynamical system also plays a
crucial role in the \emph{semantic verification} of dynamic
programming.
DP \citep{bellman1957} is a method for solving sequential decision
problems.
These problems are at the core of many applications in economics,
logistics and computer science and are, in principle, well
understood~\citep{bellman1957, de_moor1995, 
  gnesi1981dynamic, 2017_Botta_Jansson_Ionescu}.

Proving that dynamic programming is semantically correct boils down to
showing that the value function \ensuremath{\Varid{val}} that is at the core of the
backwards induction algorithm of DP is extensionally equal to a
specification \ensuremath{\Varid{val'}}.

The \ensuremath{\Varid{val}} function of DP takes \ensuremath{\Varid{n}} policies or decision rules and is
computed by iterating \ensuremath{\Varid{n}} times a monadic dynamical system similar to
the function argument of \ensuremath{\Varid{flow}} but with an additional \emph{control}
argument.
At each iteration, a \emph{reward} function is mapped on the states
and the result is reduced with a \emph{measure} function.
In this computation, the measure function is applied a number of times
that is exponential in \ensuremath{\Varid{n}}.

By contrast, \ensuremath{\Varid{val'}} is computed by applying the measure function only
once, but to a structure of a size exponential in \ensuremath{\Varid{n}} that is obtained
by adding up the rewards along all the trajectories.

The equivalence between \ensuremath{\Varid{val}} and \ensuremath{\Varid{val'}} is established by structural
induction.
As in the \ensuremath{\Varid{flowTrjLemma}} discussed above, \ensuremath{\Varid{map}} preserving extensional
equality turns out to be pivotal in applying the induction hypothesis,
see \citet{brede2020} for details.

\section{Related work}\label{section:relatedwork}

As already mentioned in section \ref{section:about}, there is a large
body of literature that relates in some form to (the treatment of)
equality in intensional type theory.
Most of that work, however, is concerned with the theoretical study of
the Martin-L\"of identity type or with the implementation of variants
of type theory and thus very different in nature from the present
paper which takes a pragmatic user-level approach.

Closest to our approach from the theoretical point of view are perhaps
works on formalization in type theory using \emph{setoids}.
These were originally introduced by Bishop \citep{Bishop1967-BISFOC-3}
for his development of constructive mathematics, and studied in
\citep{hofmann1995extensional} for the treatment of weaker notions of
equality in intensional type theory.
Setoids are sets equipped with an equivalence relation and mappings
between setoids have to take equivalent arguments to equivalent
results.
The focus of our paper can thus be seen as one special case with
extensional equality of functions as the equivalence relation of
interest and thus its preservation as coherence condition on
mappings.
The price to pay when using a full-fledged setoid approach is the
presence of a potentially huge amount of additional proof obligations,
needed to coherently deal with sets (types) and their equivalence
relations -- this often is pointedly referred to as \emph{setoid hell}
(for instance in \citet{altenkirch_setoidhell}, but it seems to have
been used colloquially in the community for much longer).

Still, there are some large developments using setoids, e.g. the CoRN
library (formalizing constructive mathematics) by \citet{CoRN_library}
and the CoLoR library (for rewriting and termination) by
\citet{CoLoR_library} in Coq where the proof assistant provides the
user with some convenient tools for dealing with setoids
\citep{sozeau2010new}.
%
%
Setoids are also used in a number of formalizations of category
theory,
e.g.~\citep{huet_saibi,megasz_coq,wiegley_coq,carette_agda,hu2021formalizing}.

\emph{Homotopy Type Theory} with \emph{univalence} \citep{hottbook}
provides function extensionality as a byproduct.
However, in most languages (notably in Coq in which the Univalent
Foundations library \cite[UniMath]{UniMath} is developed), univalence
is still an axiom and thus blocks computation.
Moreover, univalence is incompatible with the principle of
\emph{Uniqueness of Identity Proofs} which e.g.\ in Idris is built in,
and in Agda has to be disabled using a special flag.
Finally, in \emph{Cubical Type Theory} \citep{cohenetal18:cubical}
function extensionality is provable because of the presence of the
\emph{interval primitive} and thus has computational content.
Cubical type theory has recently been implemented as a special version
of Agda \citep{cubicalagda2}.
Another (similar) version of homotopy type theory is implemented in
the theorem prover Arend \citep{arend_prover}.
However, it is not clear at the present stage how long it will take
for these advances in type theory to become available in mainstream
functional programming.

On the topic of interfaces (type classes) and their laws there is
related work in specifying \citep{janssonjeuring-dataconv}, rewriting
\citep{peytonjones2001playing}, testing
\citep{jeuringHaskell12ClassLaws} and proving
\citep{arvidssonetal19:typeclasslaws} type class laws in Haskell.
The equality challenges here are often related to the semantics of
non-termination as described in the Fast and Loose Reasoning paper
\citep{danielssonetal06:fastandloose}.
In a dependently typed setting there is related work on contrasting
the power of testing and proving, including Agda code for the Functor
interface with extensional equality for the identity and composition
preservation but not preservation of extensional equality
\citep{ionescujansson:LIPIcs:2013:3899}.

\citet{Carette_2014} nicely abstract the ideas about different
(minimal) interfaces that we only exemplified using verified monads.
Regarding the relation between different representations of monads, the reader
might contrast the approach in the UniMath
\citep[\texttt{CategoryTheory.Monads.KTriplesEquiv}]{UniMath} library with our approach
in section~\ref{section:monads}. The UniMath development is part of a
full-fledged formalization of category theory and relates ``the traditional view''
with the ``Wadler view'' of a monad (as we called them) via a weak equivalence of
categories. This approach is very satisfactory from an abstract mathematical
perspective.
Our equivalence result is much less general but still practically
relevant and more lightweight: although it requires considerations
about preservation of extensional equality, it does not require
stronger axioms like univalence or a larger conceptual framework.

\section{Conclusions, outlook}
\label{section:conclusions}

In dependently typed programming in the context of Martin-Löf type
theories \citep{martinlof1984, nordstrom1990programming}, the problem of
how to specify abstract data types for verified generic programming is
still not well understood.
In this work, we have shown that requiring functors to preserve
extensional equality of arrows yields abstract data types that are
strong enough to support the verification of non-trivial monadic laws
and of generic results in domain specific languages for dynamical system
and control theory.

We have shown that such a minimalist approach can be exploited to derive
results that otherwise would require enforcing the relationships between
monadic operators -- \ensuremath{\Varid{pure}}, bind, join, Kleisli composition, etc. --
through intensional equalities or, even worse, postulating function
extensionality or similar \emph{impossible} specifications.

As a consequence we have proposed to carefully distinguish between
functors whose associated \ensuremath{\Varid{map}} can be shown to preserve extensional
equality (and identity arrows and arrow composition) and functors for
which this is not the case.
%

Current work by two of the authors shows that preservation
of extensional equality is useful in designing a verified ADT
for Applicative functors \citep{mcbride2008applicative} and that
all Traversable functors satisfy \ensuremath{\Varid{mapPresEE}}.

We conjecture that carefully distinguishing between higher order
functions that can be shown to preserve extensional equality and higher
order functions for which this is not the case can pay high dividends
(in terms of concise and correct generic implementations and avoidance
of boilerplate code) also for other abstract data types.

\section*{Acknowledgments}
The authors thank the JFP editors and reviewers, whose comments have
lead to significant improvements of the original manuscript.

The work presented in this paper heavily relies on free software, among others on Coq, Idris, Agda, GHC, git, vi, Emacs, \LaTeX\ and on the FreeBSD and Debian GNU/Linux operating systems.
It is our pleasure to thank all developers of these excellent products.
This is TiPES contribution No 38. This project has received funding from
the European Union’s Horizon 2020 research and innovation programme
under grant agreement No 820970.

\subsection*{Conflicts of Interest}
None.

\bibliographystyle{jfplike}
\bibliography{references}
\label{lastpage01}
\end{document}

%% file: macros.TeX
\def\commentbegin{\quad\{\ }
\def\commentend{\}}